\documentclass{lmcs} 
\pdfoutput=1
\usepackage[utf8]{inputenc}

\usepackage{lastpage}
\lmcsdoi{22}{4}{7}
\lmcsheading{}{\pageref{LastPage}}{}{}%
{Sep.~03,~2025}{Oct.~21,~2026}{}

\usepackage{amssymb,amsmath,amsfonts,amstext,bussproofs,ifthen,multicol,listings}
\usepackage{color}


\usepackage{hyperref}

\theoremstyle{plain} 

\newcommand{\Leadsto}{\mathrm{leadsto}}
\newcommand{\Conv}{\mathrm{conv}}
\newcommand{\Div}{\mathrm{div}}
\newcommand{\Dlf}{\mathrm{dlf}}

\newcommand{\txt}[1]{\texttt{#1}}

\begin{document}

\title[TREBL Part I: Theory]{TREBL -- A Relative Complete Temporal Event-B Logic\texorpdfstring{\\}{. }Part I: Theory}

\thanks{The research of Flavio Ferrarotti has been funded by the Federal Ministry for Innovation, Mobility and Infrastructure (BMIMI), the Federal Ministry for Economy, Energy and Tourism (BMWET), and the State of Upper Austria in the frame of the COMET Module Dependable Production Environments with Software Security (DEPS) (FFG grant no. 888338) and the SCCH competence center INTEGRATE (FFG grant no. 892418) in the COMET - Competence Centers for Excellent Technologies Programme managed by Austrian Research Promotion Agency FFG.
}
\thanks{Part of this work is supported by the ANR project EBRP: EventB-Rodin-Plus under grant no. ANR-19-CE25-0010.}

\author[K.-D. Schewe]{Klaus-Dieter Schewe\lmcsorcid{0000-0002-8309-1803}}[a] 
\author[F. Ferrarotti]{Flavio Ferrarotti\lmcsorcid{0000-0003-2278-8233}}[b]
\author[P. Rivi\`{e}re]{Peter Rivi\`{e}re\lmcsorcid{0000-0002-2644-7471}}[a,c]
\author[N. K. Singh]{Neeraj Kumar Singh\lmcsorcid{0000-0002-1124-0179}}[a]
\author[G. Dupont]{Guillaume Dupont\lmcsorcid{0000-0002-9185-0515}}[a]
\author[Y. {A\"{\i}t Ameur}]{Yamine {A\"{\i}t Ameur}\lmcsorcid{0000-0003-4582-9712}}[a]

\address{Institut Nationale Polytechnique de Toulouse / ENSEEIHT, Toulouse, France}
\email{kd.schewe@gmail.com, \{nsingh,yamine\}@enseeiht.fr, guillaume.dupont@toulouse-inp.fr}

\address{Software Competence Centre Hagenberg, Hagenberg, Austria}	
\email{flavio.ferrarotti@scch.at}  

\address{Japan Advanced Institute of Science and Technology, Ishikawa, Japan}
\email{priviere@jaist.ac.jp}

\begin{abstract}
  \noindent The verification of liveness conditions is an important aspect of state-based rigorous methods. This article addresses the extension of the logic of Event-B to a powerful logic, in which properties of traces of an Event-B machine can be expressed. However, all formulae of this logic are still interpreted over states of an Event-B machine rather than traces. The logic exploits that for an Event-B machine $M$ a state $S$ determines all traces of $M$ starting in $S$. We identify a fragment called TREBL of this logic, in which all liveness conditions of interest can be expressed, and define a set of sound derivation rules for the fragment. We further show relative completeness of these derivation rules in the sense that for every valid entailment of a formula $\varphi$ one can find a derivation, provided the machine $M$ is sufficiently refined. The decisive property is that certain variant terms must be definable in the refined machine. We show that such refinements always exist. Throughout the article several examples from the field of security are used to illustrate the theory.
\end{abstract}

\maketitle

\section*{Introduction}\label{S:one}

State-based consistency conditions such as state or transition invariants are included in rigorous state-based methods such as B~\cite{abrial:2005}, 
Abstract State Machines (ASMs)~\cite{boerger:2003}, TLA\textsuperscript{+}~\cite{lamport:2002} or Event-B~\cite{abrial:2010}, just to mention the most important ones. The verification of such conditions is well supported by appropriate logics such as the logics for Event-B~\cite{schmalz:eth2012}, TLA\textsuperscript{+}~\cite{merz:cai2003}, 
deterministic ASMs~\cite{staerk:jucs2001} and arbitrary non-deterministic ASMs~\cite{ferrarotti:amai2018}. 
As the logics are  definitional extensions of first-order logic with types, they are complete~\cite{vaananen:sep2019}, which is a valuable asset for verification.

It is important to note that in these logics not only conditions over states can be expressed, but arbitrary conditions over finite sequences of states. However, the formulae are interpreted over states. This reflects the basic property of rigorous state-based methods that a state determines the traces starting in that state. For instance, for Event-B it is easy to express the firing of a particular event by means of a transition invariant.

Besides such consistency conditions other liveness conditions are likewise important. These conditions comprise among others (conditional) {\em progress}, i.e. any state satisfying a condition $\varphi$ is always followed eventually (i.e. some time later) by a state satisfying $\psi$ (for $\varphi = \textbf{true}$ we obtain unconditional progress, which is usually called {\em existence}), or {\em persistence}, i.e. eventually a condition $\varphi$ will hold forever. {\em Fairness} is a specific progress condition, in which $\psi$ is a transition condition expressing the firing of an event. The verification of liveness conditions requires reasoning about complete traces of a specification, i.e. they are intrinsically connected to temporal logic \cite{kroeger:2008}. 

Specific temporal logics that have been introduced for the verification of desirable liveness properties of sequential and concurrent (interleaved) systems are Linear-Time Temporal Logic (LTL)~\cite{pnueli:focs1977} and Computation Tree Logic (CTL)~\cite{clarke:lop1981}. For many liveness conditions it suffices to consider only the UNTIL-fragment of LTL~\cite{manna:scp1984}. These logics, however, have been defined as temporal \emph{propositional} logics, hence their expressiveness is limited: formulae are built over a fixed, finite set of atomic propositions, and a state is simply an assignment of truth values to these propositions. As a consequence, a transition condition that relates the value of a variable ranging over an unbounded domain across two consecutive states cannot be expressed directly. For instance, if an event performs the assignment $x := x + 1$ for a variable 
$x$ ranging over the natural numbers, this cannot be captured by a single LTL or CTL formula. Indeed an atomic proposition describes a property of one state in isolation, so relating the value of $x$ before and after the transition would require a separate case for every possible value of $x$, i.e. infinitely many cases. In particular, they require extensions when used in connection with any of the rigorous methods mentioned above, the states of which are given by typed variables ranging over arbitrary, possibly infinite domains. 
The same applies to extensions of these temporal logics such as HyperLTL, HyperCTL, HyperCTL$^*$ emphasising variables for runs and quantification over them~\cite{clarkson:post2014} as well as TeamLTL and TeamCTL emphasising team semantics, i.e. the interpretation of formulae over arbitrary sets of runs~\cite{gutsfeld:lics2022}. Although these logics add quantification over traces or team semantics over sets of traces, their atomic propositions remain of the same fixed, Boolean kind, so the limitation illustrated above persists for them as well.

Hoang and Abrial started to investigate the use of the UNTIL-fragment of LTL in connection with Event-B \cite{hoang:icfem2011}. Based on their work Rivi\`{e}re et al. provided a tool support using the EB4EB meta-theory in combination with RODIN \cite{riviere:nfm2023}. Defining a logic that integrates LTL with the logic of Event-B is rather straightforward; instead of propositions, formulae that can be defined for a given Event-B machine need to be considered. In doing so, NEXT-formulae are already covered by the logic of Event-B, and hence the omission of the NEXT modal operator does no harm\footnote{Note that a similar argument is used for the logic of TLA$^+$~\cite{merz:cai2003}.}.

Naturally, this integration yields a first-order modal logic (called LTL(EB) in~\cite{ferrarotti:foiks2024}), which cannot be complete. Nonetheless, Hoang and Abrial discovered a few derivation rules for invariance, existence, progress and persistence and proved their soundness, thus showing that if certain variant terms can be derived in an Event-B machine, such liveness properties can be verified. However, the decisive remaining question is, if there exists a set of derivation rules that is also complete, at least for a well-defined fragment of LTL(EB).

In \cite{ferrarotti:foiks2024} Ferrarotti et al. approached this problem by defining a fragment $\square$LTL of LTL(EB) using the type of formulae in invariance, existence, progress and persistence conditions as defining constructors. This enabled the soundness proofs from \cite{hoang:icfem2011} to be preserved with slight generalisations. In addition, a few standard derivation rules for modal formulae were added, and it was shown that the set of all these derivation rules is {\em relative complete} for the $\square$LTL fragment, a problem that was left open in \cite{hoang:icfem2011}. The meaning of relative completeness is that all valid formulae are derivable, if appropriate variant terms can be defined in Event-B machines. Furthermore, it was possible to prove that such variants always exist in conservatively refined machines. The proof of relative completeness was possible under the assumption that the machines are tail-homogeneous, which means that the intrinsic non-determinism in Event-B is restricted. The tail-homogeneity restriction is necessary, as a linear time temporal logic does not work well together with branching traces. That is, the logic $\square$LTL is sound and relative complete for proofs on sufficiently refined, tail-homogeneous Event-B machines.

\paragraph{Rationale and Scope.}

Classical temporal logics such as LTL, CTL, CTL$^*$, etc. are complete---for instance, it is well known that LTL can be embedded into monadic first-order logic---but their propositional nature severely limits their expressiveness for the verification of liveness properties in connection with state-based rigorous methods such as  B~\cite{abrial:2005}, Abstract State Machines (ASMs)~\cite{boerger:2003}, TLA\textsuperscript{+}~\cite{lamport:2002} or Event-B~\cite{abrial:2010}. If instead we consider temporal extensions of (at least) first-order logic, we obtain significantly enlarged temporal logics, but it is well known that such temporal logics cannot be complete. This choice between Scylla and Charybdis---lack of expressiveness or incompleteness---is seriously hampering verification of conditions that concern complete traces for any rigorous systems specifications.

At the core of the problem we find that temporal formulae are interpreted over arbitrary traces, i.e. sequences of states that are defined as propositional theories over some fixed signature\footnote{That is, we assume a fixed finite set $\{ p_1, \dots, p_n \}$ of propositional variables, and a state (also called {\em world}) is given by assigning truth values to these variables.}, with only few general restrictions. In rigorous state-based methods, however, there is no interest in arbitrary traces, but in traces that are defined by a specification. Therefore, the general idea is to combine non-propositional temporal logic with a restriction of interpretations to traces defined by a machine (specified by one of these methods). Better than this, the means for the specification of state transitions can be integrated into the logic---so the logic becomes even more powerful---and temporal formulae can be simply defined as shortcuts. Consequently, the interpretation over traces can be replaced by an interpretation over states (the start states of the traces). The details of this approach will be presented and discussed in this article. 

The most important consequence of this idea is that it allows us to prove relative completeness, where ``relative'' refers to assumptions about the specified machines, which can always be satisfied. This is the key problem addressed in this article. While we develop the theory on grounds of the Event-B method \cite{abrial:2010}, the general idea applies to other state-based rigorous methods as well. In fact, the integration of update sets and their effects is an intrinsic component of the logic of Abstract State Machines \cite{ferrarotti:amai2018}, but as long as the method provides a logic, in which properties of states and state transitions can be expressed, and permits the definition of update sets, it should not be a problem to adapt the results of this article accordingly. This has already been sketched for Abstract State Machines \cite{ferrarotti:foiks2024}, and it should also be possible to define relative complete temporal logics for B, TLA$^+$ and other methods.

\paragraph{Contribution of this Article.}
As emphasised above, the key problems to be addressed in a temporal Event-B logic are expressiveness and completeness. In particular, it is desirable to preserve a key feature of the logic of Event-B, i.e. to enable reasoning over traces without having to introduce traces as first-class objects in the logic. The objective of this article is therefore to polish and extend the work on the complete fragment of LTL(EB) \cite{ferrarotti:foiks2024}, in particular to get rid of restrictions concerning the (relative) completeness results, to highlight the applicability of the resulting temporal Event-B logic. In a forthcoming second part we will show how the logic can be effectively supported by the EB4EB meta-theory in connection with the RODIN tool. As reported in \cite{riviere:nfm2023} this has already been done for the fragment of the logic considered in \cite{ferrarotti:foiks2024}.

Concerning expressiveness we pick up a key idea from related work on ASMs. Differences between states are captured by {\em update sets} and these can also be made explicit for any pairs of states of an Event-B machine. Likewise, we can define the effect of an update set in the logic of Event-B, i.e. we can introduce formulae $\psi = [X]\varphi$ that express that after firing an update set $X$ a given formula $\varphi$ holds. Such update sets can even be extended from state transitions to transitions by multiple steps, and quantification over the number of states can be enabled.

Note that this defines a very powerful logic, in which arbitrary conditions over traces can be expressed, an approach that Lamport characterised as ``evil'' \cite{lamport:pnueli2010}. However, as in \cite{ferrarotti:abz2024} we only intend to exploit this logic to define the common temporal operators as shortcuts; all derivation rules can then be expressed over a sufficiently large temporal fragment. The advantage is that as for the logic of Event-B all formulae of the temporal fragment will be interpreted over states rather than traces, which allows us to solve the completeness problem.

Therefore, a key contribution is the definition of common temporal operators (ALWAYS $\square$, EVENTUALLY $\lozenge$, UNTIL $\mathcal{U}$) in the extension of the logic of Event-B. Then temporal formulae express properties of traces of Event-B machines, but are interpreted over states, which reflects the simple fact that a state of a given Event-B machine determines the set of traces starting in that state. We can also distinguish formulae that are to hold in all traces from those that are to hold in one trace or a selected set of traces. In doing so, the temporal semantics exploited in \cite{hoang:icfem2011,ferrarotti:foiks2024} is nonetheless preserved.

We proceed defining the TempoRal Event-B Logic (TREBL) as the fragment spanned by formulae of the logic of Event-B for a given machine and selected temporal operators. For this logic derivation rules are formulated, and their soundness is proven, which is just a slight generalisation of the work by Hoang and Abrial~\cite{hoang:icfem2011}, and Ferrarotti et al.  \cite{ferrarotti:foiks2024}. Some of these derivation rules involve variant terms, i.e. terms that take values in some well-founded set.

Finally, we prove the relative completeness of the defined set of derivation rules, i.e. every valid formula can also be derived, provided that the machine is ``sufficiently refined'', meaning that all necessary variant terms can be defined. A key step in the proof is to show that it is always possible to obtain such refinements.

We demonstrate the application of TREBL and its supporting tool using challenging desired properties in security. For instance, we see that in TREBL the non-interference problem \cite{biskup:2009,mclean:jcs1992} can be expressed by a rather simple invariance condition with a state transition formula, whereas in logics such as LTL, CTL and the like transition conditions can hardly be expressed.

The research reported in this article is a significant extension of previous work \cite{ferrarotti:foiks2024}. The logic has been extended to capture not only conditions that are to hold in all traces, but also those that are to hold in just one or in a well-defined subset of traces. All proofs are included, and the restrictions in the previously obtained relative completeness result in \cite{ferrarotti:foiks2024} have been removed.

\paragraph{Organisation of the Article.}

In Section \ref{sec:trebl} we introduce TREBL. As in \cite{ferrarotti:foiks2024} we first give a brief introduction of the logic of Event-B, and highlight specific formulae for state transitions, deadlock-freeness and event firing. Then we define the extensions mentioned above concerning update sets and effects of update sets with respect to single and multiple steps. This is used to define temporal operators as shortcuts of complex formulae of the extended Event-B logic, which gives rise to the definition of the temporal fragment TREBL. We show that by comparison with the semantics used in \cite{hoang:icfem2011,ferrarotti:foiks2024} we do not lose anything, but gain the interpretability of TREBL-formulae over states rather than traces. We show the usefulness of TREBL by several sophisticated examples with preference on security conditions.

Section \ref{sec:fragment} is dedicated to the soundness of TREBL. First we introduce several types of variant terms that allow certain temporal formulae to be derived. We complement variants for convergence and divergence by variants for eventuality. We prove that there always exist refinements in which needed variant terms can be defined, which shows that the corresponding temporal formulae can be mechanically proven, provided that appropriate variants are defined. We proceed with derivation rules for TREBL and the proof of their soundness, which slightly generalises the work in \cite{hoang:icfem2011,ferrarotti:foiks2024} and adopts extensions from \cite{ferrarotti:abz2024}.

In Section \ref{sec:complete} we show the relative completeness of the set of derivation rules, which generalises the main result of \cite{ferrarotti:foiks2024} and at the same time removes the restriction that traces must be tail-homogeneous. We illustrate the theory for selected application examples.

Section \ref{sec:discussion} provides a brief discussion of previous and related work, before we conclude with a summary and valuation of the main results.

\section{The Temporal Event-B Logic}\label{sec:trebl}

The logic of Event-B is the result of integrating set theory and first-order logic (or equivalently exploiting monadic second-order logic with definable sets, i.e. Henkin semantics). Formulae in the logic are always interpreted over states. We assume familiarity with the well-known Event-B method \cite{abrial:2010}, and only recall some basic concepts fixing the notation needed in the remainder of this article. Different to our approach in the previous conference paper \cite{ferrarotti:foiks2024} we first extend the logic of Event-B in such a way that conditions over traces of a machine can be expressed, while the interpretation over states is preserved. Then common temporal operators can be expressed as shortcuts of this extended logic. Nonetheless, the semantics used in \cite{ferrarotti:foiks2024,hoang:icfem2011} is preserved. Then TREBL is defined as a fragment of this extended Event-B logic.

\subsection{A Glimpse of Event-B}

The basic schema views an Event-B \emph{machine} as offering a set of \emph{events} $E$ (operations), which are executed one at a time, i.e. only one event can be fired at any time point. Events are usually parameterised, which results in invariant-preserving state transitions. The firing of events is the only means for updating states. 

A \emph{state} in Event-B is formed by a finite tuple of variables $\bar{v}$ taking values in certain sets with (explicitly or implicitly defined) auxiliary functions and predicates. We write $val_S(v)$ for the value of the state variable $v \in \bar{v}$ in the state $S$. Sets are usually specified in an Event-B \emph{context}, but syntactical details are of no importance here. 

Each \emph{event} $e_i \in E$ has the form $\mathbf{any} \, \bar{x} \, \mathbf{where} \, G_i(\bar{x}, \bar{v}) \, \mathbf{then} \, A_i(\bar{x}, \bar{v}, \bar{v}') \, \mathbf{end}$, 
where $\bar{x}$ are the \emph{parameters}, the first-order formula $G_i(\bar{x}, \bar{v})$ is the \emph{guard} and $A_i(\bar{x}, \bar{v}, \bar{v}')$ is the \emph{action} of~$e_i$.

An event $e_i$ is \emph{enabled} in state $S$ with variables $\bar{v}$, if there are values for its parameters $\bar{x}$ that make its guard $G_i(\bar{x}, \bar{v})$ hold in $S$.  Otherwise, the event is \emph{disabled}. If all events of a machine $M$ are disabled in a state $S$, then $M$ is said to be \emph{deadlocked} in $S$. 

An Event-B {\em action} $A_i(\bar{x}, \bar{v}, \bar{v}')$ is a list of assignments of the form 
\[ v_1, \dots, v_n :\!| \, P(\bar{x}, \bar{v}, (v_1', \dots, v_n')) \; , \]
where the $v_i$ are variables that occur in the tuple $\bar{v}$ of state variables and $P$ is a \emph{before-after predicate} relating the values of $v_i$ (before the action) and $v_i'$ (after the action). The values assigned to the $v_i$ are chosen non-deterministically\footnote{It is commonly assumed that this choice is external, i.e. the values are provided by the environment and not by the machine itself.} from the set of values  $(v_1', \dots, v_n')$ that satisfy $P(\bar{x}, \bar{v}, (v_1', \dots, v_n'))$. Thus, each action $A_i(\bar{x}, \bar{v}, v')$ corresponds to a before-after predicate $P_{A_i}(\bar{x}, \bar{v}, \bar{v}')$ formed by the conjunction of all before-after predicates of the assignments in $A_i(x, \bar{v}, (v_1', \dots, v_n'))$. These predicates are defined using the language of set theory, i.e. first order logic with the binary relation-symbol $\in$. There are several dedicated notations for (typed) set operations such as set comprehension, Cartesian product, etc. We do not discuss them here as they are definable in first-order logic and well known.   

An action also allows assignments of the form $v := E(\bar{x},\bar{v})$ and $v :\in E(\bar{x},\bar{v})$, where $E(\bar{x},\bar{v})$ is an expression. The assignment $v := E(\bar{x},\bar{v})$ deterministically assigns the value of $E(\bar{x},\bar{v})$ to $v$, while $v :\in E(\bar{x},\bar{v})$  non-deterministically assigns an element of the set $E(\bar{x},\bar{v})$ to $v$. However, these two assignments are merely shortcuts of assignments in the general form $v :\!| \, P(\bar{x}, \bar{v}, v')$ with before-after predicates $v' = E(\bar{x},\bar{v})$ and $v' \in E(\bar{x},\bar{v})$, respectively.  

The values of the variables in the initial state are set by means of a special event called  $\mathit{init}$ that has no parameters nor guard. The values that variables $\bar{v}$ can take in a state are constrained by invariants $I(\bar{v})$. These invariants are to hold in every reachable state, which is achieved by proving that they are \emph{established} by the initialisation event $\mathit{init}$ and subsequently \emph{preserved} by all other events.  

Given an event $e$, we say that a state $S^\prime$ is an $e$-successor state of $S$, if $S^\prime$ is a possible
after-state of the firing of $e$ from the before-state $S$. Lifting the definition to a machine $M$, we say that $S^\prime$ is an $M$-successor state of $S$ if there exists an event $e$ of $M$ such that $S^\prime$ is an $e$-successor of $S$.

Event-B defines a \emph{trace} $\tau$ of a machine $M$ as a (possibly infinite) sequence of states $S_0, S_1, S_2, \ldots$ such that $S_0$ is an initial state satisfying the after predicate defined by the event $\mathit{init}$, and for every pair of consecutive states $S_i$, $S_{i+1}$ in $\tau$, there is an event $e$ such that $S_{i+1}$ is an $e$-successor state of $S_i$. If the computation $\tau$ ends in a state $S_n$, i.e. the computation terminates, then $M$ must be deadlocked in $S_n$. For a trace $\tau = S_0, S_1, \ldots$ let $\ell(\tau) = n$, if $\tau$ is finite and has length $n$, and let $\ell(\tau) = \omega$ if it is infinite. Furthermore, let $\tau^{(k)}$ for $0 \leq k \leq \ell(\tau)$ denote the sequence $S_k, S_{k+1}, \ldots$ obtained from $\tau$ by removing its first $k$ elements.

\subsection{The Logic of Event-B}

Given a machine $M$ we immediately obtain a logic $\mathcal{L}$, in which properties of $M$ can be expressed. If the state variables of $M$ are $\bar{v}$, then the set of basic well-formed formulae of $\mathcal{L}$ consists of the set of first-order logic formulae over the signature $\tau$ of the machine $M$ with all free variables among those in $\bar{v}$. Such formulae are interpreted in any state $S$ of $M$, where the value of the free (state) variables is interpreted by the value of these variables in $S$. That is, in $\mathcal{L}$ we can express properties of states\footnote{Actually, the logic $\mathcal{L}$ permits the expression of more than just properties of states. Also properties of state transitions and thus also of successor states can be expressed, but the formulae are nonetheless interpreted over states. With a bit of technical effort one can even express properties of finite prefixes of traces, and for the interpretation the first state in such sequences is needed.}, and such formulae are commonly referred to as the {\em state formulae} of $M$.

We adopt common notation and write $S \models \varphi$ iff the formula $\varphi \in \mathcal{L}$ holds in the state $S$. Note that this requires that the free variables of $\varphi$ are among the state variables $\bar{v}$. A particular case arises for sentences: if $\varphi$ does not contain free variables, its validity does not depend on the state $S$.

For an arbitrary set $\Phi$ of $\mathcal{L}$ formulae we write $S \models \Phi$ iff $S \models \varphi$ holds for all $\varphi \in \Phi$. Furthermore, $\Phi \models \psi$ holds (for $\psi \in \mathcal{L}$ and $\Phi \subseteq \mathcal{L}$) iff whenever $S \models \Phi$ holds, then also $S \models \psi$ holds. For the special case $\Phi = \emptyset$ all states $S$ satisfy all formulae in $\Phi$, so $\emptyset \models \psi$---abbreviated as $\models \psi$---simply means that all states $S$ satisfy $\psi$. 

We write $M \models \varphi$ iff the initial state of the machine $M$ satisfies the formula $\varphi$. Note again that for sentences $\varphi$ we have $M \models \varphi$ iff $\models \varphi$ holds, i.e. the validity of the formula is independent from the states under consideration.

Let $E$ and $\bar{v}$ denote the set of events and the sequence of state variables of an Event-B machine $M$. For each $e_i \in E$ let $G_i(\bar{x}, \bar{v})$ and $P_{A_i}(\bar{x}, \bar{v}, \bar{v}')$ denote the guard and the before-after predicate corresponding to the action $A_i(\bar{x}, \bar{v}, \bar{v}')$ of event $e_i$, respectively. Then the formula $\text{enabled}_i(\bar{x}) \equiv G_i(\bar{x}, \bar{v})$ holds in a state $S$ iff the event $e_i$ with parameters $\bar{x}$ is {\em enabled} in $S$ and depending on the feasibility of the action can be fired to produce a successor state. Analogously, the formula
\begin{gather}
\text{fired}_i(\bar{x}) \equiv G_i(\bar{x}, \bar{v}) \wedge \bigwedge_{\genfrac{}{}{0pt}{2}{e_j \in E}{j \neq i}} \neg G_j(\bar{x}, \bar{v}) \label{eq-fired}
\end{gather}
expresses that the event $e_i$ is the only event enabled in $S$ (with parameters $\bar{x}$). According to the definition of traces this implies that if $S$ is a state in the trace, then the event $e_i$ is not only enabled in $S$, but will be fired. Such a condition is essential for the expression of fairness\footnote{Note that if for an event $e_i$ that is enabled in a state $S$ with some parameters $\bar{x}$ always another event $e_j$ is also enabled with the same parameters, then we can obtain traces, in which $e_i$ is never fired.} properties.

The logic $\mathcal{L}$ also contains formulae that express properties of state transitions, which are therefore called {\em state transition formulae}. For instance, for any state formula $\varphi$ we can define the first-order formulae
\begin{gather}
\mathcal{N}\varphi(\bar{v}) \equiv \bigwedge_{e_i \in E} \forall \bar{x} \Big ( G_i(\bar{x}, \bar{v}) \rightarrow \forall \bar{v}^\prime ( P_{A_i}(\bar{x}, \bar{v}, \bar{v}^\prime) \rightarrow \varphi(\bar{v}^\prime) ) \Big ) \label{eq-next}
\end{gather}
and
\begin{gather}
\mathcal{N}^1\varphi(\bar{v}) \equiv \bigvee_{e_i \in E} \exists \bar{x} \Big ( G_i(\bar{x}, \bar{v}) \wedge \exists \bar{v}^\prime ( P_{A_i}(\bar{x}, \bar{v}, \bar{v}^\prime) \wedge \varphi(\bar{v}^\prime) ) \Big ) \; .\label{eq-next1}
\end{gather}

\noindent
The formula $\mathcal{N}\varphi$ holds in a state $S$, if the formula $\varphi$ holds in all $M$-successor states of $S$, and the formula $\mathcal{N}^1\varphi$ holds in a state $S$, if the formula $\varphi$ holds in at least one $M$-successor state of $S$.

Other important state or transition formulae in $\mathcal{L}$ that already appear in proof rules from Hoang and Abrial \cite{hoang:icfem2011} are $\Leadsto(\varphi_1, \varphi_2)$ and $\Dlf(\varphi)$. For these let $M$ be an Event-B machine with a set $E$ of events and state variables $\bar{v}$. 
For state formulae $\varphi_1$ and $\varphi_2$ we say that \emph{$M$ leads from $\varphi_1$ to $\varphi_2$} iff the following sentence holds:
\begin{gather}
\Leadsto(\varphi_1, \varphi_2) \equiv \forall \bar{v} \bigg( \varphi_1(\bar{v}) \rightarrow \bigwedge_{e_i \in E} \forall \bar{x} \bar{v}'  \Big( \varphi_1(\bar{v}) \wedge G_i(\bar{x}, \bar{v}) \wedge P_{A_i}(\bar{x}, \bar{v}, \bar{v}') \rightarrow \varphi_2(\bar{v}') \Big) \bigg).  \label{eq-leadsto}
\end{gather}

\noindent
Note that it only makes sense to have $\models \Leadsto(\varphi_1, \varphi_2)$, as the formula contains a quantification over the state variables $\bar{v}$. If $\models \Leadsto(\varphi_1, \varphi_2)$ holds, then in particular for every trace $\tau = S_0, S_1, S_2, \ldots$ of $M$ and every $0 \leq k < \ell(\tau)$ it holds that whenever $S_k \models \varphi_1$ then $S_{k+1} \models \varphi_2$. However, the formula $\Leadsto(\varphi_1, \varphi_2)$ expresses more than that, as there may exist states that satisfy $\varphi_1$, but do not appear in any trace.

Therefore, it makes sense to also define a state-dependent formula $\text{leadslocto}(\varphi_1, \varphi_2)$, which is not a sentence. We say that $M$ {\em leads from $\varphi_1$ to $\varphi_2$ in state $S$} iff the following formula holds for the state $S$ with state variables $\bar{v}$:
\begin{gather}
\text{leadslocto}(\varphi_1, \varphi_2) \equiv \varphi_1(\bar{v}) \rightarrow \bigwedge_{e_i \in E} \forall \bar{x} \bar{v}'  \Big( G_i(\bar{x}, \bar{v}) \wedge P_{A_i}(\bar{x}, \bar{v}, \bar{v}') \rightarrow \varphi_2(\bar{v}') \Big).  \label{eq-leadslocto}
\end{gather}

\noindent
We can also define the analogous sentence
\begin{gather}
\Leadsto^1(\varphi_1, \varphi_2) \equiv \forall \bar{v} \bigg( \varphi_1(\bar{v}) \rightarrow \bigvee_{e_i \in E} \forall \bar{x} \bar{v}'  \Big( G_i(\bar{x}, \bar{v}) \wedge P_{A_i}(\bar{x}, \bar{v}, \bar{v}') \rightarrow \varphi_2(\bar{v}') \Big) \bigg) \; , \label{eq-leadsto1}
\end{gather}
which expresses that whenever $\varphi_1$ holds in some state $S_k$ of $M$, then any successor state $S_{k+1}$ will satisfy $\varphi_2$, regardless, whether $S_k$ occurs in some trace $\tau$ of $M$ (with $k < \ell(\tau)$) or not,
as well as the ``local'' formula
\begin{gather}
\text{leadslocto}^1(\varphi_1, \varphi_2) \equiv \varphi_1(\bar{v}) \rightarrow \bigvee_{e_i \in E} \forall \bar{x} \bar{v}'  \Big( G_i(\bar{x}, \bar{v}) \wedge P_{A_i}(\bar{x}, \bar{v}, \bar{v}') \rightarrow \varphi_2(\bar{v}') \Big) \; , \label{eq-leadslocto1}
\end{gather}
which expresses that if $\varphi_1$ holds in the state $S$ of $M$, then there is a successor state $S^\prime$ satisfying $\varphi_2$.

For a state formula $\varphi$ we say that \emph{$M$ is deadlock-free in $\varphi$} iff the following sentence is satisfied:
\begin{gather}
\Dlf(\varphi) \equiv \varphi(\bar{v}) \rightarrow \bigvee_{e_i \in E} \exists \bar{x} \; G_i(\bar{x}, \bar{v})  . \label{eq-dlf}
\end{gather}

\noindent
It expresses that whenever a state satisfies $\varphi$, there is at least one event $e_i$ of $M$ that is enabled. 

Note that this definition is different from the one given by Hoang and Abrial \cite{hoang:icfem2011}, which involves a quantification over state variables, and thus satisfaction means that at least one event $e_i$ of $M$ is enabled in every state $S$, in particular (but not only) in all states in a trace $\tau$. Then $S_{\ell(\tau)-1} \models \neg \varphi$ holds for every finite trace $\tau = S_0, \dots, S_{\ell(\tau)-1}$ of $M$. 

\begin{exa}
Consider a simple Event-B machine $M$ with a single state variable $x \in \mathbb{N}$,
initialized to $x := 0$, and two events:

\noindent
\begin{minipage}{0.48\textwidth}
\begin{lstlisting}[mathescape=true]
$\textbf{inc}$
$\textbf{when}$ $x < N$ $\textbf{then}$ $x := x + 1$ $\textbf{end}$
\end{lstlisting}
\end{minipage}
\hfill
\begin{minipage}{0.48\textwidth}
\begin{lstlisting}[mathescape=true]
$\textbf{dec}$
$\textbf{when}$ $x > 0$ $\textbf{then}$ $x := x - 1$ $\textbf{end}$
\end{lstlisting}
\end{minipage}
\noindent
where $N \in \mathbb{N}$ is a constant. The state formulae of $\mathcal{L}$ are
first-order formulae over the variable $x$, such as $x = 0$, $x > 0$, or $x \leq N$.
 
\medskip
\noindent\emph{Next-state operator.}
The formula $\mathcal{N}(x \leq N)$ holds in a state $S$ iff every successor state
satisfies $x \leq N$. By~(\ref{eq-next}), this unfolds to:
\begin{equation*}
(x < N \rightarrow x + 1 \leq N) \;\wedge\; (x > 0 \rightarrow x - 1 \leq N).
\end{equation*}
Both conjuncts hold for all $x \in \mathbb{N}$, so $\models \mathcal{N}(x \leq N)$,
confirming that $x \leq N$ is preserved by every event of $M$.
 
\medskip
\noindent\emph{Leads-to.}
The sentence $\Leadsto(x = k,\; x = k{+}1 \vee x = k{-}1)$ expresses that from any
state with $x = k$, every successor state has $x = k \pm 1$. This holds for all
$0 < k < N$, since \textbf{inc} increments $x$ by one and \textbf{dec} decrements $x$
by one, so no event changes $x$ by more than one in a single step.
 
\medskip
\noindent\emph{Deadlock-freedom.} The machine $M$ is deadlock-free provided $N > 0$: in any state with $0 \leq x \leq N$,
either $x < N$ enables \textbf{inc}, or $x > 0$ enables \textbf{dec}, or both.
Formally, $\models \Dlf(0 \leq x \leq N)$ follows from~(\ref{eq-dlf}) under the
assumption $N > 0$. If $N = 0$, the machine deadlocks immediately in the initial
state $x = 0$, as neither event is enabled.
\end{exa}

\subsection{Invariants and Reachability}

In addition, the specification of an Event-B machine $M$ always includes an {\em invariant}, i.e. a state formula $\iota(\bar{v})$ in the logic $\mathcal{L}$. The invariant must be satisfied in all states in all traces of $M$. This is intrinsically coupled with a {\em consistency proof obligation}, which is to be proven inductively. That is, the invariant must be satisfied by the initial state of $M$, which requires
\begin{gather}
M \models \iota(\bar{v}) \; , \label{eq-consistency-pa1}
\end{gather}
and whenever a state $S$ of $M$ satisfies the invariant $\iota$, then every successor state $S^\prime$ of $S$---produced by firing one of the events of $M$---also satisfies $\iota$, i.e.
\begin{gather}
M \models \forall \bar{v} \big( \iota(\bar{v}) \rightarrow \mathcal{N} \iota(\bar{v}) \big) \; . \label{eq-consistency-pa2}
\end{gather}

\noindent
Note that the formula in (\ref{eq-consistency-pa2}) is a sentence. A machine $M$ satisfying (\ref{eq-consistency-pa1}) and (\ref{eq-consistency-pa2}) is called {\em consistent}.

In Event-B we are only interested in consistent machines, so the consistency proof obligation is integrated into common support tools such as RODIN, and a proof must be given---the common terminology is to {\em discharge} the consistency proof obligation.

However, a proof of (\ref{eq-consistency-pa1}) and (\ref{eq-consistency-pa2}) shows more we need. Condition (\ref{eq-consistency-pa2}) requires the preservation of consistency with respect to $\iota$ by all events in all states $S$ of $M$, but there may exist states $S$ satisfying $\iota$ that are not {\em reachable} from the initial state, i.e. $S$ does not occur in a trace of $M$, and for such states consistency preservation is obsolete.

Unfortunately, reachability of a state is a property that is undecidable in general\footnote{This is related to the problem that NEXT is not preserved by refinement in Event-B}. Hence there is no way to replace (\ref{eq-consistency-pa2}) by a condition that excludes non-reachable states, so the consistency proof obligation formulated above is the only feasible way to handle consistency\footnote{Note that the problem of being unable to restrict the focus to reachable states is present also for all other state-based rigorous methods.}. Phrased differently, while the logic $\mathcal{L}$ of Event-B is known to be complete, this does not cover conditions that are to hold in all reachable states. 

We have to keep this in mind, when we prove relative completeness for the temporal extension TREBL of $\mathcal{L}$. We will only consider machines $M$ that are consistent with respect to the invariant $\iota$ that is an intrinsic part of the specification of $M$. We will further assume that any state formula $\varphi$ of $\mathcal{L}$ that is to hold in all reachable states of $M$ is made part of the invariant $\iota$ (using conjunction). For instance, if we need $\text{leadslocto}(\varphi_1, \varphi_2)$ to hold in all reachable states, we assume that the formula is included in $\iota$, so by (\ref{eq-consistency-pa2}) de facto $\Leadsto(\iota\wedge\varphi_1, \varphi_2)$ will be proven.

\subsection{Multiple-Step Extension}

We want to extend the logic of Event-B in such a way that we can express conditions over traces. For this we first make the effect of firing an event $e_i$ of an Event-B machine $M$ explicit by means of update sets. If in some state $S$ the event $e_j \in E$ is fired with parameters $\bar{x}$, then the corresponding {\em update set} is the set of all pairs $(\mathbf{v}_i, v_i^\prime)$, where the $\mathbf{v}_i$ ($1 \le i \le n$) are the state variables (not their values in $S$) of $M$, and the $v_i^\prime$ are the selected values satisfying the before-after-predicate $P_{A_j}(\bar{x}, \bar{v}, \bar{v}^\prime)$ of the event $e_j$. In order to represent such sets in the logic we need new constants\footnote{Note that for the interpretation of logical formulae (in $\mathcal{L}$ and in extensions) we need to have a domain of values, but the state variables $\mathbf{v}_i$ are syntactic constants that can be used to define terms, but cannot appear in the interpretation. Update sets, however, contain locations, i.e. state variables in Event-B, and thus they must be represented by set values in the domain used for the interpretation.} $c_i$ that are in a one-to-one correspondence with the state variables $\mathbf{v}_i$ ($1 \le i \le n$). Then the formula
\begin{gather}
\text{USet}_{e_i}(\bar{x}, X) \equiv G_i(\bar{x}, \bar{v}) \wedge P_{A_i}(\bar{x}, \bar{v}, \bar{v}^\prime) \wedge \forall c, c^\prime ( X(c, c^\prime) \leftrightarrow \bigvee_{1 \le i \le n} (c = c_i \wedge c^\prime = v_i^\prime) ) \label{eq-eventuset}
\end{gather}
holds in a state $S$, if $X$ is the update set of the event $e_i$ with parameters $\bar{x}$, where $\mathbf{v}_1, \dots, \mathbf{v}_n$ are the state variables and $\bar{v}^\prime = (v_1^\prime, \dots, v_n^\prime)$ are the values selected to fulfil the before-after predicate $P_{A_i}$ of the event $e_i$. Furthermore, the formula
\begin{gather}
\text{USet}(X) \equiv \bigvee_{e_i \in E} \exists \bar{x} ( \text{USet}_{e_i}(\bar{x}, X) ) \label{eq-uset}
\end{gather}
holds in state $S$, if $X$ is a possible update set of $M$ in that state. Note that as non-determinism is permitted, there is usually more than one update set in a state $S$, but every such update set is intrinsically connected to one of the possible successor states.

In addition to these formulae that express that update sets are yielded by a machine in a state, we need formulae that express the effect of applying an update set. First, if $S$ is a state of an Event-B machine $M$ and $\Delta$ is an update set for $M$, then $S + \Delta$ denotes the well-defined state $S^\prime$, in which $val_{S^\prime}(\mathbf{v}_i) = v_i^\prime$ holds for all $(\mathbf{v}_i, v_i^\prime) \in \Delta$, and $val_{S^\prime}(\mathbf{v}_i) = val_S(\mathbf{v}_i)$ holds, if no such pair exists in $\Delta$. Here we use the common notation that $val_S(\mathbf{v}_i)$ denotes the value of the state variable $\mathbf{v}_i$ in the state $S$.

If $X$ is a set variable\footnote{Note that variables $X$ representing update sets are set variables of arity 2, but the logic may use other set variables as well.} in the logic $\mathcal{L}$ and the value of $X$ in a state $S$ under a variable assignment\footnote{As common in logic $\zeta$ assigns values in the domain to variables in formulae. Here we also have set variables of some arity, so set variables of arity 2 are assigned sets of pairs, and if the first component of all such pairs are constant $c_i$ representing the state variables $\mathbf{v}_i$, then the assigned value represents an update set.} $\zeta$ is a set of pairs $(c_i, c_i^\prime)$ with the specific constants $c_i$ introduced above and there are no two different pairs $(c_i, c_i^j) \in val_{S,\zeta}(X)$ ($j = 1,2$), then we say that $val_{S,\zeta}(X)$ {\em represents the update set} $\Delta = \{ (\mathbf{v}_i, c_i^\prime) \mid (c_i, c_i^\prime) \in val_{S,\zeta}(X) \}$. In this way we capture the subtle distinction between an update set $\Delta$, which is an extra-logical concept, as it contains the state variables $\mathbf{v}_i$, and a representing set value $val_{S,\zeta}(X)$ in the domain for the interpretation.

The effect of applying an update set is then expressible by additional modal formulae $[X] \varphi$, where $X$ is a set variable and $\varphi$ is a formula of the logic $\mathcal{L}$. For the interpretation of such a formula in a state $S$ with a variable assignment $\zeta$ we obtain \textbf{false}, if $val_{S,\zeta}(X)$ represents the update set $\Delta$ and the interpretation of $\varphi$ in the state $S + \Delta$ with assignment $\zeta$ gives \textbf{false}. In all other cases we obtain \textbf{true}. 

Note that according to this definition we obtain \textbf{true}, if $val_{S,\zeta}(X)$ represents the update set $\Delta$ and the interpretation of $\varphi$ in the state $S + \Delta$ with assignment $\zeta$ gives \textbf{false}. However, we also obtain \textbf{true}, if $val_{S,\zeta}(X)$ does not represent an update set. At first sight this appears a bit strange. However, if we do not have an update set, then there is no successor state, so it makes sense to define that any formula will hold.

These additional formulae have been adopted from the logic of non-deterministic ASMs \cite{ferrarotti:amai2018}. This is easily possible, because Event-B supports the definition of arbitrary sets. 

We can now go one step further and consider update sets that combine the updates of multiple steps. For this we use non-negative integers $k$ in formulae $\text{USet}^*(k,X)$, which are defined inductively by $\text{USet}^*(0,X) \leftrightarrow X = \emptyset$ and 
\begin{gather}
\text{USet}^*(k+1, X) \leftrightarrow \exists Y, Z \big( \text{USet}(Y) \wedge [Y] \text{USet}^*(k,Z) \wedge \text{comp}(Y,Z,X) \big) \; , \label{eq-uset*}
\end{gather}
where the formula
\begin{gather}
\text{comp}(Y,Z,X) \equiv \forall c, c^\prime \big( X(c,c^\prime) \leftrightarrow ( Y(c,c^\prime) \wedge \neg\exists c^{\prime\prime} ( Z(c, c^{\prime\prime}) ) ) \vee Z(c, c^\prime) \big) \label{eq-comp}
\end{gather}
expresses that $X$ is the sequential composition of $Y$ followed by $Z$. 

Due to the interpretation of $[Y]\varphi$ above the case of finite traces is covered. If the length of the trace is exceeded, the subformula $[Y] \text{USet}^*(k,Z)$ will be interpreted as \textbf{true}, and then $\text{USet}^*(k+1, X)$ can only be interpreted as \textbf{false}, i.e. there is no update set capturing the updates of $k+1$ steps.

We extend the logic $\mathcal{L}$ by the formulae $\text{USet}^*(k,X)$---there is no need to use $\text{USet}(X)$, as it is equivalent to $\text{USet}^*(1,X)$---as well as $[X]\varphi$, and we permit existential and universal quantification over the variables in such formulae. The resulting logic will be denoted as $\mathcal{L}^{ext}$. Note that still all formulae in $\mathcal{L}^{ext}$ are interpreted over states. Nonetheless, they can be exploited to express properties of traces. In the following we will write $M \models \varphi$ for formulae $\varphi \in \mathcal{L}^{ext}$ whenever we have $S_0 \models \varphi$ for any initial state $S_0$ of $M$.

The following is an informal summary of the interpretation of the formulae in $\mathcal{L}^{ext}$:

\begin{itemize}

\item The logic $\mathcal{L}^{ext}$ contains all formulae $\varphi$ of the logic $\mathcal{L}$. Then $S \models \varphi$ holds iff $\varphi$ is interpreted as \textbf{true}, when each state variable $\mathbf{v}$ is interpreted by its value in the state $S$.

\item The logic $\mathcal{L}^{ext}$ contains formulae $\varphi \equiv \text{USet}^*(k,X)$, where $k$ is a non-negative integer and $X$ is a set variable. Then $S \models \varphi$ holds iff there exists an update set $\Delta$ and a set value $V$ representing $\Delta$ that is assigned to $X$ such that $\Delta$ is the composed update set yielded by $k$ steps of the machine $M$ starting in state $S$. The formal definitions are given in (\ref{eq-uset}), (\ref{eq-eventuset}), (\ref{eq-uset*}) and (\ref{eq-comp}).

\item The logic $\mathcal{L}^{ext}$ contains formulae $\varphi \equiv [X]\psi$, where $X$ is a set variable and $\psi$ is a formula of $\mathcal{L}^{ext}$. Then $S \models \varphi$ holds iff there exists an update set $\Delta$ yielded in $S$ and a set value $V$ representing $\Delta$ that is assigned to $X$ such that $S + \Delta \models \psi$ holds.

\end{itemize}

\subsection{Temporal Operators}

In order to support formal verification of computer programs Pnueli introduced linear time temporal logic (LTL), which extends propositional logic with modal operators referring to time \cite{pnueli:focs1977}. There are different ways to define the basic temporal operators; the most common ones are $\square$ (always), $\lozenge$ (eventually, i.e. sometimes in the future), $\mathcal{U}$ (until), and $\mathcal{N}$ (next). Most commonly used are the first three of them, as they suffice to express the most important liveness conditions \cite{manna:scp1984}. Thus formulae are built from propositions, the usual Boolean operators $\neg, \wedge, \vee, \rightarrow$, and modal operators. However, the propositional nature of LTL is a severe restriction, while an extension of first-order logic or even second-order logic with Henkin semantics by these temporal operators leads to incompleteness.

Hoang and Abrial~\cite{hoang:icfem2011} extended LTL using Event-B state formulae instead of propositions. Following this approach, Ferrarotti et al. \cite{ferrarotti:foiks2024} defined the logic LTL(EB), the {\em well-formed formulae} of which were defined as the closure of the set $\mathcal{L}$ of basic state formulae under the usual Boolean operators $\neg, \wedge, \vee, \rightarrow$ and modal operators $\square$ (always), $\lozenge$ (eventually) and $\mathcal{U}$ (until). Note that $\lozenge \, \varphi$ is merely a shortcut for $\mathbf{true} \, \mathcal{U} \, \varphi$, as can be easily proven from the semantics we define below. As we saw above, the absence of the modal operator $\mathcal{N}$ (next) does not matter, as conditions that are to hold in successor states are already covered by the first-order logic of Event-B.

As in LTL the formulae of LTL(EB) were interpreted over traces of Event-B machines \cite{ferrarotti:foiks2024}. However, inspired by related research on modal extensions of the logic of Abstract State Machines \cite{ferrarotti:abz2024} we will now show that all temporal operators can be defined as shortcuts of formulae of the logic $\mathcal{L}^{ext}$, while the defining properties over traces remain valid. This will allow us some generalisations such as temporal formulae that hold over defined sets of traces.

\subsubsection{All-Traces Temporal Formulae}

We say that the machine $M$ satisfies the formula $\square \varphi$ (with $\varphi \in \mathcal{L}^{ext}$)---interpreted as ``$\varphi$ {\em always} holds in all traces''---iff for all traces $S_0, S_1, \dots$ of $M$ we have $S_i \models \varphi$ for all $i \ge 0$; this is well defined, because $\mathcal{L}^{ext}$ formulae are interpreted over states. We write $M \models \square \varphi$ for this. Furthermore, for $i \ge 0$ there must exist an update set $\Delta$ with $S_i = S_0 + \Delta$ such that $S_0 \models \text{USet}^*(i,X)$ with $X$ representing $\Delta$, and $S_i \models \varphi$ holds iff $S_0 \models [X]\varphi$. Therefore, the formula $\square \varphi$ is merely a shortcut for the $\mathcal{L}^{ext}$ formula
\begin{gather}
\forall n \forall X ( \text{USet}^*(n,X) \rightarrow [X]\varphi ) \; . \label{eq-always}
\end{gather}

\noindent
Note that there is no problem with finite traces. If $n$ exceeds the length of the trace, the subformula $\text{USet}^*(n,X)$ will become \textbf{false}, regardless which value is assigned to $X$, and then the implication is interpreted as \textbf{true}, so in essence only finitely many values for $n$ have to be considered.

Due to the interpretation of $[X]\varphi$ above the case of finite traces is covered. If $n$ exceeds the length of the trace, the subformula $[Y] \text{USet}^*(k,Z)$ will be interpreted as \textbf{true}, and then $\text{USet}^*(k+1, X)$ can only be interpreted as \textbf{false}, i.e. there is no update set capturing the updates of $k+1$ steps.

Analogously we say that $M$ satisfies the formula $\lozenge \varphi$ (with $\varphi \in \mathcal{L}^{ext}$)---denoting that ``$\varphi$ {\em eventually} holds in all traces''---iff for all traces $S_0, S_1, \dots$ of $M$ we have $S_i \models \varphi$ for some $i \ge 0$. We write $M \models \lozenge \varphi$ for this. We can see that $\lozenge \varphi$ is also a shortcut of an $\mathcal{L}^{ext}$ formula, which in this case is slightly more complex. 

For this, consider the negation of $\lozenge \varphi$, so there exists a trace $S_0, S_1, \dots$ of $M$ with $S_i \models \neg\varphi$ for all $i \ge 0$. If all traces are infinite, then $\forall n \exists X ( \text{USet}^*(n,X) \wedge [X]\neg\varphi )$ must hold, i.e. after arbitrarily many steps of $M$ we reach a state satisfying $\neg \varphi$. Furthermore, all previous states in the same trace must also satisfy $\neg \varphi$. For this note that every update set $\Delta$ that results from $n>0$ steps of a machine $M$ in some state $S$ can be written as sequential composition of an update set $\Delta_1$ yielded from $m < n$ steps in $S$ and an update set $\Delta_2$ yielded from $n-m$ steps in the $m$'th successor state of $S$ resulting from applying $\Delta_1$ to $S$. Sequential composition of update sets is expressed by formula $\text{comp}(X_1, X_2, X)$ that we introduced before, where $X$ and $X_i$ ($i=1,2$) represent $\Delta$ and $\Delta_i$, respectively. That is, if we decompose $X$ into $X_1$ and $X_2$ for the first $m$ and the following $n-m$ steps, respectively, we must also have that $[X_1]\neg \varphi$ holds. Combining these two considerations, we obtain the formula
\begin{gather*}
\forall n \exists X \Big( \text{USet}^*(n,X) \wedge [X]\neg\varphi \wedge \forall m \big( m<n \rightarrow \forall X_1 X_2 ( \text{USet}^*(m,X_1) \wedge \\
[X_1] \text{USet}^*(n-m,X_2) \wedge \text{comp}(X_1,X_2,X) \rightarrow [X_1]\neg\varphi ) \big) \Big) \; . 
\end{gather*}

\noindent
In order to capture also finite traces we can replace $\text{USet}^*(n,X)$ herein by the formula $\text{U}_c^*(n, X)$, which we define as a shortcut for
\[ \text{USet}^*(n,X) \vee \exists n' \big( n'<n \wedge \text{USet}^*(n',X) \wedge [X] \; \forall \bar{x} \bigwedge_{e_i \in E} \neg G_i(\bar{x}, \bar{v}) \big) \; .\]

\noindent
Taking again the negation we see that $\lozenge \varphi$ is a shortcut for the following $\mathcal{L}^{ext}$ formula
\begin{gather}
\exists n \forall X \bigg( (\text{U}_c^*(n,X) \wedge [X]\neg\varphi ) \rightarrow  
\exists m  \big( m < n \wedge \exists X_1 X_2 ( \text{USet}^*(m,X_1) \wedge \notag\\
\qquad  [X_1]\text{USet}^*(n-m,X_2) \wedge \text{comp}(X_1, X_2, X) \wedge [X_1]\varphi ) \big) \bigg) \; .\label{eq-eventually}
\end{gather}

\noindent
We can also express UNTIL-formulae in $\mathcal{L}^{ext}$. $M$ satisfies the formula $\varphi \mathcal{U} \psi$ (with $\varphi, \psi \in \mathcal{L}^{ext}$)---denoting that ``$\varphi$ holds {\em until} $\psi$ holds in all traces''---iff for all traces $S_0, S_1, \dots$ of $M$ there exists some $i \ge 0$ with $S_i \models \psi$ and $S_j \models \varphi$ for all $j < i$. We write $M \models \varphi \mathcal{U} \psi$ for this. 

Phrased differently, whenever in a trace we have a state $S_m$ that does not satisfy $\varphi$ or that satisfies $\psi$ (reachable from $S_0$ by means of some $m$-step update set $X$), then there must exist an earlier state $S_{n_2}$ (reachable by means of some $n_2$-step update set $Y$ with $n_2 \le m$) satisfying $\psi$ such that all earlier states $S_{n_1}$ (reachable by means of some $n_1$-step update set $X_1$ with $n_1 < n_2$ such that $Y$ is the composition of $X_1$ and some $X_2$) satisfy $\varphi$. Hence $\varphi \mathcal{U} \psi$ is the $\mathcal{L}^{ext}$ formula
\begin{gather}
\forall m  \forall X \big( \text{USet}^*(m, X) \wedge [X](\neg\varphi  \vee \psi) \rightarrow \exists n_2 \exists Y \big( n_2 \le m \wedge \qquad\qquad \notag\\
\text{USet}^*(n_2,Y) \wedge [Y] \psi \wedge \forall n_1 \forall X_1, X_2 ( n_1 < n_2 \wedge \text{USet}^*(n_1, X_1) \wedge \notag\\
\qquad\qquad [X_1]\text{USet}^*(n_2 - n_1, X_2) \wedge \text{comp}(X_1, X_2, Y) \rightarrow [X_1]\varphi ) \big) \big) \; . \label{eq-until}
\end{gather}

\subsubsection{One-Trace Temporal Formulae}

The temporal $\mathcal{L}^{ext}$ formulae $\square \varphi$ (ALWAYS $\varphi$), $\lozenge \varphi$ (EVENTUALLY $\varphi$) and $\varphi \mathcal{U} \psi$ ($\varphi$ UNTIL $\psi$) defined above are interpreted over states $S$ and they hold in all traces of a machine $M$ starting in this state $S$. Temporal logics such as CTL and CTL$^*$ \cite{clarke:lop1981} also consider formulae that are to hold in just one trace. We show that these one-trace analogues of temporal logic formulae can also be defined in $\mathcal{L}^{ext}$.

We say that $M$ satisfies the formula $\square^1 \varphi$ (with $\varphi \in \mathcal{L}^{ext}$)---interpreted as ``$\varphi$ always holds in some trace''---iff there exists a trace $S_0, S_1, \dots$ of $M$ with $S_i \models \varphi$ for all $i \ge 0$. We write $M \models \square^1 \varphi$ for this. Then $\neg \square^1 \neg \varphi$ should be equivalent to $\lozenge \varphi$. Hence $\square^1 \varphi$ is a shortcut for the $\mathcal{L}^{ext}$ formula
\begin{gather}
\forall n \exists X \bigg( \text{U}_c^*(n,X) \wedge [X]\varphi \wedge \forall m  \big( m < n \rightarrow \forall X_1 X_2 ( \text{USet}^*(m,X_1) \wedge \notag\\
 \quad [X_1]\text{USet}^*(n-m,X_2) \wedge \text{comp}(X_1, X_2, X) \rightarrow [X_1]\varphi ) \big) \bigg) \; . \label{eq-always1}
\end{gather}

\noindent
In the same way we say that $M$ satisfies the formula $\lozenge^1 \varphi$ (with $\varphi \in \mathcal{L}^{ext}$)---denoting that ``$\varphi$ eventually holds in some trace''---iff there exists a trace $S_0, S_1, \dots$ of $M$ with $S_i \models \varphi$ for some $i \ge 0$. We write $M \models \lozenge^1 \varphi$ for this. Then $\neg \lozenge^1 \neg \varphi$ should be equivalent to $\square \varphi$, hence $\lozenge^1 \varphi$ is a shortcut for the formula
\begin{gather}
\exists n \exists X ( \text{USet}^*(n,X) \wedge [X]\varphi ) \; . \label{eq-eventually1}
\end{gather}

\noindent
Finally, we can express that ``$\varphi$ holds until $\psi$ in some trace'' by the $\mathcal{L}^{ext}$ formula
\begin{gather}
\exists n \exists X \big( \text{USet}^*(n, X) \wedge
[X] \psi \wedge \forall m \forall X_1 X_2 ( m<n \wedge \text{USet}^*(m, X_1) \wedge \notag\\
\qquad\qquad [X_1]\text{USet}^*(n-m, X_2) \wedge \text{comp}(X_1, X_2, Y) \rightarrow [X_1]\varphi ) \big) \; , \label{eq-until1}
\end{gather}
and we write $M \models \varphi \mathcal{U}^1 \psi$ for this. The formula expresses that there exists a run $S_0, S_1, \dots$ of $M$ and some $i \ge 0$ with $S_i \models \psi$ and $S_j \models \varphi$ for all $j < i$.

All modal operators $\square$, $\lozenge$, $\mathcal{U}$, $\square^1$, $\lozenge^1$ and $\mathcal{U}^1$ can be freely combined in $\mathcal{L}^{ext}$. For instance $\lozenge \square^1 \varphi$ expresses that in all traces $S_0, S_1, \dots$ we reach a state $S_k$ such that in one continuation of the prefix $S_0, \dots, S_k$ all states satisfy $\varphi$. Such formulae can now be interpreted over states. There may be some doubts, whether there are many applications, where such general conditions occur, but technically it is possible to handle them.

Different to the work in \cite{ferrarotti:foiks2024} we now can also capture universal and existential quantification over traces without leaving the Event-B frame of first-order logic and set theory. 

\subsubsection{Temporal Formulae over Definable Sets of Traces}

Here we want to go one step further and consider also trace quantifiers, by means of which we can define formulae that are to hold over well-defined subsets of the set of traces. This has been inspired by temporal logics with team semantics \cite{gutsfeld:lics2022}. The sets of traces we are interested in are to be characterised by a temporal formula, which according to the semantics of $\mathcal{L}^{ext}$ must hold in the inititial state(s) of the traces. The general study of temporal formulae over definable sets of traces is beyond the scope of this article; we will only consider sets of traces defined by temporal formulae of the form $\lozenge\square \chi$ for some $\mathcal{L}^{ext}$ formula $\chi$. That is we consider traces that either terminate with a state $S_m$ satisfying $\chi$ or are infinite with a tail of states $S_\ell$ ($\ell \ge k$ for some $k$) all satisfying $\chi$. For short, we consider traces $S_0, S_1, \dots$, for which there exists some $k$ such that $S_\ell$ satisfies $\chi$ for all $\ell \ge k$. In these cases we say that the trace satisfies the selection condition $\langle \chi \rangle$.

We will write $\square^\chi \varphi$ for an $\mathcal{L}^{ext}$ formula that is intended to hold in the initial state $S_0$ of a machine $M$, if for all traces $S_0, S_1, \dots$ that satisfy the selection condition $\langle \chi \rangle$ we have $S_i \models \varphi$ for all $i \ge 0$. Analogously, we write $\lozenge^\chi \varphi$ for an $\mathcal{L}^{ext}$ formula that is intended to hold in the initial state $S_0$ of a machine $M$, if for all traces $S_0, S_1, \dots$ that satisfy the selection condition $\langle \chi \rangle$ we have $S_i \models \varphi$ for at least one $i \ge 0$.\footnote{Note that by means of these formulae we define quantifiers for arbitrary $\mathcal{L}^{ext}$ formulae $\chi$, by means of which the sets of traces satisfying the selection condition $\langle \chi \rangle$ are defined, and with these quantifiers we restrict the common temporal operators $\square$ and $\lozenge$.}

In order to define $\square^\chi \varphi$ we need to modify the defining formula (\ref{eq-always}) for $\square \varphi$. Herein the subformula $\text{USet}^*(n,X) \rightarrow [X]\varphi$ expresses that if after $n$ steps we reach a state $S_n$ and the corresponding update set is $X$, then applying $X$ leads to a successor state satisfying $\varphi$. We need to restrict this to states that appear in a trace terminating with a state satisfying $\chi$ or having an infinite tail of states all satisfying $\chi$. This is the case, if there is an $m$-step successor state corresponding to some update set $Y$ and there is a continuation of this trace with states all satisfying $\chi$. With this we see that $\square^\chi \varphi$ is a shortcut for the $\mathcal{L}^{ext}$ formula
\begin{gather}
\forall n \forall X \Big( \text{USet}^*(n,X) \wedge \exists m \exists Y \big( [X]\text{USet}^*(m,Y) \wedge [Y]\square^1 \chi \big) \rightarrow [X]\varphi \Big) \; , \label{eq-alwaysteam}
\end{gather}
where the subformula $\square^1 \chi$ is defined by (\ref{eq-always1}).

A formula $\neg \lozenge^\chi \neg \varphi$ holds, if there exists a trace $S_0, S_1, \dots$ satisfying the selection condition $\langle \chi \rangle$ such that $S_i \models \varphi$ holds for all $i \ge 0$. Consider again the case that $\chi$ holds for traces $S_0, S_1, \dots$, for which there exists some $k$ such that $S_\ell$ satisfies $\chi$ for all $\ell \ge k$. In this case we need to modify the defining formula (\ref{eq-always1}) for $\square^1 \varphi$ in such a way that the state reached by means of the $n$-step update set $X$ appears in a trace satisfying the selection condition $\langle\chi\rangle$. Hence, $\neg \lozenge^\chi \neg \varphi$ is a shortcut of the $\mathcal{L}^{ext}$ formula
\begin{gather*}
\forall n \exists X \bigg( \text{U}_c^*(n,X) \wedge [X]\varphi \wedge \forall m  \big( m < n \rightarrow \forall X_1 X_2 ( \text{USet}^*(m,X_1) \wedge \\
\quad \qquad [X_1]\text{USet}^*(n-m,X_2) \wedge \text{comp}(X_1, X_2, X) \rightarrow [X_1]\varphi ) \big) \wedge \\ 
\exists m \exists Y \big( [X]\text{USet}^*(m,Y) \wedge [Y]\square^1 \chi \big) \bigg) \; ,
\end{gather*}
and by taking the negation we see that $\lozenge^\chi \varphi$ is a shortcut of the $\mathcal{L}^{ext}$ formula
\begin{gather}
\exists n \forall X \bigg(\text{U}_c^*(n,X)  \wedge \forall m  \big( m < n \rightarrow \forall X_1 X_2 ( \text{USet}^*(m,X_1) \wedge \notag \\
\quad \qquad [X_1]\text{USet}^*(n-m,X_2) \wedge \text{comp}(X_1, X_2, X) \rightarrow [X_1] \neg \varphi ) \big) \wedge \notag \\ 
\exists m \exists Y \big( [X]\text{USet}^*(m,Y) \wedge [Y]\square^1 \chi \big) \rightarrow [X] \varphi\bigg) \; . \label{eq-eventuallyteam}
\end{gather}

\subsection{Definition of TREBL}

We now define the logic TREBL as a fragment of $\mathcal{L}^{ext}$, in which important liveness conditions can be expressed. Of particular interest are the following conditions:

\begin{description}

\item[Invariance] An {\em invariance condition} is given by an $\mathcal{L}^{ext}$ formula of the form $\square \, \varphi$ for some $\mathcal{L}$ formula $\varphi$. Such a condition is satisfied by a machine $M$ iff all states $S$ in all traces of $M$ satisfy $\varphi$.\footnote{Note that invariance conditions have been an intrinsic part of Event-B (and likewise all other rigorous methods) since the very beginnings, and for invariance alone there would be no need to consider a temporal logic.}

\item[Existence] An {\em existence condition} is given by an $\mathcal{L}^{ext}$ formula of the form $\square \lozenge \varphi$ for some $\mathcal{L}$ formula $\varphi$. Such a condition is satisfied by a machine $M$ iff in every infinite trace of $M$ there is an infinite subsequence of states satisfying $\varphi$, and every finite trace of $M$ terminates in a state satisfying $\varphi$.

\item[Progress] A {\em progress condition} is given by an $\mathcal{L}^{ext}$ formula of the form $\square (\varphi \rightarrow\lozenge \psi)$ for some $\mathcal{L}$ formulae $\varphi$ and $\psi$. Such a condition is satisfied by a machine $M$ iff in every trace of $M$ a state $S_i$ satisfying $\varphi$ is followed some time later by a state $S_j$ ($j \ge i$) satisfying $\psi$. A progress condition with $\varphi = \textbf{true}$ degenerates to an existence condition.

\item[Persistence] A {\em persistence condition} is given by an $\mathcal{L}^{ext}$  formula of the form $\lozenge \square \, \varphi$ for some $\mathcal{L}$ formula $\varphi$. Such a condition is satisfied by a machine $M$ iff every infinite trace of $M$ ends with an infinite sequence of states satisfying $\varphi$, i.e. for some $k$ we have $S_i \models \varphi$ for all $i \ge k$, and every finite trace of $M$ terminates in a state satisfying $\varphi$.

\item[Reachability] A {\em reachability condition} is given by an $\mathcal{L}^{ext}$ formula of the form $\lozenge \, \varphi$ for some $\mathcal{L}$ formula $\varphi$. Such a condition is satisfied by a machine $M$ iff in all traces of $M$ there exists a state $S$ satisfying $\varphi$.

\end{description}

\noindent
For all four kinds of liveness conditions\footnote{In \cite{ferrarotti:foiks2024} an existence condition was also called an {\em unconditional progress condition}, and a progress condition was called {\em conditional progress condition} to emphasise that a progress condition with $\varphi = \textbf{true}$ degenerates to an existence condition.} we can add variants by replacing one or both of the temporal operator by the operators in (\ref{eq-always1}) and (\ref{eq-eventually1}) that refer to just one trace. It also does not do any harm to generalise from $\mathcal{L}$ formulae to $\mathcal{L}^{ext}$ formulae except for the formulae $\varphi$ in progress conditions. 

Then it makes sense to define TREBL as the closure of the $\mathcal{L}$ under a restrictive set of temporal constructors, which basically have the form of the formulae above, so we get the following definition.

\begin{defi}\label{def-trebl}\rm

The set $\mathcal{LT}$ of {\em well-formed formulae} of TREBL is the smallest set inductively defined by the following rules:   

\begin{itemize}

\item For $\varphi \in \mathcal{L}$ we also have $\varphi \in \mathcal{LT}$;

\item For $\varphi \in \mathcal{LT}$, $\psi \in \mathcal{L}$, $X \in \{ \square, \square^1 \}$ and $Y \in \{ \lozenge, \lozenge^1 \}$ the $\mathcal{L}^{ext}$ formulae $X \varphi$, $Y \varphi$, $XY \varphi$, $YX \varphi$ and $X (\psi \rightarrow Y \varphi )$ are formulae in $\mathcal{LT}$;

\item For $\varphi \in \mathcal{LT}$ the quantified $\mathcal{L}^{ext}$ formulae $\forall x \varphi$ and $\exists x \varphi$ are formulae in $\mathcal{LT}$.
    
\end{itemize}

\end{defi}

\noindent
Note that TREBL is defined as the closure of the Event-B logic $\mathcal{L}$ under certain temporal operators. We only need the temporal operators, but not the predicates \text{USet}, \text{USet}$^*$ and $[ \cdot ]$ that were used to define them. These predicates are only used as a means that allows temporal formulae to be defined as shortcuts\footnote{According to Lamport \cite{lamport:pnueli2010} the direct use of time or step variables leads to very complicated formulae and therefore should be discouraged. Indeed, all defining formulae (\ref{eq-always}) -- (\ref{eq-eventuallyteam}) are very complex, but this complexity does not prevent the existence of a rather simple set of derivation rules, and the complexity will not impact on the application; it will only be relevant for the proof of soundness and relative completeness.} for formulae in $\mathcal{L}^{ext}$ that are interpreted over states rather than traces. This even allows us to include arbitrary nesting of the defining operators. We already mentioned before that this generality may be not needed for most applications, but technically, it does do no harm. We will see this later, when we discuss derivation rules for nested formulae.

Further note that for the definition of TREBL we only used the operators occurring in the common liveness conditions above, but we did not use $\mathcal{U}$-formulae or definable sets of traces. Definable sets of traces will nonetheless become relevant in the proofs in the next section.

In the next section we will present a sound set of derivation rules for $\mathcal{LT}$, which modify and extend the derivation rules discovered by Hoang and Abrial \cite{hoang:icfem2011}. We then show that together with derivation rules for first-order logic with types we obtain a relative complete set of derivation rules, where {\em relative completeness} refers to the existence of derivations, if certain variant terms can be defined for the Event-B machine under consideration. Fortunately, we can also show that for sufficiently refined machines the existence of such variant terms can always be guaranteed. In the completeness proof we will eliminate the restriction that machines must be {\em tail-homogeneous}, which was assumed in \cite{ferrarotti:foiks2024}.

\subsection{Examples}

We now look at some examples and their formalisation in TREBL. First we consider non-interference of traces, an example often classified as a 2-safety hyperproperty (as it compares pairs of traces). Then we give two examples of liveness conditions.


\begin{exa}\label{bsp-noninterference}

\emph{Non-Interference} of traces is a relevant security property when multiple users with different access rights collaborate concurrently~\cite{biskup:2009}. We assume a set of users $u_1, \dots, u_n$, each permitted to handle only events classified by
linearly ordered labels $\ell_1 < \dots < \ell_n$, respectively. The system satisfies non-interference if, for each user $u_i$, the observable behavior of the system is unaffected by events classified at higher security levels ($\ell_j$ for $j > i$), even when $u_i$ has full knowledge of the system's specification and reasoning capabilities.

We define an Event-B machine for the multi-level stack example from~\cite{mclean:jcs1992} assuming a set of users $U = \{u_1, \dots, u_k\}$, a set of stacks $P = \{ p_1, \dots, p_n \}$, a set of data values~$D$, and a linearly ordered set of clearance levels $L = \{ \ell_1 < \dots < \ell_n \}$. The Event-B constants $\txt{user\_cl} \in U \to L$ and $\txt{stack\_cl} \in P \to L$ assign to each user and each stack a corresponding clearance level. The variable $\txt{op\_cl} \in L$ records the clearance level of the most recently executed operation, and is initialized to the minimum clearance level
$\ell_1$. The variables $\txt{top} \in P \to \mathbb{N}$, $\txt{store} \in P \times \mathbb{N} \to D$, and $\txt{answer} \in U \to D$ represent, respectively, the current stack pointer, stack element storage, and last read result for each user; they are initially empty.

To prevent users with higher-level clearances from passing information to users with lower-level clearances, users are forbidden to:

\begin{enumerate}
\item read stacks with clearance levels above their own clearance, and
\item push or pop stacks with clearance levels below their own clearance.
\end{enumerate}

\noindent
Each user with sufficient security clearance can execute \txt{push}, \txt{pop},
\txt{peek}, and \txt{peek0} events, whose guards and actions enforce these
constraints. The full Event-B machine~$M$ is given below.

\begin{multicols}{2}
\begin{minipage}{\linewidth}
\begin{lstlisting}[mathescape=true]
$\textbf{sets}$ $U;\ P;\ L;\ D$
$\textbf{constants}$
  $\txt{user\_cl}$  $\text{// user clearance}$ 
  $\txt{stack\_cl}$ $\text{// stack clearance}$
  $\ell_1$ $\text{// minimum clearance level}$
\end{lstlisting}
\end{minipage}

\begin{minipage}{\linewidth}
\begin{lstlisting}[mathescape=true]
$\textbf{variables}$
  $\txt{top}$    $\text{// current stack pointer}$
  $\txt{store}$  $\text{// stack element storage}$
  $\txt{answer}$ $\text{// last read result}$
  $\txt{op\_cl}$ $\text{// current operation clearance}$
\end{lstlisting}
\end{minipage}
\end{multicols}

\begin{multicols}{3}
\begin{minipage}{\linewidth}
\begin{lstlisting}[mathescape=true]
$\textbf{axioms}$
  $\txt{user\_cl} \in U \to L$
  $\txt{stack\_cl} \in P \to L$
  $\ell_1 = \mathrm{min}(L)$
\end{lstlisting}
\end{minipage}

\begin{minipage}{\linewidth}
\begin{lstlisting}[mathescape=true]
$\textbf{invariants}$
  $\txt{op\_cl} \in L$
  $\txt{top} \in P \to \mathbb{N}$
  $\txt{store} \in P \times \mathbb{N} \to D$
  $\txt{answer} \in U \to D$
\end{lstlisting}
\end{minipage}

\begin{minipage}{\linewidth}
\begin{lstlisting}[mathescape=true]
$\textbf{init}$
  $\txt{op\_cl} := \ell_1$
  $\txt{top}    := \emptyset$
  $\txt{store}  := \emptyset$
  $\txt{answer} := \emptyset$
\end{lstlisting}
\end{minipage}
\end{multicols}

\begin{multicols}{2}
\begin{minipage}{\linewidth}
\begin{lstlisting}[mathescape=true]
$\textbf{push}$
$\textbf{any}$ $u,\ p,\ d$
$\textbf{when}$
  $u \in U;\ p \in P;\ d \in D$
  $\txt{user\_cl}(u) \leq \txt{stack\_cl}(p)$
$\textbf{then}$
  $\txt{op\_cl} := \txt{user\_cl}(u)$
  $\txt{top}(p) := \txt{top}(p) + 1$
  $\txt{store}(p,\,\txt{top}(p){+}1) := d$
$\textbf{end}$
\end{lstlisting}
\end{minipage}

\begin{minipage}{\linewidth}
\begin{lstlisting}[mathescape=true]
$\textbf{pop}$
$\textbf{any}$ $u,\ p$
$\textbf{when}$
  $u \in U;\ p \in P$
  $\txt{top}(p) > 0$
  $\txt{user\_cl}(u) \leq \txt{stack\_cl}(p)$
$\textbf{then}$
  $\txt{op\_cl} := \txt{user\_cl}(u)$
  $\txt{top}(p) := \txt{top}(p) - 1$
$\textbf{end}$
\end{lstlisting}
\end{minipage}
\end{multicols}

\begin{multicols}{2}
\begin{minipage}{\linewidth}
\begin{lstlisting}[mathescape=true]
$\textbf{peek}$
$\textbf{any}$ $u,\ p$
$\textbf{when}$
  $u \in U;\ p \in P$
  $\txt{top}(p) > 0$
  $\txt{user\_cl}(u) \geq \txt{stack\_cl}(p)$
$\textbf{then}$
  $\txt{op\_cl} := \txt{user\_cl}(u)$
  $\txt{answer}(u) :=
      \txt{store}(p,\,\txt{top}(p))$
$\textbf{end}$
\end{lstlisting}
\end{minipage}

\begin{minipage}{\linewidth}
\begin{lstlisting}[mathescape=true]
$\textbf{peek0}$
$\textbf{any}$ $u,\ p$
$\textbf{when}$
  $u \in U;\ p \in P$
  $\txt{top}(p) = 0$
  $\txt{user\_cl}(u) \geq \txt{stack\_cl}(p)$
$\textbf{then}$
  $\txt{op\_cl} := \txt{user\_cl}(u)$
  $\txt{answer}(u) := \emptyset$
$\textbf{end}$
\end{lstlisting}
\end{minipage}
\end{multicols}

\noindent
The guards enforce the access control policy: \textbf{push} and \textbf{pop} require $\txt{user\_cl}(u) \leq \txt{stack\_cl}(p)$, ensuring that a user may only write to stacks at or above their own clearance level. Conversely, \textbf{peek} and \textbf{peek0} require $\txt{user\_cl}(u) \geq \txt{stack\_cl}(p)$, ensuring that
a user may only read from stacks at or below their own clearance level. In every event, $\txt{op\_cl}$ is updated to $\txt{user\_cl}(u)$, recording the clearance level of the most recently performed operation.

Non-interference requires that the observable output $\txt{answer}(u)$ of any user $u$ is never affected by operations performed at a strictly higher clearance level, and similarly that the state of any stack $p$ is not modified by operations at a clearance level strictly above $\txt{stack\_cl}(p)$. Formally, the machine $M$ must satisfy the invariances $\square\varphi_{\mathrm{push}}$, $\square\varphi_{\mathrm{pop}}$, and $\square\varphi_{\mathrm{peek}}$, where, using primed variables to denote successor-state
values:
\begin{align}
\varphi_{\mathrm{push}} &\;\equiv\;
  \forall \ell\,\bigl(\txt{op\_cl}' = \ell \;\rightarrow\;
  \forall p\,(\txt{stack\_cl}(p) < \ell
    \rightarrow \txt{top}'(p) = \txt{top}(p) \;\wedge{}\notag\\
  &\phantom{{}\equiv{}\forall \ell\,(\txt{op\_cl}' = \ell \;\rightarrow{}\forall p\,(}
    \txt{store}'(p,\txt{top}'(p)) =
    \txt{store}(p,\txt{top}(p)))\bigr),
  \label{eq-ni-push}\\[4pt]
\varphi_{\mathrm{pop}} &\;\equiv\;
  \forall \ell\,\bigl(\txt{op\_cl}' = \ell \;\rightarrow\;
  \forall p\,(\txt{stack\_cl}(p) < \ell
    \rightarrow \txt{top}'(p) = \txt{top}(p) \;\wedge{}\notag\\
  &\phantom{{}\equiv{}\forall \ell\,(\txt{op\_cl}' = \ell \;\rightarrow{}\forall p\,(}
    \txt{store}'(p,\txt{top}'(p)) =
    \txt{store}(p,\txt{top}(p)))\bigr),
  \label{eq-ni-pop}\\[4pt]
\varphi_{\mathrm{peek}} &\;\equiv\;
  \forall \ell\,\bigl(\txt{op\_cl}' = \ell \;\rightarrow\;
  \forall u\,(\txt{user\_cl}(u) < \ell
    \rightarrow \txt{answer}'(u) = \txt{answer}(u))\bigr).
  \label{eq-ni-peek}
\end{align}

\noindent
The invariants are organised according to what each class of operation can actually modify. \textbf{push} and \textbf{pop} never touch $\txt{answer}$, so $\varphi_{\mathrm{push}}$ and $\varphi_{\mathrm{pop}}$ only constrain the stack state: after such an operation at clearance level $\ell$, the top pointer and stored content of every stack with a lower clearance level must be unchanged. Dually, \textbf{peek} and \textbf{peek0} never touch $\txt{top}$ or $\txt{store}$, so $\varphi_{\mathrm{peek}}$ only constrains $\txt{answer}$: after a read at clearance level $\ell$, the answer of every user with a lower clearance level must be unchanged. Together, the three invariants ensure that no information can flow from higher to lower clearance levels across any
single step. Since every event updates $\txt{op\_cl}$ to $\txt{user\_cl}(u)$ and the guards prevent any event from accessing resources outside the permitted clearance range, this per-step condition lifts 
by a straightforward induction on the number of steps along the trace to the global non-interference property.

Note that $\square\varphi_{\mathrm{push}}$, $\square\varphi_{\mathrm{pop}}$, and $\square\varphi_{\mathrm{peek}}$ are in general \emph{trace properties}, since they relate each state to its immediate successor via the primed variables. However, the primed variables refer to values that are definable in the next state, so the formulae above can be rewritten into equivalent formulae in $\mathcal{L}$ using the next-operator~$\mathcal{N}$. Alternatively, one can introduce auxiliary variables into the Event-B machine to record the previous values of $\txt{op\_cl}$, $\txt{top}$,
$\txt{store}$, and $\txt{answer}$, thereby reducing each trace property to a \emph{safety property} expressible as a plain state predicate. In either case, making these properties part of the invariant of the (possibly enriched) machine gives rise to consistency proof obligations, so we in fact prove more than the temporal invariances $\square\varphi_{\mathrm{push}}$, $\square\varphi_{\mathrm{pop}}$, and $\square\varphi_{\mathrm{peek}}$ alone.

\end{exa}


\begin{exa}\label{bsp-fairness}

{\em Fairness} for an Event-B machine means that every event that is enabled must eventually be fired. This can be expressed by quantified progress conditions
\[ \forall \bar{x} \big( \square ( \text{enabled}_i(\bar{x}) \rightarrow \lozenge \text{fired}_i(\bar{x}) ) \big) \]
for all events $e_i$. A weaker form would be to require the progress conditions
\[ \square \big( \exists \bar{x} \; \text{enabled}_i(\bar{x}) \rightarrow \lozenge  \exists \bar{y} \; \text{fired}_i(\bar{y}) \big) \]
for all events $e_i$, where the enabled event will eventually be fired with different parameters.

\end{exa}

\begin{exa}\label{bsp-cocon}

Bounded deducibility was introduced by Popescu et al. in \cite{popescu:afp2014} as a means to express privacy constraints in concurrent systems. In a nutshell, we assume an {\em observation infrastructure}, which is defined by a partial function $O$, which may assign an observation $O(S) \in \mathcal{O}$ to a state $S$. Then a trace $S_0, S_1, \dots$ gives rise to a sequence $\textit{Obs\/}_i$ of lists of observations. Intuitively, an event that is fired in state $S_i$ either appends $O(S_i)$ to the value of a variable \textit{Obs\/}---which is the empty list in the initial state---or leaves the value of \textit{Obs\/} unchanged. Informally, the list $\textit{Obs\/}_i$ captures what an agent---a potential attacker---can see from the prefix $S_0, \dots, S_i$ of the trace. We further assume a {\em secrecy infrastructure}, which is defined by a partial function $G$, which may assign a secret $G(S) \in \mathcal{G}$ to a state $S$. Then a trace $S_0, S_1, \dots$ also gives rise to a sequence $\textit{Sec\/}_i$ of lists of secrets.

A {\em bounded deducibility constraint} (BD) comprises a {\em declassification trigger} $\psi$ and a {\em declassification bound} $B$. $\psi$ is a formula that can be evaluated in a state, though it is tacitly assumed that it is restricted to a formula that does not refer to any successor or predecessor state. $B$ is a binary relation on lists of secrets. For BD security we consider {\em eligible prefixes} $S_0, \dots , S_p$ of traces such that all states in the prefix satisfy $\neg \psi$. Then the machine $M$ is {\em BD secure} iff whenever we have $(\textit{Sec\/}_p, G) \in B$ there exists an eligible prefix $S_0^\prime, \dots, S_q^\prime$ of some trace with $\textit{Obs\/}_p = \textit{Obs\/}^\prime_q$ and $\textit{Sec\/}^\prime_q = G$. Informally phrased, for every trace that might reveal some secrets there exists an indistinguishable trace (by means of the observations), in which the secrets are bounded and thus not revealed\footnote{In essence, BD security requires a relation between different traces to hold, which is beyond the fragment of $\mathcal{L}^{ext}$ we are dealing within this article. Nonetheless, we should stress that the version of TREBL investigated in this article seems not to be a maximal fragment of $\mathcal{L}^{ext}$ with respect to having a sound and relative complete set of derivation rules. However, for the sake of keeping this article reasonably compact, we postpone further extensions to TREBL to future work. Here we leave it as an open conjecture, that an extension capturing also BD security is possible.}.

In \cite{popescu:jar2021,popescu:afp2021} Popescu et al. used a simple conference management system CoCon (inspired by EasyChair) as an application example for BD security. As a simple example consider the privacy requirement that a group of users should learn nothing about the content of a paper beyond the last submitted version, unless one of them becomes a co-author of that paper. This constraint was named PAP$_2$ in \cite{popescu:jar2021}; together with other constraints it merely ensures that PC members should only see the last version of any paper and only, after the submission phase has been closed.

This requirement can be captured by a BD constraint as shown in \cite{popescu:jar2021}; we omit the details here. However, in TREBL the requirement can be captured in a different way using a transition invariant. For this we assume a set UIDs of valid user identifiers and a set paperIDs of valid paper identifiers. Furthermore, assume functions \textit{authors\/}, \textit{content\/} and \textit{final\/}, which assign to a paper identifier $pid$ and a version number $i$ the set of authors, the content and a Boolean value indicating, if the version is the final version. Then let $\psi$ be the formula
\begin{gather*}
\bigwedge_{e \in E} \forall uid\; pid\; i\; c \big( uid \in \text{UIDs} \wedge pid \in \text{paperIDs} \wedge uid \notin \textit{authors\/}(pid, i) \wedge \\ 
\textit{content\/}(pid, i) = c \wedge \neg \textit{final\/}(pid, i) \rightarrow c \notin \textit{output\/}_e^\prime(uid) \; ,
\end{gather*}
where \textit{output\/}$_e(uid)$ refers to the output of event $e$ made to the user with the identifier $uid$. Then $\square \psi$ captures the requirement PAP$_2$ without referring to a BD constraint.

\end{exa}

\section{Sound Derivation Rules}\label{sec:fragment}

Our aim is to develop a proof system for TREBL based on the derivation rules in \cite{hoang:icfem2011}, the modifications in \cite{ferrarotti:foiks2024} and the extensions in \cite{ferrarotti:abz2024}. This will support mechanical reasoning about liveness properties for Event-B machines. 

Let $\mathfrak{R}$ be a set of axioms and inference rules. A formula $\varphi \in \mathcal{L}^{ext}$ is \emph{derivable} from a set $\Phi$ of $\mathcal{L}^{ext}$ formulae using $\mathfrak{R}$ (denoted as $\Phi \vdash_{\mathfrak{R}} \varphi$, or simply $\Phi \vdash \varphi$ when $\mathfrak{R}$ is clear from the context) iff there is a deduction from formulae in $\Phi$ to $\varphi$ that only applies axioms and inference rules from $\mathfrak{R}$. We write $\vdash \varphi$ as a shortcut for $\emptyset \vdash \varphi$.

As we are interested in the derivation of formulae that are to hold for a given Event-B machine $M$ we consider derivations that depend on the specification of $M$, i.e. we refer to the events of $M$, their guards and their actions. In this case we write $M \vdash \varphi$.

\subsection{Rules for Invariance}

For invariance conditions we can exploit a simple induction principle. In order to establish invariance of a formula $\varphi \in \mathcal{L}$ it suffices to show that $\varphi$ holds in the initial state of any trace $S_0, S_1, \dots$, and furthermore that for every $1 \leq k < \ell(\sigma)$ whenever $S_k \models \varphi$ holds, then also $S_{k+1} \models \varphi$ holds. In addition, invariance of a state formula $\varphi$ implies invariance of a weaker state formula $\psi$. This gives rise to the following lemma.

\begin{lem}\label{lem-invariance}

For formulae $\varphi, \psi \in \mathcal{L}$ the derivation rules
\begin{center}
\begin{minipage}{6cm}
\begin{prooftree}
\AxiomC{$M \vdash \varphi$}
\AxiomC{\hspace*{-5mm}$M \vdash \Leadsto(\varphi,\varphi)$}
\LeftLabel{$\mathrm{INV}_1$:}
\BinaryInfC{$M \vdash \square \varphi$}
\end{prooftree}
\end{minipage}
\qquad
\begin{minipage}{5.5cm}
\begin{prooftree}
\AxiomC{$\vdash \varphi \rightarrow \psi$}
\AxiomC{\hspace*{-5mm}$M \vdash \square \varphi$}
\LeftLabel{$\mathrm{INV}_2$:}
\BinaryInfC{$M \vdash \square \psi$}
\end{prooftree}
\end{minipage}   
\end{center}
are sound for the derivation of invariance properties in TREBL.

\end{lem}

\proof

Let $S_0, S_1, \dots$ be an arbitrary trace of $M$. We proceed by induction on $k$. If $M \models \varphi$ holds, then $S_0 \models \varphi$ holds by definition, which gives the induction base. If for arbitrary $k$ we have $S_k \models \varphi$, then with $M \models \Leadsto(\varphi,\varphi)$ we immediately get $S_{k+1} \models \varphi$, which completes the induction step and shows the soundness of $\mathrm{INV}_1$.

If $M \models \square \varphi$ holds, then we have $S_i \models \varphi$ for all $i \ge 0$. As also $\models \varphi \rightarrow \psi$ holds, we get $S_i \models \psi$ for all $i \ge 0$, hence $M \models \square \psi$, which shows the soundness of $\mathrm{INV}_2$.\qed

For the invariant $\iota$ of $M$, the consistency proof obligation corresponds to the antecedents of rule $\mathrm{INV}_1$ for the case $\varphi \equiv \iota$, and the consequence expresses what we really want to have, the satisfaction of $\iota$ in all reachable states. We also said that for any invariant $\psi$, a formula in $\mathcal{L}$, we assume that it is made a part of the invariant $\iota$. Then $\vdash \iota \rightarrow \psi$ becomes a triviality. For a consistent machine $M$ we also get $M \vdash \square \iota$ and hence $M \vdash \square \psi$. Therefore, with the general assumptions made the discharging of the consistency proof obligation suffices for invariance proofs.

If the antecedent $M \vdash \Leadsto(\varphi,\varphi)$ in the lemma is replaced by the weaker one-trace analog $M \vdash \Leadsto^1(\varphi,\varphi)$, we obtain invariance of $\varphi$ in one trace instead of all traces, hence the following modified lemma.

\begin{lem}\label{lem-inv1}

For formulae $\varphi, \psi \in \mathcal{L}$ the derivation rules
\begin{center}
\begin{minipage}{6cm}
\begin{prooftree}
\AxiomC{$M \vdash \varphi$}
\AxiomC{\hspace*{-5mm}$M \vdash \Leadsto^1(\varphi,\varphi)$}
\LeftLabel{$\mathrm{INV}_1^1$:}
\BinaryInfC{$M \vdash \square^1 \varphi$}
\end{prooftree}
\end{minipage}
\qquad
\begin{minipage}{5.5cm}
\begin{prooftree}
\AxiomC{$\vdash \varphi \rightarrow \psi$}
\AxiomC{\hspace*{-5mm}$M \vdash \square^1 \varphi$}
\LeftLabel{$\mathrm{INV}_2^1$:}
\BinaryInfC{$M \vdash \square^1 \psi$}
\end{prooftree}
\end{minipage}
\end{center}
are sound for the derivation of one-trace invariance properties in TREBL.

\end{lem}

\proof

We use induction on $k$ to construct a trace $S_0, S_1, \dots$ with $S_i \models \varphi$ for all $i \ge 0$. If we assume $M \models \varphi$, then by definition we have $S_0 \models \varphi$ for the initial state $S_0$. If $S_k \models \varphi$, then as we assume $M \models \Leadsto^1(\varphi, \varphi)$ there exists a successor state $S_{k+1}$ with $S_{k+1} \models \varphi$, which gives the desired trace and shows the soundness of the rule $\mathrm{INV}_1^1$.

Next let $S_0, S_1, \dots$ be a trace of $M$ with $S_i \models \varphi$ for all $i \ge 0$. If we assume $M \models \square^1 \varphi$, such a trace exists. If we further assume $\models \varphi \rightarrow \psi$, we immediately get also $S_i \models \psi$ for all $i \ge 0$, which shows $M \models \square^1 \psi$ and hence the soundness of the rule $\mathrm{INV}_2^1$.\qed

As the antecedents in the rules INV$_1$ and INV$_1^1$ are formulae of $\mathcal{L}$, Lemmata \ref{lem-invariance} and \ref{lem-inv1} reflect the known fact that invariance proofs do not require any investigation of temporal logic. Note that for the soundness of rule INV$_1$ we have $M \models \varphi$ iff the initial state of $M$ satisfies $\varphi$. This will change for existence, persistence and reachability, and will require rather sophisticated arguments for progress conditions.

\subsection{Variant Terms and Formulae}

We will first concentrate on existence, persistence and reachability formulae in TREBL, in which subformulae are in $\mathcal{L}$ rather than in TREBL. In \cite{hoang:icfem2011} the notions of {\em convergence} and {\em divergence} have been introduced, which differ from existence and persistence conditions only by the treatment of finite traces. For the sake of consistent terminology\footnote{The terminology seems indeed to be counter-intuitive, because in the case of ``convergence'' we may have a trace with infinitely many swaps between states satisfying $\varphi$ and states satisfying $\neg\varphi$, whereas in the case of ``divergence'' we may have traces with infinite tails of states  all satisfying $\varphi$.} we preserve these notions using the specific predicates $\Conv(\varphi)$ and $\Div(\varphi)$ for convergence and divergence formulae, respectively, and formulate derivation rules for them. These derivation rules exploit {\em variant formulae} in $\mathcal{L}$ defined by certain terms in the Event-B machine $M$. We also preserve the terminology of {\em c-variants} and {\em d-variants} used in \cite{ferrarotti:foiks2024} for these. 

As TREBL contains also formulae that are to hold in one trace rather than in all traces, we generalise the investigation of variants for the modified existence and persistence formulae. We also introduce {\em e-variants} to support proofs of reachability formulae.

In addition to the proof of soundness of these derivation rules we show that it is always possible to refine a machine $M$ conservatively to a machine $M^\prime$, in which the needed variant-terms can be defined. These results (proven in our previous work \cite{ferrarotti:foiks2024} for the case of existence and persistence for all-trace temporal formulae) will be the key for relative completeness.

\subsubsection{Variants for Existence Proofs}

For a formula $\varphi \in \mathcal{L}$ a machine $M$ satisfies the existence formula $\square\lozenge \neg \varphi$ iff for every trace $\tau = S_0, S_1, \dots$ of $M$ there is no $k$ such that $S_\ell \models \varphi$ holds for all $\ell \ge k$. In this case we call $M$ to be {\em convergent} in $\varphi$ and we use the notation $\Conv(\varphi)$ for this \cite{ferrarotti:foiks2024,hoang:icfem2011}.

Hoang and Abrial showed that $\Conv(\varphi)$ can be derived by means of {\em variant} terms \cite[p.~460]{hoang:icfem2011}. For this suppose that $t(\bar{v})$ is a well-formed term that takes values in some well-founded set\footnote{For instance, $V$ could be the set $\mathbb{N}$ of natural numbers or the set $\text{HF}(A)$ of hereditarily finite sets over some finite set $A$, as used in \cite{ferrarotti:foiks2024}.} $V$ with minimum 0, and let $\text{var}_c(t,\varphi)$ denote the sentence
\begin{gather}
\forall \bar{v} \bigwedge_{e_i \in E} \forall \bar{x} \bar{v}' \Big( \varphi(\bar{v}) \wedge G_i(\bar{x}, \bar{v}) \rightarrow \big( t(\bar{v}) \neq 0 \wedge ( P_{A_i}(\bar{x}, \bar{v}, \bar{v}') \rightarrow t(\bar{v}') < t(\bar{v}) ) \big) \Big) . \label{eq-conv}
\end{gather}

\noindent
We call the first-order sentence $\text{var}_c(t,\varphi)$ a {\em c-variant formula}, and the term $t(\bar{v})$ a {\em c-variant}. Then $\text{var}_c(t,\varphi)$ is satisfied iff for all states $S$, whenever $\varphi$ holds and an event $e_i$ is enabled, we have that $t(\bar{v})$ evaluates to a non-zero element in $V$, and an execution of $e_i$ decreases $t(\bar{v})$. Hoang and Abrial showed the following:

\begin{lem}\label{lem-conv-rule}

If $\text{var}_c(t,\varphi)$ is a valid formula for $\varphi \in \mathcal{L}$ and no trace of $M$ terminates in a state satisfying $\varphi$, then $M$ satisfies the convergence formula $\Conv(\varphi)$, in other words, the derivation rule
\begin{center}
\begin{minipage}{14.5cm}
\begin{prooftree}
\AxiomC{$M \vdash \text{var}_c(t,\varphi)$}
\AxiomC{\hspace*{-5mm}$M \vdash \, \square \Dlf(\varphi)$}
\BinaryInfC{$M \vdash \Conv(\varphi)$}
\end{prooftree}
\end{minipage}
\end{center}
is sound.

\end{lem}

\noindent
Note that c-variant formulae are sentences, and we request the decrease of the variant in all states satisfying $\varphi$, not just in reachable states. If we omitted the universal quantification over $\bar{v}$, we would instead get a state formula, and we could prove the lemma using an invariance formula expressing the satisfaction in all reachable states satisfying $\varphi$. Then it would become necessary to add this state formula to the invariant. With Lemma \ref{lem-conv} we will show that we can always establish the stronger condition in (\ref{eq-conv}).

\proof

If $M \models \square \Dlf(\varphi)$ holds, then every finite trace of $M$ terminates in a state satisfying $\neg\varphi$.
If $M \models \text{var}_c(t,\varphi)$, but $M \not\models \Conv(\varphi)$, then by the definitions of temporal formulae and $\text{var}_c(t,\varphi)$ there is an infinite subsequence $S_i, S_{i+1}, \dots$ of states of some trace $\tau$ of $M$ such that for the values $b_j$ of $t(\bar{v})$ in $S_j$ we have $b_i > b_{i+1} > \dots$. As the order $\le$ on $V$ is well-founded, this is not possible.\qed


As $\Conv(\varphi)$ is merely another notation for an existence formula $\square\lozenge \neg\varphi$, an immediate consequence of Lemma \ref{lem-conv-rule} is that for $\varphi \in \mathcal{L}$ the derivation rule
\begin{center}
\begin{minipage}{14.5cm}
\begin{prooftree}
\AxiomC{$M \vdash \text{var}_c(t,\neg \varphi)$}
\AxiomC{\hspace*{-5mm}$M \vdash \, \square \Dlf(\neg \varphi)$}
\LeftLabel{{\rm E}: \quad}
\BinaryInfC{$M \vdash \square\lozenge \, \varphi$}
\end{prooftree}
\end{minipage} 
\end{center}
is also sound.

The derivation rule E allows us to reduce proofs of existence conditions to proofs of formulae in first-order logic, if we can find appropriate c-variant terms for $M$. We will show that this is always possible in some conservative refinement of $M$. Let us write $\text{var}_c(M, t, \varphi)$ for the sentence $\text{var}_c(t,\varphi)$ to emphasise that the set of events $E$ is taken from the Event-B machine $M$. 

\begin{lem}\label{lem-conv}

Let $M$ be an Event-B machine with state variables $\bar{v} = (v_0, ..., v_m)$ satisfying a convergence property $\mathit{conv}(\varphi)$. Then there exists an Event-B machine $M'$ with state variables $\bar{v}' = (\bar{v}, \bar{w}, l, s, u)$, where $\bar{v}$, $\bar{w} = (w_0, \ldots, w_m)$, $l$, $s$ and $u$ are pairwise different variables, and a term $t(\bar{v}, \bar{w}, l, s, u)$ such that: 

\begin{enumerate}

\item The formula $\mathit{var}_c(M, t, \varphi)$ is valid for $M'$.  
 
\item For each trace $\sigma$ of $M$ there exists a trace $\sigma'$ of $M'$ with $\sigma = \sigma'|_{s=1,\bar{v}}$ and vice versa, where $\sigma'|_{s=1,\bar{v}}$ is the sequence of states resulting from the selection of those with $s = 1$ and projection to the state variables $\bar{v}$.

\end{enumerate}

\end{lem}

\noindent
The full proof has been outsourced to Appendix \ref{appA}, because it was already published in the extended version of \cite{ferrarotti:foiks2024}. The main idea in it can be sketched briefly as follows. The machine $M'$ uses a copy $\bar{w}$ of the state variables, and events are first executed on them. Furthermore, $M'$ uses a status variable $s$. As long as the machine $M'$ is in a state satisfying $\varphi$ and $s=0$ holds, the tuple $\bar{w}$ is added to a list $l$. In case of a state satisfying $\neg\varphi$ the machine will switch to status $s=1$. As there can only be finite subsequences of states satisfying $\varphi$, such a status switch will occur, unless the machine runs into a deadlock. When the status $s=1$ is reached, the collected tuples in the list $\ell$ will be one-by-one copied to the original state variables $\bar{v}$, until finally the machine switches back to status $s=0$. In this way all firing of events are first executed on a copy $\bar{w}$, and the length of the list $l$ defines the desired variant term, which is reset by each status switch from $s=0$ to $s=1$. In addition, a few subtle cases concerning deadlocks need to be considered separately, for which another status variable $u$ is used.

\subsubsection{Local Convergence}

For TREBL we also need to investigate all diversified existence formulae $XY \varphi$ with $X \in \{ \square, \square^1 \}$, $Y \in \{ \lozenge, \lozenge^1 \}$ and (for now) $\varphi \in \mathcal{L}$. We will proceed analogously using a one-trace counterpart of convergence, for which we define a predicate $\text{conv}^1(\varphi)$ for $\varphi \in \mathcal{L}$. 

It is tempting to define that $M$ satisfies $\text{conv}^1(\varphi)$ iff there exists a trace $S_0, S_1, \dots$ of $M$ such that there is no $k$ with $S_\ell \models \varphi$ for all $\ell \ge k$. If such a trace were finite, the last state $S_k$ would satisfy $\neg\varphi$; if the claimed trace were infinite, there would be infinitely many states $S_i$ with $S_i \models \neg \varphi$. However, as we will see soon, this definition would be too weak for the required proofs by means of variants.

Instead, we must require that there is no ``fantail'' of states all satisfying $\varphi$. That is, we say that $M$ is {\em locally convergent} in $\varphi$ iff there exists a trace $S_0, S_1, \dots$ of $M$ such that there is no $k$ with $S_{\ell}^\prime \models \varphi$ for all $\ell \ge k$ and all traces $S_0^\prime, S_1^\prime, \dots$ of $M$ with $S_i^\prime = S_i$ for all $i \le k$. We write $M \models \text{conv}^1(\varphi)$, if $M$ is locally convergent in $\varphi$.

We will now show that local convergence can be proven by means of variant terms---in this case we will call them {\em lc-variants}---and this will give rise to derivation rules for all diversified existence formulae that use at least one one-trace temporal operator. For this suppose that $t(\bar{v})$ is again a well-formed term that takes values in some well-founded set $V$ with minimum 0, and let $\text{var}_c^1(t,\varphi)$ denote the sentence
\begin{gather}
\forall \bar{v} \Big( \varphi(\bar{v}) \rightarrow t(\bar{v}) \neq 0 \wedge \bigvee_{e_i \in E} \exists \bar{x} \bar{v}' \big( G_i(\bar{x}, \bar{v}) \wedge P_{A_i}(\bar{x}, \bar{v}, \bar{v}') \wedge t(\bar{v}') < t(\bar{v}) \big) \Big) \; . \label{eq-conv1}
\end{gather}

\noindent
We call the first-order sentence $\text{var}_c^1(t,\varphi)$ an {\em lc-variant formula}, and the term $t(\bar{v})$ is called an {\em lc-variant}. Then $\text{var}_c^1(t,\varphi)$ is satisfied iff for all states $S$, whenever $\varphi$ holds, then $t(\bar{v})$ evaluates to a non-zero element in $V$ and there is an enabled event $e_i$, the execution of which decreases $t(\bar{v})$. We can show the following:

\begin{lem}\label{lem-conv1-rule}

If $\text{var}_c^1(t,\varphi)$ is a valid formula for $\varphi \in \mathcal{L}$, then $M$ satisfies the local convergence formula $\Conv^1(\varphi)$, in other words, the derivation rule
\begin{center}
\begin{minipage}{14.5cm}
\begin{prooftree}
\AxiomC{$M \vdash \text{var}_c^1(t,\varphi)$}
\UnaryInfC{$M \vdash \Conv^1(\varphi)$}
\end{prooftree}
\end{minipage}   
\end{center}
is sound.

\end{lem}

\proof

Assume that $M \models \text{var}_c^1(t,\varphi)$, but $M \not\models \Conv^1(\varphi)$. Then for all traces $S_0, S_1, \dots$ of $M$ there is some $k$ such that $S_\ell^\prime \models \varphi$ for all $\ell \ge k$ and all traces $S_0^\prime, S_1^\prime, \dots$ that coincide on the first $k+1$ states with $S_0, \dots, S_k$. According to the definition of $\text{var}_c^1(t,\varphi)$ we obtain for each such trace a strictly decreasing sequence $v_k > v_{k+1} > \dots$ of values in $V$. As $V$ is well-founded, this is impossible.\qed


Let us now look at the diversified one-trace existence formulae. A formula $\square^1 \lozenge^1 \varphi$ is equivalent to $\neg \lozenge\square \neg\varphi$, and the subformula $\lozenge\square \neg\varphi$ expresses that all traces of $M$ have a $\neg\varphi$ fantail, i.e. $\square^1 \lozenge^1 \varphi$ is equivalent to $\Conv^1(\neg\varphi)$. Therefore, a direct consequence of Lemma \ref{lem-conv1-rule} is the soundness of the derivation rule
\begin{center}
\begin{minipage}{14.5cm}
\begin{prooftree}
\AxiomC{$M \vdash \text{var}_c^1(t,\neg\varphi)$}
\LeftLabel{{\rm E$^{11}$}: \quad}
\UnaryInfC{$M \vdash \square^1 \lozenge^1 \varphi$}
\end{prooftree}
\end{minipage}  
\end{center}

\noindent
Analogously, for a formula $\square \lozenge^1 \varphi$ we must have the absence of $\neg\varphi$ fantails for all traces. In particular, traces cannot terminate in states satisfying $\neg\varphi$. Therefore, another direct consequence of Lemma \ref{lem-conv1-rule} is the soundness of the derivation rule
\begin{center}
\begin{minipage}{14.5cm}
\begin{prooftree}
\AxiomC{$M \vdash \text{var}_c^1(t,\neg\varphi)$}
\AxiomC{$M \vdash \square \Dlf(\neg\varphi)$}
\LeftLabel{{\rm E$^{01}$}: \quad}
\BinaryInfC{$M \vdash \square \lozenge^1 \varphi$}
\end{prooftree}
\end{minipage}   
\end{center}

\noindent
Thirdly, for a formula $\square^1 \lozenge \varphi$ we must have the absence of $\neg\varphi$ fantails for some trace, and in addition in every trace we must have a state satisfying $\varphi$. Therefore, another direct consequence of Lemma \ref{lem-conv1-rule} is the soundness of the derivation rule
\begin{center}
\begin{minipage}{14.5cm}
\begin{prooftree}
\AxiomC{$M \vdash \text{var}_c^1(t,\neg\varphi)$}
\AxiomC{$M \vdash \lozenge \varphi$}
\LeftLabel{{\rm E$^{10}$}: \quad}
\BinaryInfC{$M \vdash \square^1 \lozenge \varphi$}
\end{prooftree}
\end{minipage}  
\end{center}

\noindent
To complete the investigation of diversified existence formulae we show the analogue of Lemma \ref{lem-conv}, i.e. the existence of variant terms for local convergence in a conservatively refined Event-B machine. The proof of the following lemma will be mostly identical to the proof of Lemma \ref{lem-conv} and will be given in Appendix \ref{appA}.

\begin{lem}\label{lem-conv1}

Let $M$ be an Event-B machine with state variables $\bar{v} = (v_0, ..., v_m)$ satisfying a local convergence property $\mathit{conv}^1(\varphi)$. Then there exists an Event-B machine $M'$ with state variables $\bar{v}' = (\bar{v}, \bar{w}, l, s, u, c)$, where $\bar{v}$, $\bar{w} = (w_0, \ldots, w_m)$, $l$, $s$, $u$ and $c$ are pairwise different variables, and a term $t(\bar{v}, \bar{w}, l, s, u, c)$ such that: 

\begin{enumerate}

\item The formula $\mathit{var}_c^1(M, t, \varphi)$ is valid for $M'$.  
 
\item For each trace $\sigma$ of $M$ there exists a trace $\sigma'$ of $M'$ with $\sigma = \sigma'|_{s=1,\bar{v}}$ and vice versa, where $\sigma'|_{s=1,\bar{v}}$ is the sequence of states resulting from the selection of those with $s = 1$ and projection to the state variables $\bar{v}$.

\end{enumerate}

\end{lem}

\subsubsection{Variants for Persistence Proofs}

Analogously, $M$ satisfies a persistence formula $\lozenge\square\, \varphi$ with $\varphi \in \mathcal{L}$ iff for every trace $\tau = S_0, S_1, \dots$ of $M$ there is a $k$ such that $S_{k'} \models \varphi$ for every $k' \geq k$. This implies that finite traces must terminate in a state $S_{\ell(\tau)-1}$ satisfying $\varphi$. Disregarding         restriction for finite traces leads to divergence. We say that $M$ {\em is divergent} in $\varphi$---notation: $M \models \Div(\varphi)$---iff for every infinite trace $\tau = S_0, S_1, \dots$ of $M$ there is a $k$ such that $S_{k'} \models \varphi$ for every $k' \geq k$. 

The trivial extension $\tilde{M}$ is defined by adding an event that fires in a state $S$ iff no other event can be fired, i.e. $S \models \neg \Dlf(\textbf{true})$ holds, and then it does nothing. Formally, the guard of this dummy event is $\neg \Dlf(\textbf{true})$, and the action is the list of assignments $\mathbf{v}_i := \mathbf{v}_i$ for all state variables $\mathbf{v}_i$. Note that if $\tilde{M} \models \Div(\varphi)$ holds, then also $M \models \Div(\varphi)$ holds, but the converse is false in general. However, if $M \models \Div(\varphi)$ holds and all finite traces terminate in a state satisfying $\varphi$, then also $\tilde{M} \models \Div(\varphi)$ follows.

Hoang and Abrial showed that $\Div(\varphi)$ can also be derived by means of {\em variant} terms \cite[p.~460]{hoang:icfem2011}. For this suppose again that $t(\bar{v})$ is a well-formed term that takes values in some well-founded set $V$ with minimum 0, and let $\text{var}_d(t,\varphi)$ denote the sentence
\begin{gather}
\forall \bar{v} \bigg( \bigwedge_{e_i \in E} \forall  \bar{x} \bar{v}' \Big( \neg \varphi(\bar{v}) \wedge G_i(\bar{x}, \bar{v}) \rightarrow \big( t(\bar{v}) \neq 0 \wedge ( P_{A_i}(\bar{x}, \bar{v}, \bar{v}') \rightarrow t(\bar{v}') < t(\bar{v}) ) \big) \Big) \notag\\
\wedge \Big( \varphi(\bar{v}) \wedge G_i(\bar{x}, \bar{v}) \wedge P_{A_i}(\bar{x}, \bar{v}, \bar{v}') \rightarrow t(\bar{v}') \le t(\bar{v}) ) \Big) \bigg) \; .\label{eq-div}
\end{gather}

\noindent
We call the first-order sentence $\text{var}_d(t,\varphi)$ a {\em d-variant formula}, and the term $t(\bar{v})$ is called a {\em d-variant}. Then $\text{var}_d(t,\varphi)$ is satisfied iff for all states $S$, whenever $\neg\varphi$ holds and an event $e_i$ is enabled, we have that $t(\bar{v})$ evaluates to a non-zero element of $V$, and an execution of $e_i$ decreases $t(\bar{v})$, and whenever $\varphi$ holds and an event $e_i$ is enabled, we have that an execution of $e_i$ does not increase $t(\bar{v})$. Hoang and Abrial showed the following:

\begin{lem}\label{lem-div-rule}

If $\text{var}_d(t,\varphi)$ is a valid formula, then $M$ is divergent in $\varphi$, in other words, the derivation rule
\begin{center}
\begin{minipage}{14.5cm}
\begin{prooftree}
\AxiomC{$M \vdash \text{var}_d(t,\varphi)$}
\UnaryInfC{$M \vdash \Div(\varphi)$}
\end{prooftree}
\end{minipage}  
\end{center}
is sound.

\end{lem}

\proof

If $M \models \text{var}_d(t,\varphi)$, but $M \not\models \Div(\varphi)$, then by the definition of $\Div(\varphi)$ there is an infinite trace $\sigma = S_0, S_1, \dots$ of $M$ such that for all $i$ there is some $k_i \ge i$ with $S_{k_i} \models \neg\varphi$. Let $b_j$ denote the value of $t(\bar{v})$ in $S_j$. Then by the definition of $\text{var}_d(t,\varphi)$ we have $b_i \ge b_{i+1}$ and $b_{k_i} > b_{k_i + 1}$ for all $i$. As the order $\le$ on $V$ is well-founded, this is not possible.\qed


As $\Div(\varphi) \wedge \square \Dlf(\neg \varphi)$ is equivalent to the existence formula $\lozenge\square \,\varphi$, an immediate consequence of Lemma \ref{lem-div-rule} is that for $\varphi \in \mathcal{L}$ the derivation rule
\begin{center}
\begin{minipage}{14.5cm}
\begin{prooftree}
\AxiomC{$M \vdash \, \text{var}_d(t,\varphi)$}
\AxiomC{$M \vdash \, \square \Dlf(\neg \varphi)$}
\LeftLabel{$\mathrm{P}$: \quad}
\BinaryInfC{$M \vdash \lozenge \square \, \varphi$}
\end{prooftree}
\end{minipage}
\end{center}
is sound for the derivation of persistence formulae.

The derivation rule P allows us to reduce proofs of persistence formulae to proofs of formulae in first-order logic, if we can find appropriate d-variant terms. We will show that this is always possible in a conservative refinement of the machine $M$. Let $\text{var}_d(M, t,\varphi)$ denote the formulae $\text{var}_d(t,\varphi)$ to emphasise the dependence of the specification of $M$. 

\begin{lem}\label{lem-div}

Let $M$ be an Event-B machine with state variables $\bar{v} = (v_0, ..., v_m)$ satisfying a divergence property $\mathit{div}(\varphi)$. Then there exists an Event-B machine $M'$ with state variables $\bar{v}' = (\bar{v}, \bar{c}, \bar{b})$, where $\bar{c} = (c_0, \ldots, c_m)$, $\bar{b} = (b_0, \ldots, b_m)$ and $\bar{v}$ are pairwise different variables, and a term $t(\bar{v}, \bar{c}, \bar{b})$ such that: 

\begin{enumerate}

\item The formula $\mathit{var}_d(M', t, \varphi)$ is valid for $M'$;
 
\item For each trace $\sigma$ of $M$ there exists a trace $\sigma'$ of $M'$ with $\sigma = \sigma'|_{\bar{v}}$ and vice versa, where $\sigma'|_{\bar{v}}$ is the sequence of states resulting from the projection to the state variables $\bar{v}$.

\end{enumerate}

\end{lem}

\noindent
Again before entering into the proof let us sketch the key idea as follows. For $V$ we encode all values as hereditarily finite sets in $\text{HF}(A)$, which is naturally ordered, and the order is well-founded. In $M'$ for each state variable $v_i$ we add a state variable $c_i$, which we initialise as $\emptyset$, and into which a new value $v'$ is inserted, whenever the machine makes a step in a state satisfying $\neg\varphi$. Then the tuple $(c_0, \dots, c_m)$ is a monotone increasing sequence of tuples of sets. As in every trace there are only finitely many states satisfying $\neg\varphi$, there exists a maximum tuple that will never be exceeded. In $M'$ we initialise $(b_0, \dots, b_m)$ by these maximum set values, and never update these state variables. As the order $\le$ on $\text{HF}(A)$ subsumes set inclusion, we can see that $\#(b_0, \dots, b_m) - \#(c_0, \dots, c_m)$ defines the desired variant term. The full proof is given in Appendix \ref{appA}. It also appeared in the extended version of \cite{ferrarotti:foiks2024}.

\subsubsection{Local Divergence}

For TREBL we also need to investigate all diversified persistence formulae $YX \varphi$ with $X \in \{ \square, \square^1 \}$, $Y \in \{ \lozenge, \lozenge^1 \}$ and $\varphi \in \mathcal{L}$. We first introduce a one-trace analogue of divergence, for which we define a predicate $\text{div}^1(\varphi)$ for $\varphi \in \mathcal{L}$. We say that $M$ is {\em locally divergent} in $\varphi$ (notation: $M \models \text{div}^1(\varphi)$ iff there exists a trace $S_0, S_1, \dots$ of $M$ and some $k$ such that $S_\ell \models \varphi$ holds for all $\ell \ge k$.

We will now show that local divergence can also be proven by means of variant terms---in this case we will call them {\em ld-variants}---and this will give rise to a derivation rule for one of the diversified persistence formulae that uses both one-trace temporal operators. For this suppose that $t(\bar{v})$ is again a well-formed term that takes values in some well-founded set $V$ with minimum 0, and let $\text{var}_d^1(t,\varphi)$ denote the sentence
\begin{gather}
\forall \bar{v} \bigg( \Big( \neg \varphi(\bar{v}) \rightarrow t(\bar{v}) \neq 0 \wedge \exists \bar{x} \bar{v}' \bigvee_{e_i \in E} \big( G_i(\bar{x}, \bar{v}) \wedge P_{A_i}(\bar{x}, \bar{v}, \bar{v}') \wedge t(\bar{v}') < t(\bar{v}) \big) \Big) \wedge \notag\\
\Big( \big( \varphi(\bar{v}) \wedge \exists \bar{x} \bigvee_{e_i \in E} G_i(\bar{x}, \bar{v}) \big) \rightarrow \exists \bar{x} \bar{v}' \bigvee_{e_i \in E} \big( G_i(\bar{x}, \bar{v}) \wedge P_{A_i}(\bar{x}, \bar{v}, \bar{v}') \wedge t(\bar{v}') \le t(\bar{v}) \big) \Big) \bigg) \; .\label{eq-div1}
\end{gather}

\noindent
We call the first-order sentence $\text{var}_d^1(t,\varphi)$ an {\em ld-variant formula}, and the term $t(\bar{v})$ is called an {\em ld-variant}. We can show the following:

\begin{lem}\label{lem-div1-rule}

If $\text{var}_d^1(t,\varphi)$ is a valid formula, then $M$ is locally divergent in $\varphi$, in other words, the derivation rule
\begin{center}
\begin{minipage}{14.5cm}
\begin{prooftree}
\AxiomC{$M \vdash \text{var}_d^1(t,\varphi)$}
\UnaryInfC{$M \vdash \Div^1(\varphi)$}
\end{prooftree}
\end{minipage}    
\end{center}
is sound.

\end{lem}

\proof

If $M \models \text{var}_d^1(t,\varphi)$, but $M \not\models \Div^1(\varphi)$, then by the definition of $\Div^1(\varphi)$ any trace $S_0, S_1, \dots$ of $M$ is either finite terminating in a state $S_k$ with $S_k \models \neg\varphi$ or is infinite with infinitely many $S_i$ satisfying $S_i \models \neg\varphi$. The first case is excluded, because by the definition of $\text{var}_d^1(t,\varphi)$ there exists an event $e_i$ that can be fired in $S_k$. In the second case the definition of $\text{var}_d^1(t,\varphi)$ implies that there exists an infinite decreasing sequence of values in $V$. As the order $\le$ on $V$ is well-founded, this is not possible.\qed


The diversified persistence formula $\lozenge^1 \square^1 \varphi$ holds for $M$ iff there exists a trace that is either finite and terminates in a state satisfying $\varphi$ or it is infinite and there are only finitely many states in the trace satisfying $\neg\varphi$. That is, the formula is equivalent to $\text{div}^1(\varphi)$. Hence, a direct consequence of Lemma \ref{lem-div1-rule} is the soundness of the derivation rule
\begin{center}
\begin{minipage}{14.5cm}
\begin{prooftree}
\AxiomC{$M \vdash \, \text{var}_d^1(t,\varphi)$}
\LeftLabel{$\mathrm{P}^{11}$: \quad}
\UnaryInfC{$M \vdash \lozenge^1 \square^1  \, \varphi$}
\end{prooftree}
\end{minipage}   
\end{center}

\noindent
We can again show that ld-variants exist in sufficiently refined machines, which will imply relative completeness. The proof of the following lemma is given in Appendix \ref{appA}.

\begin{lem}\label{lem-div1}

Let $M$ be an Event-B machine with state variables $\bar{v} = (v_0, ..., v_m)$ satisfying a local divergence property $\mathit{div}^1(\varphi)$. Then there exists an Event-B machine $M'$ with state variables $\bar{v}' = (\bar{v}, \bar{c}, \bar{b})$, where $\bar{c} = (c_0, \ldots, c_m)$, $\bar{b} = (b_0, \ldots, b_m)$ and $\bar{v}$ are pairwise different variables, and a term $t(\bar{v}, \bar{c}, \bar{b})$ such that: 

\begin{enumerate}

\item The formula $\mathit{var}_d^1(M', t, \varphi)$ is valid for $M'$;
 
\item For each trace $\sigma$ of $M$ there exists a trace $\sigma'$ of $M'$ with $\sigma = \sigma'|_{\bar{v}}$ and vice versa, where $\sigma'|_{\bar{v}}$ is the sequence of states resulting from the projection to the state variables $\bar{v}$.

\end{enumerate}

\end{lem}

\noindent
This leaves the cases of diversified persistence formulae, in which one temporal operator refers to all traces and the other one to just one trace. $M$ satisfies $\lozenge\square^1 \varphi$ iff in every trace $S_0, S_1, \dots$ of $M$ there exists a $k$ such that there is a trace $S_0^\prime, S_1^\prime, \dots$ with $S_i^\prime = S_i$ for all $i \le k$ and $S_\ell \models \varphi$ for all $\ell \le k$. In particular, all finite traces end in a state $S_k$ satisfying $\varphi$. 

Phrased differently, every trace has a prefix $S_0, \dots, S_k$, and in the ``fantail'' starting with $S_k$ there is one trace with only finitely many states $S_\ell$ satisfying $\neg\varphi$, and in case of a finite trace terminating in a state satisfying $\varphi$. Therefore, let $t_1$ and $t_2$ be terms that take values in well-founded sets $V_1$ and $V_2$, respectively, and let the minimal elements in these sets be denoted as $0$. Then define the sentence $\text{var}_d^{p1}(t, \varphi)$ with $t = (t_1, t_2)$ as
\begin{gather}
\forall \bar{v} \Bigg( \Big( t_1(\bar{v}) \neq 0 \rightarrow \Big( \neg\varphi \rightarrow \bigvee_{e_i \in E} \exists \bar{x} \bar{v}^\prime \big( G_i(\bar{x}, \bar{v}) \wedge P_{A_i}(\bar{x}, \bar{v}, \bar{v}^\prime) \big) \wedge \notag\\
\bigwedge_{e_i \in E} \forall \bar{x} \bar{v}^\prime
\big( G_i(\bar{x}, \bar{v}) \wedge P_{A_i}(\bar{x}, \bar{v}, \bar{v}^\prime) \rightarrow t_1(\bar{v}^\prime) < t_1(\bar{v}) \big) \Big) \notag\\
\wedge \Big( \varphi(\bar{v}) \rightarrow \bigwedge_{e_i \in E} \forall \bar{x} \bar{v}^\prime \big( G_i(\bar{x}, \bar{v}) \wedge P_{A_i}(\bar{x}, \bar{v}, \bar{v}^\prime) \rightarrow t_1(\bar{v}^\prime) \le t_1(\bar{v}) \big) \Big) \Big) \notag\\
\wedge \Bigg( t_1(\bar{v}) = 0 \rightarrow 
\bigg( \neg \varphi(\bar{v}) \rightarrow t_2(\bar{v}) \neq 0 \wedge \bigvee_{e_i \in E} \exists \bar{x} \bar{v}^\prime
\big( G_i(\bar{x}, \bar{v}) \wedge P_{A_i}(\bar{x}, \bar{v}, \bar{v}^\prime) \wedge t_2(\bar{v}^\prime) < t_2(\bar{v})  \big) \bigg) \notag\\
\wedge \bigg( \varphi(\bar{v}) \wedge \bigvee_{e_i \in E} \exists \bar{x} G_i(\bar{x}, \bar{v}) \rightarrow \bigvee_{e_i \in E} \exists \bar{x} \bar{v}^\prime
\big( G_i(\bar{x}, \bar{v}) \wedge P_{A_i}(\bar{x}, \bar{v}, \bar{v}^\prime) \wedge t_2(\bar{v}^\prime) \le t_2(\bar{v})  \big) \bigg) \Bigg) \Bigg) \; . \label{eq-pdiv1}
\end{gather}

\noindent
We call this sentence a {\em pld-variant formula} and the pair term $t = (t_1, t_2)$ a {\em pld-variant}\footnote{This notation is motivated by the formula being a modification of $\text{var}_d^1(t, \varphi)$ with prefixes for all traces.}. We can show the following lemma:

\begin{lem}\label{lem-divp1-rule}

For $t = (t_1, t_2)$ and $\varphi \in \mathcal{L}$ the derivation rule
\begin{center}
\begin{minipage}{14.5cm}
\begin{prooftree}
\AxiomC{$M \vdash \text{var}_d^{p1}(t,\varphi)$}
\LeftLabel{$\mathrm{P}^{01}$: \quad}
\UnaryInfC{$M \vdash \lozenge \square^1  \, \varphi$}
\end{prooftree}
\end{minipage}   
\end{center}
is sound.
\end{lem}

\proof

Assume that $M$ satisfies $\text{var}_d^{p1}(t,\varphi)$ with $t = (t_1, t_2)$. As $V_1$ is well-founded, there cannot be a trace with infinitely many states satisfying $\neg\varphi$; in such states any firing of an event strictly decreases the value of $t_1$. Thus, in every trace we either have a state $S_k$ with $S_\ell \models \varphi$ for all $\ell \ge k$ or $S_k \models t_1(\bar{v}) = 0$.

For a state $S_k$ satisfying $t_1(\bar{v}) = 0$ there exists a continuation of the trace, which either terminates in a state satisfying $\varphi$ or in which the values of the term $t_2$ decrease, and they decrease strictly for each state satisfying $\neg\varphi$. As $V_2$ is well-founded, there cannot exist a trace with a strictly decreasing sequence of values for the term $t_2$, which implies that we must reach a state $S_k$ with $S_\ell \models \varphi$ for all $\ell \ge k$. Hence $M$ satisfies $\lozenge \square^1 \varphi$ as claimed.\qed


Analogously, $M$ satisfies $\lozenge^1 \square \varphi$ iff there exists one trace $S_0, S_1, \dots$ and some $k$ such that $S_\ell^\prime \models \varphi$ holds for all $\ell \ge k$ and all traces $S_0^\prime, S_1^\prime, \dots$ with $S_i^\prime = S_i$ for all $i \le k$. Informally phrased, there exists a trace with a fantail of states all satisfying $\varphi$. Therefore, let $t_1$ and $t_2$ be terms that take values in well-founded sets $V_1$ and $V_2$, respectively, and let the minimal elements in these sets be denoted as $0$. Then define the sentence $\text{var}_d^p(t, \varphi)$ with $t = (t_1, t_2)$ as
\begin{gather}
\forall \bar{v} \bigg( \Big( t_1(\bar{v}) \neq 0 \rightarrow \exists \bar{x} \bar{v}^\prime \bigvee_{e_i \in E} \big( G_i(\bar{x}, \bar{v}) \wedge P_{A_i}(\bar{x}, \bar{v}, \bar{v}^\prime) \notag\\
\wedge ( \neg \varphi(\bar{v}) \rightarrow t_1(\bar{v}^\prime) < t_1(\bar{v}) ) \wedge ( \varphi(\bar{v}) \rightarrow t_1(\bar{v}^\prime) \le t_1(\bar{v}) ) \big) \Big) \notag\\
\wedge \Big( t_1(\bar{v}) = 0 \rightarrow \bigwedge_{e_i \in E} \forall \bar{x} \bar{v}^\prime \notag\\
\Big( \neg \varphi(\bar{v}) \wedge G_i(\bar{x}, \bar{v}) \rightarrow \big( t_2(\bar{v}) \neq 0 \wedge ( P_{A_i}(\bar{x}, \bar{v}, \bar{v}^\prime) \rightarrow t_2(\bar{v}^\prime) < t_2(\bar{v}) ) \big) \Big) \notag\\
\wedge \Big( \varphi(\bar{v}) \wedge G_i(\bar{x}, \bar{v}) \wedge P_{A_i}(\bar{x}, \bar{v}, \bar{v}^\prime) \rightarrow t_2(\bar{v}^\prime) \le t_2(\bar{v}) \Big) \Big) \bigg) \; . \label{eq-pdiv}
\end{gather}

\noindent
We call this sentence a {\em pd-variant formula} and the pair term $t = (t_1, t_2)$ a {\em pd-variant}\footnote{This notation is motivated by the formula being a modification of $\text{var}_d(t, \varphi)$ with a prefix for some trace.}. We can show the following lemma:

\begin{lem}\label{lem-divp-rule}

For $t = (t_1, t_2)$ and $\varphi \in \mathcal{L}$ the derivation rule
\begin{center}
\begin{minipage}{14.5cm}
\begin{prooftree}
\AxiomC{$M \vdash \text{var}_d^p(t,\varphi)$}
\LeftLabel{$\mathrm{P}^{10}$: \quad}
\UnaryInfC{$M \vdash \lozenge^1 \square  \, \varphi$}
\end{prooftree}
\end{minipage}
\end{center}
is sound.

\end{lem}

\proof

Assume that $M$ satisfies $\text{var}_d^p(t, \varphi)$ with pd-variant $t = (t_1, t_2)$. As $V_1$ is well-founded, there cannot exist a trace, in which the values of $t_1(\bar{v})$ form a strictly decreasing sequence. Therefore, the first part of (\ref{eq-pdiv}) implies that there exists a trace with a finite prefix of states satisfying $t_1(\bar{v}) \neq 0$ or there exists a trace $S_0, S_1, \dots$ and a $k$ such that $S_\ell \models \varphi$ holds for all $\ell \ge k$.

In the former case for every trace $S_0, S_1, \dots$ continuing this prefix there must exist a $k$ such that $S_\ell \models \varphi$ holds for all $\ell \ge k$. Otherwise the second part of (\ref{eq-pdiv}) would yield an infinite strictly decreasing sequence of values in $V_2$. As $V_2$ is well-founded, this is impossible.\qed


\begin{rem}

Negations of existence formulae are persistence formulae and vice versa, so the question arises why for persistence four different types of variants are needed, whereas for existence we only needed two. It is possible to define also prefix versions for convergence and local convergence, but these will be equivalent to $\Conv(\varphi)$ or $\Conv^1(\varphi)$.

\end{rem}

\noindent
To conclude the investigation of persistence formulae we address the existence of pld- and pd-variants in refined machines, and prove the analog of Lemmata \ref{lem-div} and \ref{lem-div1} for these cases.

\begin{lem}\label{lem-pdiv}

Let $M$ be an Event-B machine with state variables $\bar{v} = (v_0, ..., v_m)$ satisfying a diversified persistence condition $\lozenge^1 \square \varphi$ or $\lozenge \square^1 \varphi$, respectively. Then there exists an Event-B machine $M'$ with state variables $\bar{v}' = (\bar{v}, \bar{c}, \bar{b}, \bar{p}, \bar{q}, s)$, where $\bar{c} = (c_0, \ldots, c_m)$, $\bar{b} = (b_0, \ldots, b_m)$, $\bar{p} = (p_0, \ldots, p_m)$, $\bar{q} = (q_0, \ldots, q_m)$ and $\bar{v}$ are pairwise different variables and $s$ is a fresh Boolean-valued variable,and terms $t_1(\bar{v}, \bar{p}, \bar{q})$ and $t_2(\bar{v}, \bar{c}, \bar{b})$ such that: 

\begin{enumerate}

\item The formula $\mathit{var}_d^p(M', t, \varphi)$ (or $\mathit{var}_d^{p1}(M', t, \varphi)$, respectively) with $t = (t_1, t_2)$ is valid for $M'$;
 
\item For each trace $\sigma$ of $M$ there exists a trace $\sigma'$ of $M'$ with $\sigma = \sigma'|_{\bar{v}}$ and vice versa, where $\sigma'|_{\bar{v}}$ is the sequence of states resulting from the projection to the state variables $\bar{v}$.

\end{enumerate}

\end{lem}

\proof

We exploit the same idea as for the proofs of Lemmata \ref{lem-div} and \ref{lem-div1}. However, for each state variable $v_i$ we now add two state variables $c_i$ and $p_i$, both initialised as $\emptyset$, and we use an additional status variable $s$ that is initialised to $0$. Into $p_i$ we insert a new value $v^\prime$ whenever $v_i$ is updated to $v^\prime$ in a state satisfying $\neg\varphi \wedge s=0$, and we insert $v^\prime$ into $c_i$ whenever $v_i$ is updated to $v^\prime$ in a state satisfying $\neg\varphi \wedge s=1$. Furthermore, we initialise $\bar{q}$ to a tuple of maximum values that are never exceeded in the prefix(es) of the trace(s) that are guaranteed by $\lozenge^1 \square \varphi$ or $\lozenge \square^1 \varphi$, respectively, and we initialise $\bar{b}$ to a tuple of maximum values that are never exceeded in the postfix(es) of the trace(s) that are guaranteed by $\lozenge^1 \square \varphi$ or $\lozenge \square^1 \varphi$, respectively. In case an insertion into $p_i$ exceeds $q_i$, the status variable $s$ is changed to $1$.

We define the variant term $t = (t_1, t_2)$ with $t_1(\bar{v}, \bar{p}, \bar{q}) = (q_0 - p_0, \dots, q_m - p_m)$ and  $t_2(\bar{v}, \bar{c}, \bar{b}) = (b_0 - c_0, \dots, b_m - c_m)$. Then the detailed arguments to show conditions (i) and (ii) are exactly the same as in the proofs of Lemmata \ref{lem-div} and \ref{lem-div1}.\qed


\subsubsection{Variants for Reachability Proofs}

To complete our investigation of proofs of liveness conditions by means of variant formulae we now look at reachability formulae $\lozenge \varphi$ and $\lozenge^1 \varphi$. In the former case for every trace $S_0, S_1, \dots$ there must exist some minimal $k$ such that all states $S_i$ with $i < k$ satisfy $\neg\varphi$. Therefore, let $t_1$ be a term that takes values in some well-founded set $V$ with minimal element $0$, and assume that one state variable $b$ takes Boolean values. In particular, $b = v_i$ and $b^\prime = v_i^\prime$ hold for some $i$. Then we define the sentence $\text{var}_e(t, \varphi)$ with $t = (t_1, b)$ as
\begin{gather}
\forall \bar{v} \bigg( \bigwedge_{e_i \in E} \forall \bar{x} \bar{v}' \Big( \varphi(\bar{v}) \wedge b=0 \wedge G_i(\bar{x}, \bar{v}) \rightarrow \qquad\qquad \notag\\
\big( t_1(\bar{v}) \neq 0 \wedge ( P_{A_i}(\bar{x}, \bar{v}, \bar{v}') \rightarrow t_1(\bar{v}') < t_1(\bar{v}) \wedge b^\prime = 0 ) \big) \Big) \bigg) \; . \label{eq-reach}
\end{gather}

\noindent
We call this sentence an {\em e-variant formula} and the pair term $t = (t_1, b)$ an {\em e-variant}. We can show the following lemma:

\begin{lem}\label{lem-reach-rule}

If $\square \text{var}_e(t,\neg\varphi)$ with $t = (t_1, b)$ is a valid formula, then every trace of $M$ with an initial state satisfying $b=0$ reaches a state satisfying $\varphi$, in other words, the derivation rule
\begin{center}
\begin{minipage}{14.5cm}
\begin{prooftree}
\AxiomC{$M \vdash b = 0$}
\AxiomC{$M \vdash \text{var}_e(t, \neg\varphi)$}
\LeftLabel{$\mathrm{R}$: \quad}
\BinaryInfC{$M \vdash \lozenge \, \varphi$}
\end{prooftree}
\end{minipage}   
\end{center}
is sound.

\end{lem}

\proof

If $M \models \text{var}_e(t, \neg\varphi) \wedge b=0$ holds, but $M \not\models \lozenge \varphi$, then by definition of $\text{var}_e(t, \neg\varphi)$ there is a trace $S_0, S_1, \dots$ with $S_i \models \neg\varphi \wedge b=0$ for all $i$, and for the values $b_i$ of $t(\bar{v})$ in $S_i$ we have $b_i > b_{i+1} > \dots$. As the order $\le$ on $V$ is well-founded, this is not possible.\qed


For a one-trace reachability formula $\lozenge^1 \varphi$ we also use a pair term $t = (t_1, b)$ as above, which we call {\em e1-variant}, and define an {\em e1-variant formula} as
\begin{gather}
\forall \bar{v} \Big( \varphi(\bar{v}) \wedge b=0 \rightarrow t_1(\bar{v}) \neq 0 \notag\\
\wedge \bigvee_{e_i \in E} \exists \bar{x} \bar{v}^\prime \big( G_i(\bar{x}, \bar{v}) \wedge P_{A_i}(\bar{x}, \bar{v}, \bar{v}^\prime) \wedge t_1(\bar{v}^\prime) < t_1(\bar{v}) \wedge b^\prime = 0 \big) \Big) \; .
\label{eq-reach1}
\end{gather}

\noindent
The following lemma is the analog of Lemma \ref{lem-reach-rule} for the case of reachability of a state satisfying $\varphi$ in one trace.

\begin{lem}\label{lem-reach1-rule}

If $\text{var}_e^1(t,\neg\varphi)$ with $t = (t_1, b)$ is a valid formula, then there exists a trace of $M$ with an initial state satisfying $b=0$ that reaches a state satisfying $\varphi$, in other words, the derivation rule
\begin{center}
\begin{minipage}{14.5cm}
\begin{prooftree}
\AxiomC{$M \vdash b = 0$}
\AxiomC{$M \vdash \text{var}_e^1(t, \neg\varphi)$}
\LeftLabel{$\mathrm{R}^1$: \quad}
\BinaryInfC{$M \vdash \lozenge^1 \, \varphi$}
\end{prooftree}
\end{minipage}
\end{center}
is sound.
\end{lem}

\proof

If $M \models \text{var}_e^1(t, \neg\varphi) \wedge b=0$ holds, then by definition either $M \models \varphi$---and hence immediately $M \models \lozenge^1 \varphi$---or there is a trace $S_0, S_1, \dots$  and a $k$ such that $S_i \models \neg\varphi \wedge b=0$ holds for all $i \le k$, and for the values $b_i$ of $t(\bar{v})$ in $S_i$ we have $b_i > b_{i+1} > \dots$. As the order $\le$ on $V$ is well-founded, there must exists a maximal $k$ with this property, and by definition we get $S_k \models \varphi$, hence $M \models \lozenge^1 \varphi$.\qed


To conclude the investigation of reachability formulae we show that in the case of reachability of a state $S$ satisfying $\varphi$ in all traces or in one trace, respectively, we can always define a conservative refinement, in which an e-variant or e1-variant can be defined. The proof of the following lemma is given in Appendix \ref{appA}.

\begin{lem}\label{lem-reach}

Let $M$ be an Event-B machine with state variables $\bar{v} = (v_0, ..., v_m)$ satisfying a reachability property $\lozenge \varphi$ or a one-trace reachability property $\lozenge^1 \varphi$. Then there exists an Event-B machine $M'$ with state variables $\bar{v}' = (\bar{v}, \bar{w}, l, s, u, b)$, where $\bar{v}$, $\bar{w} = (w_0, \ldots, w_m)$, $l$, $s$, $u$ and $b$ are pairwise different variables, and a term $t_1(\bar{v}, \bar{w}, l, s, u)$ such that: 

\begin{enumerate}

\item The formula $\mathit{var}_e(M, t, \varphi)$ (or $\mathit{var}_e(M, t, \varphi)$, respectively) with $t = (t_1, b)$ is valid for $M'$.  
 
\item For each trace $\sigma$ of $M$ there exists a trace $\sigma'$ of $M'$ with $\sigma = \sigma'|_{s=1,\bar{v}}$ and vice versa, where $\sigma'|_{s=1,\bar{v}}$ is the sequence of states resulting from the selection of those with $s = 1$ and projection to the state variables $\bar{v}$.

\end{enumerate}

\end{lem}

\noindent
Lemmata \ref{lem-invariance} -- \ref{lem-reach} above imply that adding the derivation rules INV$_1$, INV$_2$, INV$_1^1$, INV$_2^1$, E, E$^{10}$, E$^{01}$, E$^{11}$, P, P$^{10}$, P$^{01}$, P$^{11}$, R and R$^1$ to a sound and complete set of derivation rules for first-order logic (with set types) yields a sound set of derivation rules for the derivation of all-traces and one-trace invariance and reachability formulae, all versions of existence formulae and all versions of existence formulae, provided that the Event-B machines are conservatively refined to guarantee the existence of the required variants. We use the notion of {\em relative completeness} here to emphasise that the required refinements depend on the formulae that are to be proven. Of course, for a proof of an existence, persistence or reachability property the appropriate variant must be defined; the lemmata above only guarantee that this is always possible. In these cases we say that $M$ is {\em sufficiently refined}. Note that the extensions exploited in the Lemmata \ref{lem-conv}, \ref{lem-conv1}, \ref{lem-div}, \ref{lem-div1}, \ref{lem-pdiv} and \ref{lem-reach} above are indeed $(1,n)$-refinements. 

\subsection{Progress Derivation Rules}

The thorough investigation of different types of variants in the previous subsection enables the derivation of existence, persistence and reachability conditions in case the subformulae are in $\mathcal{L}$\footnote{An extension to nested temporal operators as foreseen in the definition of TREBL will be dealt with later. In fact, we will see that this nesting does not add much expressive power, so it is questionnable whether it is useful to include it in TREBL, but on the other side they do not increase the difficulty of the proofs either.}. This leaves us with the handling of progress formulae $\square (\varphi \rightarrow \lozenge \psi)$ and their diversification resulting from replacing one or both of the temporal operators by their one-trace analog.

Hoang and Abrial showed the soundness of two derivation rules for progress conditions \cite{hoang:icfem2011}. However, these rules contain arbitrary UNTIL-formulae that are not covered by TREBL. Therefore, we combined the rules into a single sound derivation rule \cite{ferrarotti:foiks2024}. With this rule it was also possible to achieve relative completeness, but only under the assumption that machines are tail-homogeneous \cite{ferrarotti:foiks2024}.

We will now go one step further and remove restrictions in the relative completeness result in our previous work. For this we need to exploit some $\mathcal{L}^{ext}$ formulae of the form $\square^\chi \varphi$ with specific formulae $\varphi$, which is inspired by related work in the context of ASMs \cite{ferrarotti:abz2024}. In a second step we will introduce {\em simultaneous} and {\em conditional} variants, which will then allow us to dispense with formulae that refer to subsets of the set of traces.

\begin{lem}\label{lem-progress}

For formulae $\varphi, \psi, \chi \in \mathcal{L}$ the derivation rules
\begin{center}
\begin{minipage}{14.5cm}
\begin{prooftree}
\AxiomC{$M \vdash \lozenge^{\neg\psi} \square \neg\chi$}
\AxiomC{\hspace*{-5mm}$M \vdash \Leadsto(\chi \wedge \neg \psi, \chi \vee \psi)$}
\AxiomC{\hspace*{-5mm}$M \vdash \square^{\neg\psi} (\varphi \wedge \neg \psi \rightarrow \chi)$}
\TrinaryInfC{$M \vdash \square (\varphi \rightarrow \lozenge \psi)$}
\end{prooftree}
\end{minipage}
and\\
\begin{minipage}{14.5cm}
\begin{prooftree}
\AxiomC{$M \vdash \lozenge^{\neg\psi} \square \neg\chi$}
\AxiomC{\hspace*{-5mm}$M \vdash \square^{\neg\psi} \text{\rm leadslocto}(\chi \wedge \neg \psi, \chi \vee \psi)$}
\AxiomC{\hspace*{-5mm}$M \vdash \square^{\neg\psi} (\varphi \wedge \neg \psi \rightarrow \chi)$}
\TrinaryInfC{$M \vdash \square (\varphi \rightarrow \lozenge \psi)$}
\end{prooftree}
\end{minipage}   
\end{center}
are sound for the derivation of progress properties in TREBL.

\end{lem}

\proof

A trace $\tau$ of $M$ that satisfies $\lozenge\square \neg\psi$ either has finite length and terminates in a state $S_k$ satisfying $\neg\psi$ or has an infinite tail of states all satisfying $\neg\psi$, i.e. there exists some $k$ such that $S_\ell \models \neg\psi$ holds for all $\ell \ge k$. Hence a trace not satisfying $\lozenge\square \neg\psi$ is either finite terminating in a state satisfying $\psi$ or contains infinitely many states $S_i$ satisfying $\psi$. In such a trace whenever there is a state $S_i$ satisfying $\varphi$ there also exists a state $S_j$ with $j \ge i$ satisfying $\psi$, so the condition for $\square (\varphi \rightarrow \lozenge\psi)$ is satisfied.

Therefore, we can concentrate on traces satisfying $\lozenge\square \neg\psi$. Assuming that $M$ satisfies the antecedents of the rule, for such traces $S_0, S_1, \dots$ of $M$ every state $S_i$ satisfies $\lozenge \square \neg\chi$, $\text{leadslocto}(\chi \wedge \neg \psi, \chi \vee \psi)$ and $\varphi \wedge \neg \psi \rightarrow \chi$. Consider a state $S_k$ with $S_k \models \varphi$. If also $S_k \models \psi$ holds, the condition for $\square (\varphi \rightarrow \lozenge\psi)$ is satisfied by definition.

Otherwise, for $S_k \models \neg \psi$ the third antecedent of the rule implies $S_k \models \chi$. The first antecedent of the rule implies that there exists some $m$ such that $S_{m^\prime} \models \neg \chi$ holds for all $m^\prime \ge m$. 

Let $m$ be minimal with this property, in particular $m > k$. Then the second antecedent of the rule implies that there exists some $\ell < m$ such that $\sigma^{(\ell')} \models \chi \wedge \neg \psi$ holds for all $k \le \ell' \le \ell$. If we take $\ell$ maximal with this property, we must have $S_\ell \models \psi$ by the definition of transition formulae. This shows again that the condition for $\square (\varphi \rightarrow \lozenge\psi)$ is satisfied, hence $M \models \square (\varphi \rightarrow \lozenge \psi)$ as claimed.\qed


Note that the antecedents of the second derivation rule in Lemma \ref{lem-progress} are slightly weaker than those of the first derivation rule. We will need this second rule for the proof of relative completeness, whereas in examples it may be easier to use the first rule, in which the second antecedent does not refer to a formula over a defined subset of the set of traces.

\subsubsection{Simultaneous Variants}

The derivation rule in Lemma \ref{lem-progress} involves antecedents that are not covered by TREBL. We will therefore investigate rules for the derivation of the antecedents with antecedents in TREBL.

First consider the antecedent $\lozenge^{\neg\psi} \square \neg\chi$, but for reasons of generality consider formulae of the form $\lozenge^\psi \square \chi$ instead. Such a formula requires that traces either converge in $\psi$ or otherwise terminate in a state satisfying $\chi$ or diverge in $\chi$. Therefore, we consider a term $t$ that can serve simultaneously for convergence of $\psi$ and divergence of $\chi$, i.e. we need to have $\text{var}_c(t, \psi) \vee \text{var}_d(t, \chi)$. Equivalently, $t$ is a term taking values in a well-founded set $V$, and we define the {\em simultaneous cd-variant formula} $\text{svar}_{cd}(t, (\psi,\chi))$ as 
\begin{gather}
\forall \bar{v} \Bigg( \psi(\bar{v}) \rightarrow \bigg( \neg\chi(\bar{v}) \rightarrow \exists \bar{x} \bar{v}' \, \Big( \bigvee_{e_i \in E}  (G_i(\bar{x}, \bar{v}) \wedge P_{A_i}(\bar{x}, \bar{v}, \bar{v}')) \Big) \bigg) \wedge \notag\\
\bigwedge_{e_i \in E} \forall \bar{x} \bar{v}' \bigg( G_i(\bar{x}, \bar{v}) \rightarrow \Big( \neg\chi(\bar{v}) \rightarrow \big( t(\bar{v}) \neq 0 \wedge \big( P_{A_i}(\bar{x}, \bar{v}, \bar{v}') \rightarrow t(\bar{v}') < t(\bar{v}) \big) \big) \Big) \notag\\
\wedge \Big( \chi(\bar{v}) \wedge P_{A_i}(\bar{x}, \bar{v}, \bar{v}') \rightarrow t(\bar{v}') \le t(\bar{v}) \Big) \bigg) \Bigg) \; . \label{eq-svar}
\end{gather}

\begin{lem}\label{lem-svar}

For formulae $\psi, \chi \in \mathcal{L}$ the derivation rule
\begin{center}
\begin{minipage}{14.5cm}
\begin{prooftree}
\AxiomC{$M \vdash \text{svar}_{cd}(t, (\psi, \chi))$}
\UnaryInfC{$M \vdash \lozenge^{\psi} \square \chi$}
\end{prooftree}
\end{minipage}
\end{center}
is sound.

\end{lem}

\noindent
An immediate consequence of this lemma is that in the derivation rules in Lemma \ref{lem-progress} the first antecedent can be replaced by $\square \text{svar}_{cd}(t, (\neg\psi, \neg\chi))$, which is an invariance formula and hence is also a TREBL formula.

\proof

Assume that $M \models \text{svar}_{cd}(t, (\psi, \chi))$ holds for an appropriate term $t$. Let $\tau = S_0, S_1, \dots$ be a trace of $M$. As (\ref{eq-svar}) does not impose any restriction on states satisfying $\neg\psi$, the trace may be finite terminating in a state satisfying $\neg\psi$ or there may be infinitely many states $S_i$ with $S_i \models \neg\psi$. Such traces do not satisfy the selection condition $\langle \chi \rangle$.

Traces that satisfy the selection condition either terminate in a state $S_k$ with $S_k \models \psi$ or there exists a $k$ such that $S_\ell \models \psi$ holds for all $\ell \ge k$. The first condition in the formula (\ref{eq-svar}) implies that a trace can only terminate in a state $S_k$ satisfying $\psi$, if also $S_k \models \chi$ holds; otherwise, there would exists an event that is enabled in $S_k$. Therefore, we can concentrate on infinite traces, for which we fix $k$ with $S_\ell \models \psi$ for all $\ell \ge k$.

Consider the sequence of states $S_k, S_{k+1}, \dots$. The second condition in (\ref{eq-svar}) yields an infinite, decreasing sequence $b_{k}, b_{k+1}, \dots$ of the values of the term $t$ in the sequence $S_k, S_{k+1}, \dots$, which is only possible, if there are only finitely many states $S_i$ with $i \ge k$ and $S_i \models \neg\chi$. Otherwise we would get an infinite, strictly decreasing subsequence contradicting the well-foundedness of $V$.\qed


As for other types of variant terms we can show that simultaneous cd-variant term always exist in sufficiently refined machines. As before we emphasise the dependence of the simultaneous variant formula on the machine by adding the machine name as a parameter to the variant formula. The proof of the following lemma is given in Appendix \ref{appA}.

\begin{lem}\label{lem-simultaneous}

Let $M$ be an Event-B machine with state variables $\bar{v} = (v_0, ..., v_m)$ satisfying $\lozenge^\psi \square \chi$ with $\psi, \chi \in \mathcal{L}$. Then there exists an Event-B machine $M'$ with state variables $\bar{v}' = (\bar{v}, \bar{c}, \bar{b})$, where $\bar{c} = (c_0, \ldots, c_m)$, $\bar{b} = (b_0, \ldots, b_m)$ and $\bar{v}$ are pairwise different variables, and a term $t(\bar{v}, \bar{c}, \bar{b})$ such that: 

\begin{enumerate}

\item The formula $\text{svar}_{cd}(M', t, (\psi,\chi))$ is valid for $M'$;
 
\item For each trace $\sigma$ of $M$ there exists a trace $\sigma'$ of $M'$ with $\sigma = \sigma'|_{\bar{v}}$ and vice versa, where $\sigma'|_{\bar{v}}$ is the sequence of states resulting from the projection to the state variables $\bar{v}$.

\end{enumerate}

\end{lem}

\noindent
The proof is given in Appendix \ref{appA}. The basic idea is to construct $M'$ as in the proof on Lemma \ref{lem-div} with the only difference that the values $b_0, \dots, b_m$ are the maximum set values in traces satisfying the selection condition $\langle \psi \rangle$.

\subsubsection{Conditional Variants}

Consider the second antecedent $\square^{\neg\psi} \text{leadslocto}(\chi \wedge \neg\psi, \chi \vee \psi)$ and the third antecedent $\square^{\neg\psi} (\varphi \wedge \neg\psi \rightarrow \chi)$ in the second derivation rule in Lemma \ref{lem-progress}. In both cases we are dealing with invariance conditions---a transition invariant in case of the second antecedent and a state invariant in case of the third antecedent---restricted to traces satisfying $\lozenge\square \neg\psi$. That is, as soon as a trace reaches a state violating the invariant, the trace should either converge in $\neg\psi$, i.e. contain infinitely many states satisfying $\psi$, or terminate in a state satisfying $\psi$. The convergence can be expressed analogously to the case of c-variants, but in this case we need to add the condition that the invariance is violated. While this would refer to states in a trace, we also need to add some form of bookkeeping, i.e. if there has been a violation of the invariance condition in an earlier state of a trace, this should also be taken into consideration. This can be simply done by requiring a Boolean-valued state variable, which at most once in a trace changes its value, so that in traces without any such change the invariance condition holds, while the other traces converge in $\neg\psi$ or terminate in a state satisfying $\psi$.

We therefore define a {\em conditional c-variant} by a pair $t = (t_1,b)$ of terms, where $t_1$ takes values in some well-founded set $V$ and $b$ is a Boolean-valued state-variable, which initially should be 0. The corresponding {\em conditional c-variant formula} $\text{cvar}_c(t, \psi, \theta)$ for the invariant $\theta = \text{leadslocto}(\varphi_1, \varphi_2)$ is then defined as
\begin{gather}
\forall \bar{v} \Bigg( \Bigg( \psi(\bar{v}) \rightarrow \bigg( b=0 \rightarrow \bigwedge_{e_i \in E} \forall \bar{x} \bar{v}^\prime \Big( \varphi_1(\bar{v}) \wedge G_i(\bar{x}, \bar{v}) \wedge P_{A_i}(\bar{x}, \bar{v}, \bar{v}^\prime) \notag\\
\rightarrow \big( \varphi_2(\bar{v}^\prime) \wedge b^\prime = 0 \big) \vee \big( t_1(\bar{v}) \neq 0 \wedge t_1(\bar{v}^\prime) < t_1(\bar{v}) \wedge b^\prime = 1 \big) \Big) \bigg)  \notag\\
\wedge \bigg( b=1 \rightarrow \bigvee_{e_i \in E} \exists \bar{x} \bar{v}^\prime \big( G_i(\bar{x}, \bar{v}) \wedge P_{A_i}(\bar{x}, \bar{v}, \bar{v}^\prime) \big) \notag \\
\wedge \bigwedge_{e_i \in E} \forall \bar{x} \bar{v}^\prime \Big( G_i(\bar{x}, \bar{v}) \wedge P_{A_i}(\bar{x}, \bar{v}, \bar{v}^\prime) \rightarrow \big( t_1(\bar{v}) \neq 0 \wedge t_1(\bar{v}^\prime) < t_1(\bar{v}) \wedge b^\prime = 1 \big) \Big) \bigg) \Bigg) \notag\\
\wedge \bigwedge_{e_i \in E} \forall \bar{x} \bar{v}^\prime \Big( \neg\psi(\bar{v}) \wedge G_i(\bar{x}, \bar{v}) \wedge P_{A_i}(\bar{x}, \bar{v}, \bar{v}^\prime) \rightarrow b^\prime = b \Big) \Bigg) \; .
\label{eq-cvar2}
\end{gather}

\noindent
Analogously, for the invariant $\theta = \varphi \wedge \neg\psi \rightarrow \chi$ the corresponding conditional c-variant formula $\text{cvar}_c(t, \psi, \theta)$ is defined as
\begin{gather}
\forall \bar{v} \Bigg( \Bigg( b=0 \rightarrow \bigwedge_{e_i \in E} \forall \bar{x} \bar{v}^\prime \bigg( G_i(\bar{x}, \bar{v}) \wedge P_{A_i}(\bar{x}, \bar{v}, \bar{v}^\prime) \rightarrow \notag\\
\Big( \theta(\bar{v}^\prime) \wedge b^\prime = 0 \Big) \vee \Big( b^\prime = 1 \wedge \big( \psi(\bar{v}) \rightarrow t_1(\bar{v}) \neq 0 \wedge t_1(\bar{v}^\prime) < t_1(\bar{v}) \big) \Big) \bigg) \Bigg) \wedge \notag\\
\wedge \Bigg( b=1 \rightarrow \bigg( \psi(\bar{v}) \rightarrow \bigvee_{e_i \in E} \exists \bar{x} \bar{v}^\prime \Big( G_i(\bar{x}, \bar{v}) \wedge P_{A_i}(\bar{x}, \bar{v}, \bar{v}^\prime) \Big) \bigg) \wedge \notag\\
\bigwedge_{e_i \in E} \forall \bar{x} \bar{v}^\prime \bigg( G_i(\bar{x}, \bar{v}) \wedge P_{A_i}(\bar{x}, \bar{v}, \bar{v}^\prime) \rightarrow \notag\\ 
\Big( b^\prime = 1 \wedge \big( \psi(\bar{v}) \rightarrow t_1(\bar{v}) \neq 0 \wedge t_1(\bar{v}^\prime) < t_1(\bar{v}) \big) \Big) \bigg) \Bigg) \Bigg) \; . \label{eq-cvar3}
\end{gather}

\noindent
For later use also consider the case of a conditional c-variant formula $\text{cvar}_c(t, \psi, \theta)$ for the invariant $\theta = \text{leadslocto}^1(\varphi_1, \varphi_2)$, which is defined as
\begin{gather*}
\forall \bar{v} \Bigg( \Bigg( \psi(\bar{v}) \rightarrow \bigg( b=0 \wedge \varphi_1(\bar{v}) \rightarrow \bigvee_{e_i \in E} \exists \bar{x} \bar{v}^\prime \Big(  \wedge G_i(\bar{x}, \bar{v}) \wedge P_{A_i}(\bar{x}, \bar{v}, \bar{v}^\prime) \wedge \varphi_2(\bar{v}^\prime) \wedge b^\prime = 0 \Big) \notag\\
\wedge \bigwedge_{e_i \in E} \forall \bar{x} \bar{v}^\prime \Big( \big( G_i(\bar{x}, \bar{v}) \wedge P_{A_i}(\bar{x}, \bar{v}, \bar{v}^\prime) \wedge \neg\varphi_2(\bar{v}^\prime) \big) \notag\\ 
\rightarrow \big( b^\prime = 1 \wedge t_1(\bar{v}) \neq 0 \wedge t_1(\bar{v}^\prime) < t_1(\bar{v}) \big) \Big) \bigg) \notag
\end{gather*}
\begin{gather}
\wedge \bigg( b=1 \rightarrow \bigvee_{e_i \in E} \exists \bar{x} \bar{v}^\prime \big( G_i(\bar{x}, \bar{v}) \wedge P_{A_i}(\bar{x}, \bar{v}, \bar{v}^\prime) \big) \notag\\
\wedge \bigwedge_{e_i \in E} \forall \bar{x} \bar{v}^\prime \Big( G_i(\bar{x}, \bar{v}) \wedge P_{A_i}(\bar{x}, \bar{v}, \bar{v}^\prime) \rightarrow \big( t_1(\bar{v}) \neq 0 \wedge t_1(\bar{v}^\prime) < t_1(\bar{v}) \wedge b^\prime = 1 \big) \Big) \bigg) \Bigg) \notag\\
\wedge \bigwedge_{e_i \in E} \forall \bar{x} \bar{v}^\prime \Big( \neg\psi(\bar{v}) \wedge G_i(\bar{x}, \bar{v}) \wedge P_{A_i}(\bar{x}, \bar{v}, \bar{v}^\prime) \rightarrow b^\prime = b \Big) \Bigg) \; .
\label{eq-cvar1}
\end{gather}

\begin{lem}\label{lem-cvar}

Let $\varphi, \psi, \chi \in \mathcal{L}$ and let $\theta$ be either $\text{leadslocto}(\chi \wedge \neg\psi, \chi \vee \psi)$, $\text{leadslocto}^1(\chi \wedge \neg\psi, \chi \vee \psi)$ or $\varphi \wedge \neg\psi \rightarrow \chi$. For $t = (t_1,b)$ the derivation rule
\begin{center}
\begin{minipage}{14.5cm}
\begin{prooftree}
\AxiomC{$M \vdash (b=0 \leftrightarrow \theta)$}
\AxiomC{$M \vdash \text{cvar}_c(t, \neg\psi, \theta)$}
\BinaryInfC{$M \vdash \square^{\neg\psi} \theta$}
\end{prooftree}
\end{minipage}
\end{center}
is sound.

\end{lem}

\proof

Assume first that $M \models \text{cvar}_c(t, \neg\psi, \text{leadslocto}(\chi \wedge \neg\psi, \chi \vee \psi))$ holds. Consider an arbitrary trace $\tau = S_0, S_1, \dots$ of $M$. Then (\ref{eq-cvar2}) implies that there is some $k$ such that for the first $k$ states $S_i$ of $\tau$, i.e. $0 \le i < k$ ($k$ may be 0), we have $S_i \models b=0 \wedge \text{leadslocto}(\chi \wedge \neg\psi, \chi \vee \psi)$. Let $k$ be maximal with this property.

For $k = \omega$ all states of $\tau$ satisfy $\text{leadslocto}(\chi \wedge \neg\psi, \chi \vee \psi)$. The same holds for $k = \ell(\tau) < \omega$. Otherwise we get $S_\ell \models b=1$ for all $\ell \ge k$, and $\tau$ can only terminate in a state satisfying $\psi$. Furthermore, there cannot exist an infinite subsequence $S_m, S_{m+1}, \dots$ of states with $S_i \models \neg\psi$ for all $i \ge m \ge k$, because otherwise we would obtain an infinite, strictly decreasing sequence $b_m, b_{m+1}, \dots$ of values in $V$ contradicting that $V$ is well-founded. That is, the trace must contain infinitely many states satisfying $\psi$ and thus does not fulfil the selection condition $\langle \neg\psi \rangle$ for the traces, which shows that $M \models \square^{\neg\psi} \text{leadslocto}(\chi \wedge \neg\psi, \chi \vee \psi)$ must hold.

For the analogous case that $M \models \text{cvar}_c(t, \neg\psi, \text{leadslocto}^1(\chi \wedge \neg\psi, \chi \vee \psi))$ holds, (\ref{eq-cvar1}) implies the existence of a trace $\tau = S_0, S_1, \dots$ such that $S_i \models b=0 \wedge \text{leadslocto}^1(\chi \wedge \neg\psi, \chi \vee \psi)$ holds for all $i$. Furthermore, for all traces not satisfying this condition there exists some $k$ with $S_\ell \models b=1$ for all $\ell > k$. Then there cannot exist an infinite subsequence $S_m, S_{m+1}, \dots$ of states with $S_i \models \neg\psi$ for all $i \ge m \ge k$, because otherwise we would obtain an infinite, strictly decreasing sequence $b_m, b_{m+1}, \dots$ of values in $V$ contradicting that $V$ is well-founded. Hence the trace does not fulfil the selection condition $\langle \neg\psi \rangle$, which shows that $M \models \square^{\neg\psi} \text{leadslocto}^1(\chi \wedge \neg\psi, \chi \vee \psi)$ must hold.

Next assume that $M \models \text{cvar}_c(t, \neg\psi, \varphi \wedge \neg\psi \rightarrow \chi)$ holds, and again let $\tau = S_0, S_1, \dots$ be an arbitrary trace of $M$. Further assume that in the initial state $S_0$ we either have $S_0 \models b=0 \wedge (\varphi \wedge \neg\psi \rightarrow \chi)$ or $S_0 \models b=1 \wedge \varphi \wedge \neg\psi \wedge \neg\chi$. From (\ref{eq-cvar3}) we obtain prefixes $S_0, \dots, S_k$ of $\tau$ with $S_i \models b=0 \wedge (\varphi \wedge \neg\psi \rightarrow \chi)$ for all $0 \le i < k$. Here $k$ may be 0, in which case the prefix is empty. Let $k$ be maximal with this property.

For $k = \omega$ or $k = \ell(\tau) < \omega$ all states of $\tau$ satisfy $\varphi \wedge \neg\psi \rightarrow \chi$. Otherwise we get $S_\ell \models b=1$ for all $\ell \ge k$, and $\tau$ can only terminate in a state satisfying $\psi$. Same as in the first case there cannot exist an infinite subsequence $S_m, S_{m+1}, \dots$ of states with $S_i \models \neg\psi$ for all $i \ge m \ge k$. Hence $M \models \square^{\neg\psi} (\varphi \wedge \neg\psi \rightarrow \chi)$ must hold.\qed


An immediate consequence of Lemma \ref{lem-cvar} is that in the second derivation rule in Lemma \ref{lem-progress} the second and third antecedents can be replaced by the antecedents of the rules in the lemma, which are formulae in $\mathcal{L}$ and hence are also TREBL formulae. That is, with $\varphi, \psi, \chi \in \mathcal{L}$, a simultaneous cd-variant $t^\prime$, a conditional c-variant $t = (t_1,b)$, $\gamma_1 = \text{cvar}_c(t, \neg\psi,  \text{leadslocto}(\chi \wedge \neg\psi, \chi \vee \psi))$ and $\gamma_2 = \text{cvar}_c(t, \neg\psi, \varphi \wedge \neg\psi \rightarrow \chi)$ the following is a sound derivation rule for TREBL:
\begin{center}
\begin{minipage}{14.5cm}
\begin{prooftree}
\AxiomC{$M \vdash (b=0 \leftrightarrow (\varphi \wedge \neg\psi \rightarrow \chi))$}
\AxiomC{$M \vdash \text{svar}_{cd}(t^\prime, (\neg \psi, \neg \chi))$}
\AxiomC{\hspace*{-5mm}$M \vdash \gamma_1$}
\AxiomC{\hspace*{-5mm}$M \vdash \gamma_2$}
\LeftLabel{$\mathrm{PR}$: \;}
\QuaternaryInfC{$M \vdash \square (\varphi \rightarrow \lozenge \psi)$}
\end{prooftree}
\end{minipage}
\end{center}

\noindent
To complete the investigation of conditional variants in the derivation rule PR for progress formulae in TREBL we show that the required conditional variants always exists in a sufficiently refined machine $M'$. The proof idea is analogous to the proof of Lemma \ref{lem-conv} with a few modifications. The proof of the following lemma is given in Appendix \ref{appA}.

\begin{lem}\label{lem-conditional}

Let $M$ be an Event-B machine with state variables $\bar{v} = (v_0, ..., v_m)$ satisfying $\square^{\neg\psi} \theta$, where $\theta$ is either $\text{leadslocto}(\chi \wedge \neg\psi, \chi \vee \psi)$, $\text{leadslocto}^1(\chi \wedge \neg\psi, \chi \vee \psi)$ or $\varphi \wedge \neg\psi \rightarrow \chi$. Then there exists an Event-B machine $M'$ with state variables $\bar{v}' = (\bar{v}, \bar{w}, l, s, u, b)$, where $\bar{v}$, $\bar{w} = (w_0, \ldots, w_m)$, $l$, $s$, $u$ and $b$ are pairwise different variables, and a term $t_1(\bar{v}, \bar{w}, l, s, u)$ such that: 

\begin{enumerate}

\item The formula $\mathit{cvar}_c(M, t, \neg\psi, \theta)$ with $t = (t_1,b)$ is valid for $M'$.  
 
\item For each trace $\sigma$ of $M$ there exists a trace $\sigma'$ of $M'$ with $\sigma = \sigma'|_{s=1,\bar{v}}$ and vice versa, where $\sigma'|_{s=1,\bar{v}}$ is the sequence of states resulting from the selection of those with $s = 1$ and projection to the state variables $\bar{v}$.

\end{enumerate}

\end{lem}

\subsubsection{Diversified Progress Formulae}

We now consider the remaining diversified progress formulae $X (\varphi \rightarrow Y \psi)$ with $X \in \{ \square, \square^1 \}$, $Y \in \{ \lozenge, \lozenge^1 \}$ and $\varphi, \psi \in \mathcal{L}$. First consider the case $Y = \lozenge$ and $X = \square^1$. In this case we can simply modify one antecedent of the second rule in Lemma \ref{lem-progress}, requesting validity not for a set of traces defined by $\neg\psi$, but only for a single trace. This should give us three sound derivation rules for the derivation of progress formulae of the form $\square^1 (\varphi \rightarrow \lozenge \psi)$.

\begin{lem}\label{lem-progress10}

For formulae $\varphi, \psi, \chi \in \mathcal{L}$ the following derivation rules are sound for the derivation of one-trace progress properties in TREBL:
\begin{center}
\begin{minipage}{14.5cm}
\begin{prooftree}
\AxiomC{\hspace*{-5mm}$M \vdash \lozenge^1 \square (\neg\psi \wedge \neg\chi)$}
\AxiomC{\hspace*{-5mm}$M \vdash \square^{\neg\psi} \text{\rm leadslocto}(\chi \wedge \neg \psi, \chi \vee \psi)$}
\AxiomC{\hspace*{-5mm}$M \vdash \square^{\neg\psi} (\varphi \wedge \neg \psi \rightarrow \chi)$}
\TrinaryInfC{$M \vdash \square^1 (\varphi \rightarrow \lozenge \psi)$}
\end{prooftree}
\end{minipage}\\
\begin{minipage}{14.5cm}
\begin{prooftree}
\AxiomC{$M \vdash \lozenge^{\neg\psi} \square \neg\chi$}
\AxiomC{\hspace*{-5mm}$M \vdash \square^1 \text{\rm leadslocto}(\chi \wedge \neg \psi, \chi \vee \psi)$}
\AxiomC{\hspace*{-5mm}$M \vdash \square^{\neg\psi} (\varphi \wedge \neg \psi \rightarrow \chi)$}
\TrinaryInfC{$M \vdash \square^1 (\varphi \rightarrow \lozenge \psi)$}
\end{prooftree}
\end{minipage}\\
\begin{minipage}{14.5cm}
\begin{prooftree}
\AxiomC{$M \vdash \lozenge^{\neg\psi} \square \neg\chi$}
\AxiomC{\hspace*{-5mm}$M \vdash \square^{\neg\psi} \text{\rm leadslocto}(\chi \wedge \neg \psi, \chi \vee \psi)$}
\AxiomC{\hspace*{-5mm}$M \vdash \square^1 (\varphi \wedge \neg \psi \rightarrow \chi)$}
\TrinaryInfC{$M \vdash \square^1 (\varphi \rightarrow \lozenge \psi)$}
\end{prooftree}
\end{minipage}
\end{center}
\end{lem} 

\proof

Consider the first rule and assume that $M$ satisfies the three antecedents. The first antecedent yields a trace $S_0, S_1, \dots$ and some $k$ such that $S_\ell \models \neg\psi \wedge \neg\chi$ holds for all $\ell \ge k$. In particular, the trace satisfies the selection condition for the other antecedents. Let $i$ be such that $S_i \models \varphi$ holds. If no such $i$ exists or if $S_i \models \psi$ holds, we immediately get $M \models \square^1 (\varphi \rightarrow \lozenge \psi)$. Therefore, assume $S_i \models \neg\psi$, hence also $S_i \models \chi$ holds by the third antecedent. Then the second antecedent implies that in every continuing trace with a $\neg\psi$ tail the following states satisfy $\chi$ until we reach a state satisfying $\psi$, which by the first antecedent must exist.

Next consider the second rule and assume that $M$ satisfies its three antecedents. The second antecedent yields a trace $S_0, S_1, \dots$ with $S_i \models \text{\rm leadslocto}(\chi \wedge \neg \psi, \chi \vee \psi)$ for all $i$. We can assume without loss of generality that the trace does not terminate in a state satisfying $\psi$ or contains infinitely many states satisfying $\psi$, because in such cases we immediately get $M \models \square^1 (\varphi \rightarrow \lozenge \psi)$. As for the first rule we can further assume some $k$ with $S_k \models \varphi \wedge \neg\psi$, hence also $S_k \models \chi$ by the third antecedent. Then by second antecedent the following states $S_{k+1}, \dots$ satisfy $\chi$ until we reach a state $S_j$ with $S_j \models \psi$, and such $j$ must exist due to the first antecedent.

Finally consider the third rule and let $M$ satisfy its three antecedents. The third antecedent yields a trace $S_0, S_1, \dots$ with $S_i \models \varphi \wedge \neg \psi \rightarrow \chi$ for all $i$. Again, we can assume without loss of generality that the trace does not terminate in a state satisfying $\psi$ or contains infinitely many states satisfying $\psi$, because in such cases $M \models \square^1 (\varphi \rightarrow \lozenge \psi)$ follows immediately. Then without loss of generality take $k$ with $S_k \models \varphi \wedge \neg\psi$, hence also $S_k \models \chi$. The second antecedent implies that the following states satisfy $\chi$ until we reach a state $S_j$ with $S_j \models \psi$, and such $j$ must exist due to the first antecedent.\qed


Using the results on conditional variants we can replace antecedents in the rules of Lemma \ref{lem-progress10} by the corresponding simultaneous cd-variant formula (\ref{eq-svar}) or the conditional c-variant formulae (\ref{eq-cvar2}) and (\ref{eq-cvar3}), respectively, which give us three derivation rules PR$^{10}_1$, PR$^{10}_2$ and PR$^{10}_3$ for the derivation of one-trace progress formulae of the form $\square^1 (\varphi \rightarrow \lozenge \psi)$ with $\varphi, \psi \in \mathcal{L}$. The antecedents of these rules are either in $\mathcal{L}$ or they are one-trace invariance or persistence conditions.

Next consider the case $Y = \lozenge^1$ and $X = \square$. In this case we can simply modify the second antecedent of the second rule in Lemma \ref{lem-progress} using $\text{leadslocto}^1(\chi \wedge \neg\psi, \chi \vee \psi)$ instead of $\text{leadslocto}(\chi \wedge \neg\psi, \chi \vee \psi)$.

\begin{lem}\label{lem-progress01}

For formulae $\varphi, \psi, \chi \in \mathcal{L}$ the derivation rule
\begin{center}
\begin{minipage}{14.5cm}
\begin{prooftree}
\AxiomC{$M \vdash \lozenge^{\neg\psi} \square \neg\chi$}
\AxiomC{\hspace*{-5mm}$M \vdash \square^{\neg\psi} \text{\rm leadslocto}^1(\chi \wedge \neg \psi, \chi \vee \psi)$}
\AxiomC{\hspace*{-5mm}$M \vdash \square^{\neg\psi} (\varphi \wedge \neg \psi \rightarrow \chi)$}
\TrinaryInfC{$M \vdash \square (\varphi \rightarrow \lozenge^1 \psi)$}
\end{prooftree}
\end{minipage}
\end{center}
is sound for the derivation of progress properties in TREBL.

\end{lem}

\proof

Without loss of generality as in the proof of Lemma \ref{lem-progress} we can concentrate on traces satisfying $\lozenge\square \neg\psi$. Assuming that $M$ satisfies the antecedents of the rule, for such traces $S_0, S_1, \dots$ of $M$ every state $S_i$ satisfies $\lozenge \square \neg\chi$, $\text{leadslocto}^1(\chi \wedge \neg \psi, \chi \vee \psi)$ and $\varphi \wedge \neg \psi \rightarrow \chi$. Consider a state $S_k$ with $S_k \models \varphi$. If also $S_k \models \psi$ holds, the condition for $\square (\varphi \rightarrow \lozenge^1 \psi)$ is satisfied by definition.

Otherwise, for $S_k \models \neg \psi$ the third antecedent of the rule implies $S_k \models \chi$. The first antecedent of the rule implies that in at least one such trace there exists some $m$ such that $S_{m^\prime} \models \neg \chi$ holds for all $m^\prime \ge m$. 

Let $m$ be minimal with this property, in particular $m > k$. Then the second antecedent of the rule implies that there exists some $\ell < m$ such that $S_{\ell'} \models \chi \wedge \neg \psi$ holds for all $k \le \ell' \le \ell$. If we take $\ell$ maximal with this property, we must have $S_\ell \models \psi$ by the definition of transition formulae. This shows again that $M \models \square (\varphi \rightarrow \lozenge \psi)$ holds as claimed.\qed


Combining Lemma \ref{lem-progress01} and the results in Lemmata \ref{lem-svar} and \ref{lem-cvar} we obtain a sound derivation rule for progress conditions of the form $\square (\varphi \rightarrow \lozenge^1 \psi)$. That is, with $\varphi, \psi, \chi \in \mathcal{L}$, a simultaneous cd-variant $t^\prime$, a conditional c-variant $t = (t_1,b)$, $\gamma_1 = \text{cvar}_c(t, \neg\psi,  \text{leadslocto}^1(\chi \wedge \neg\psi, \chi \vee \psi))$ and $\gamma_2 = \text{cvar}_c(t, \neg\psi, \varphi \wedge \neg\psi \rightarrow \chi)$ the following is a sound derivation rule for TREBL:
\begin{center}
\begin{minipage}{14.5cm}
\begin{prooftree}
\AxiomC{$M \vdash (b=0 \leftrightarrow (\varphi \wedge \neg\psi \rightarrow \chi))$}
\AxiomC{$M \vdash \text{svar}_{cd}(t^\prime, (\psi, \chi))$}
\AxiomC{\hspace*{-5mm}$M \vdash \gamma_1$}
\AxiomC{\hspace*{-5mm}$M \vdash \gamma_2$}
\LeftLabel{\hspace*{-5mm}$\mathrm{PR}^{01}$: \;}
\QuaternaryInfC{$M \vdash \square (\varphi \rightarrow \lozenge^1 \psi)$}
\end{prooftree}
\end{minipage}
\end{center}

\noindent
The existence of the required simultaneous and conditional variants in a sufficiently refined machine has already been shown in Lemmata \ref{lem-simultaneous} and \ref{lem-conditional}.

This leaves the case of progress formulae of the form $\square^1 (\varphi \rightarrow \lozenge^1 \psi)$. For these we can simply modify one antecedent of the second rule in Lemma \ref{lem-progress01}, requesting validity not for a set of traces defined by $\neg\psi$, but only for a single trace. This should give us three sound derivation rules.

\begin{lem}\label{lem-progress11}

For formulae $\varphi, \psi, \chi \in \mathcal{L}$ the following derivation rules are sound for the derivation of one-trace progress properties in TREBL:
\begin{center}
\hspace*{-3mm}\begin{minipage}{14.5cm}
\begin{prooftree}
\AxiomC{$M \vdash \lozenge^1 \square (\neg\psi \wedge \neg\chi)$}
\AxiomC{\hspace*{-5mm}$M \vdash \square^{\neg\psi} \text{\rm leadslocto}^1(\chi \wedge \neg \psi, \chi \vee \psi)$}
\AxiomC{\hspace*{-5mm}$M \vdash \square^{\neg\psi} (\varphi \wedge \neg \psi \rightarrow \chi)$}
\TrinaryInfC{$M \vdash \square^1 (\varphi \rightarrow \lozenge^1 \psi)$}
\end{prooftree}
\end{minipage}\\
\hspace*{-3mm}\begin{minipage}{14.5cm}
\begin{prooftree}
\AxiomC{$M \vdash \lozenge^{\neg\psi} \square \neg\chi$}
\AxiomC{\hspace*{-5mm}$M \vdash \square^1 \text{\rm leadslocto}^1(\chi \wedge \neg \psi, \chi \vee \psi)$}
\AxiomC{\hspace*{-5mm}$M \vdash \square^{\neg\psi} (\varphi \wedge \neg \psi \rightarrow \chi)$}
\TrinaryInfC{$M \vdash \square^1 (\varphi \rightarrow \lozenge^1 \psi)$}
\end{prooftree}
\end{minipage}\\
\hspace*{-3mm}\begin{minipage}{14.5cm}
\begin{prooftree}
\AxiomC{$M \vdash \lozenge^{\neg\psi} \square \neg\chi$}
\AxiomC{\hspace*{-5mm}$M \vdash \square^{\neg\psi} \text{\rm leadslocto}^1(\chi \wedge \neg \psi, \chi \vee \psi)$}
\AxiomC{\hspace*{-5mm}$M \vdash \square^1 (\varphi \wedge \neg \psi \rightarrow \chi)$}
\TrinaryInfC{$M \vdash \square^1 (\varphi \rightarrow \lozenge^1 \psi)$}
\end{prooftree}
\end{minipage}
\end{center}
\end{lem} 

\proof

Consider the first rule and assume that $M$ satisfies the three antecedents. The first antecedent yields a trace $S_0, S_1, \dots$ and some $k$ such that $S_\ell \models \neg\psi \wedge \neg\chi$ holds for all $\ell \ge k$. In particular, the trace satisfies the selection condition for the other antecedents. Let $i$ be such that $S_i \models \varphi$ holds. If no such $i$ exists or if $S_i \models \psi$ holds, we immediately get $M \models \square^1 (\varphi \rightarrow \lozenge^1 \psi)$. Therefore, assume $S_i \models \neg\psi$, hence also $S_i \models \chi$ holds by the third antecedent. Then the second antecedent implies that there is a continuing trace with $(\neg\psi \wedge \neg\chi)$-tail, in which the following states satisfy $\chi$ until we reach a state satisfying $\psi$, and such a state must exist due to the first antecedent.

Next consider the second rule and assume that $M$ satisfies its three antecedents. The second antecedent yields a trace $S_0, S_1, \dots$ with $S_i \models \text{\rm leadslocto}^1(\chi \wedge \neg \psi, \chi \vee \psi)$ for all $i$. We can assume without loss of generality that the trace does not terminate in a state satisfying $\psi$ or contains infinitely many states satisfying $\psi$, because in such cases we immediately get $M \models \square^1 (\varphi \rightarrow \lozenge^1 \psi)$. As for the first rule we can further assume some $k$ with $S_k \models \varphi \wedge \neg\psi$, hence also $S_k \models \chi$ by the third antecedent. Then by second antecedent the following states $S_{k+1}, \dots$ satisfy $\chi$ until we reach a state $S_j$ with $S_j \models \psi$, and such $j$ must exist due to the first antecedent.

Finally consider the third rule and let $M$ satisfy its three antecedents. The third antecedent yields a trace $S_0, S_1, \dots$ with $S_i \models \varphi \wedge \neg \psi \rightarrow \chi$ for all $i$. Again, we can assume without loss of generality that the trace does not terminate in a state satisfying $\psi$ or contains infinitely many states satisfying $\psi$, because in such cases $M \models \square^1 (\varphi \rightarrow \lozenge^1 \psi)$ follows immediately. Then without loss of generality take $k$ with $S_k \models \varphi \wedge \neg\psi$, hence also $S_k \models \chi$. The second antecedent implies that the following states satisfy $\chi$ until we reach a state $S_j$ with $S_j \models \psi$, and such $j$ must exist due to the first antecedent.\qed


Using the results on conditional variants we can replace antecedents in the rules of Lemma \ref{lem-progress11} by the corresponding simultaneous cd-variant formula (\ref{eq-svar}) or the conditional c-variant formulae (\ref{eq-cvar1}) and (\ref{eq-cvar3}), respectively, which give us three derivation rules PR$^{11}_1$, PR$^{11}_2$ and PR$^{11}_3$ for the derivation of one-trace progress formulae of the form $\square^1 (\varphi \rightarrow \lozenge^1 \psi)$ with $\varphi, \psi \in \mathcal{L}$. The antecedents of these rules are either in $\mathcal{L}$ or they are one-trace invariance or persistence conditions.

To complete our investigation of progress formulae, we consider the following derivation rule:
\begin{center}
\begin{minipage}{14.5cm}
\begin{prooftree}
\AxiomC{$M \vdash X Y \psi$}
\LeftLabel{$\mathrm{PR}_0$: \;}
\UnaryInfC{$M \vdash X (\varphi \rightarrow Y \psi)$}
\end{prooftree}
\end{minipage}
\end{center}

\noindent
This rule is obviously sound for $X \in \{ \square, \square^1 \}$ and $Y \in \{ \lozenge, \lozenge^1 \}$; it complements the derivation rules for all variants of progress formulae covering cases, where a progress condition holds, because a stronger existence condition holds.

However, for the case of $X = \square^1$ the antecedent in the rule PR$_0$ is too strong. For the proof of Theorem \ref{thm-complete} later in this section we will need a stronger rule with weaker antecedents. We observe that $M \models \square^1 (\varphi \rightarrow Y \psi)$ already holds, if there is a trace $S_0, S_1, \dots$ and some $k$ such that $S_k \models \square^1 Y \psi$ holds and $S_i \models \neg \varphi$ holds for all $i < k$; in other words there may be a prefix of states satisfying $\neg\varphi$ that can be ignored.

Therefore, with a Boolean-valued state variable $b$ we define a {\em prefix formula} $\text{pre}(\varphi, b)$ as
\begin{gather}\label{eq-prefix}
\forall \bar{v} \bigg( \bigwedge_{e_i \in E} \forall \bar{x} \bar{v}^\prime \Big( G_i(\bar{x}, \bar{v}) \wedge P_{A_i}(\bar{x}, \bar{v}, \bar{v}^\prime) \rightarrow
\big( ( \varphi(\bar{v}) \rightarrow b^\prime = b ) \wedge (\neg  \varphi(\bar{v}) \rightarrow b^\prime = 1 ) \big) \Big) \bigg).
\end{gather}

\noindent
Consider a machine satisfying $\text{pre}(\neg\varphi, b)$, and assume that in the initial state we have $\varphi$ iff $b=1$ holds. The value of $b$ can be updated at most once from 0 to 1, and this is only possible, when we reached a state satisfying $\varphi$. Thus, the formula captures that there exists a (possible empty) prefix of states all satisfying $\neg\varphi$. With this we obtain the following two derivation rules:
\begin{center}
\begin{minipage}{14.5cm}
\begin{prooftree}
\AxiomC{$M \vdash (b=1 \leftrightarrow \varphi)$}
\AxiomC{$M \vdash \text{pre}(\neg\varphi, b)$}
\AxiomC{$M \vdash \square^1 \lozenge (b=0 \vee \psi)$}
\LeftLabel{$\mathrm{PR}^{10}_0$: \;}
\TrinaryInfC{$M \vdash \square^1 (\varphi \rightarrow \lozenge \psi)$}
\end{prooftree}
\end{minipage}\\
\begin{minipage}{14.5cm}
\begin{prooftree}
\AxiomC{$M \vdash (b=1 \leftrightarrow \varphi)$}
\AxiomC{$M \vdash \text{pre}(\neg\varphi, b)$}
\AxiomC{$M \vdash \square^1 \lozenge^1 (b=0 \vee \psi)$}
\LeftLabel{$\mathrm{PR}^{11}_0$: \;}
\TrinaryInfC{$M \vdash \square^1 (\varphi \rightarrow \lozenge^1 \psi)$}
\end{prooftree}
\end{minipage}
\end{center}

\noindent
The soundness of these two rules is easy to see. Furthermore, for a machine $M$ with a trace $S_0, S_1, \dots$, which terminates in a state satisfying $\psi$ or contains infinitely many $S_i$ with $S_i \models \psi$ it is straightforward to construct a refinement $M^\prime$ with a Boolean state variable $b$ such that in the initial state we have $b=1$ iff $\varphi$ holds, and $M^\prime$ satisfies $\text{pre}(\neg\varphi, b)$ as well as $\square^1 Y (b=0 \vee \psi)$.

\subsection{Nested Temporal Operators}

So far we dealt with sound derivation rules for TREBL formulae of the form $X \varphi$, $Y \varphi$, $XY \varphi$, $YX \varphi$ and $X (\psi \rightarrow Y \varphi)$ with $X \in \{ \square, \square^1 \}$, $Y \in \{ \lozenge, \lozenge^1 \}$, and $\varphi, \psi \in \mathcal{L}$, but the definition of TREBL also allows formulae with $\varphi \in \mathcal{LT}$. Sound derivation rules for nested temporal formulae are dealt with in Appendix \ref{appB}.


\subsection{Examples}

We describe a system with a single producer and a single consumer who communicate through a shared buffer. The buffer acts as a temporary storage area but has a limited capacity, $N$.
The producer's role is to add items to the buffer, while the consumer's role is to remove them. In this model, the consumer must first make a ``request'' to signal its intent.
\begin{multicols}{3}
\begin{minipage}{\linewidth}
\begin{lstlisting}[mathescape=true]
$\textbf{variables}$
$\txt{buf}$ $\text{// items in buffer}$
$\txt{req}$ $\text{// consumer request?}$
\end{lstlisting}
\end{minipage}

\begin{minipage}{\linewidth}
\begin{lstlisting}[mathescape=true]
$\textbf{invariants}$
$\text{inv1:}$ $0 \le \txt{buf} \le N$
$\text{inv2:}$ $\txt{req} \in \{0, 1\}$
\end{lstlisting}
\end{minipage}

\begin{minipage}{\linewidth}
\begin{lstlisting}[mathescape=true]
$\textbf{init}$
$\textbf{begin}$
    $\txt{buf} := 0$
    $\txt{req} := 0$
$\textbf{end}$
\end{lstlisting}
\end{minipage}

\end{multicols}

\begin{multicols}{3}
\begin{minipage}{\linewidth}
\begin{lstlisting}[mathescape=true]
$\textbf{produce}$
$\textbf{when}$
    $\txt{buf} < N$
$\textbf{then}$
    $\txt{buf} := \txt{buf}+1$
$\textbf{end}$
\end{lstlisting}
\end{minipage}

\begin{minipage}{\linewidth}
\begin{lstlisting}[mathescape=true]
$\textbf{consumer\_request}$
$\textbf{when}$
    $\txt{req} = 0$ 
$\textbf{then}$
    $\txt{req} := 1$
$\textbf{end}$
\end{lstlisting}
\end{minipage}

\begin{minipage}{\linewidth}
\begin{lstlisting}[mathescape=true]
$\textbf{consumer\_fulfill}$
$\textbf{when}$
    $\txt{req} = 1$
    $\txt{buf} > 0$
$\textbf{then}$
    $\txt{buf} := \txt{buf}-1$
    $\txt{req} := 0$
$\textbf{end}$
\end{lstlisting}
\end{minipage}
\end{multicols}
\vspace*{-1cm}
\subsubsection*{Liveness Properties:}
This model demonstrates the following types of ``long-term guarantees'', which ensure the system makes meaningful progress and does not get stuck:
\begin{itemize}
\item \emph{Progress (Guaranteed Service).}
This property ensures that a request made by the consumer will eventually be fulfilled. That is, a pending request will not be ignored indefinitely. 
\[\Box(\texttt{req} = 1 \rightarrow \Diamond (\texttt{req} = 0)) \]
\item \emph{Existence (Recurring Availability).}
This property guarantees that the producer is not permanently blocked, ensuring the system does not reach a permanent state of gridlock where the buffer is full forever.
\[ \Box \Diamond (\texttt{buf} < N) \]
\end{itemize}

\subsubsection*{Proof Construction for $ \Box(\texttt{req} = 1 \rightarrow \Diamond (\texttt{req} = 0)) $ }

    We use the rule PR for progress. First, we instantiate $\varphi$, $\psi$, and $\chi$ with the following concrete formulae from our bounded buffer model.
\begin{itemize}
    \item The condition that triggers the need for progress is $\varphi \equiv \txt{req} = 1$.
    \item The goal state we must eventually reach is $\psi \equiv \txt{req} = 0$.
    \item The intermediate condition, $\chi$, is chosen to be the state where progress becomes inevitable. In this simple example, this coincides with the need for progress, i.e., $\chi \equiv \txt{req} = 1$.
\end{itemize}

\noindent
Next, we define the simultaneous cd-variant $t' = N + 1 -\txt{buf} + 2 \cdot \txt{req}$ and the conditional c-variant $t=(t_1,b)$, where $
t_1 \;=\; N-\txt{buf}$ and $b \;=\; 0$.

Finally, we verify the premises of PR rule:
\begin{itemize}
    \item \emph{Premise 1:} $M \vdash (b=0 \leftrightarrow (\varphi \wedge \neg\psi \rightarrow \chi))$.
By the definition of $b$ we get that $b = 0$ is always true. Since furthermore $(\varphi \wedge \neg \psi \rightarrow \chi) \equiv \txt{req} = 1 \rightarrow \txt{req} = 1$ is a tautology, we get that Premise 1 is verified.

    \item \emph{Premise 2:} $M \vdash \text{svar}_{cd}(t^\prime, (\neg \psi, \neg \chi))$. 
For the simultaneous cd-variant $t' = N + 1 -\txt{buf}+ 2 \cdot \txt{req}$, we must check that $\text{svar}_{cd}(t^\prime, (\txt{req} = 1, \txt{req} = 0))$ (cf.~\eqref{eq-svar}) allays holds. Since this formulae is trivially true whenever $\txt{req} = 0$, we only need to check the case when $\txt{req} = 1$. That is, we need to prove that whenever $\txt{req} = 1$, (i)~there is an event that can be fired and that (ii)~every event that can be fired strictly decreases $t'$. Clearly~(i) holds since either \textbf{produce} or \textbf{consumer\_fulfill} can be fired whenever $\txt{req} = 1$. Regarding~(ii), if \textbf{produce} is fired, then $N-\txt{buf}' < N-\txt{buf}$ and $\txt{req}' = \txt{req}$, which strictly decreases $t'$. If instead \textbf{consumer\_fulfill} is fired, then $\txt{buf}' = \txt{buf} - 1$ but at the same time $\txt{req}' = 0$. The former increases $t'$ by $1$, but the latter decreases it by $2$, which strictly  decreases $t'$. Finally, the event \textbf{consumer\_request} cannot be fired whenever $\txt{req} = 1$.

\item \emph{Premise 3} $M \vdash \gamma_1$. We must check that  
\[\text{cvar}_c(t,\; \txt{req} = 1,\; \text{leadslocto}(\txt{req} = 1, \; \txt{req} = 1 \vee \txt{req} = 0)),\]
defined as in~\eqref{eq-cvar2}, always holds. This boils down to checking that: (i) whenever $\txt{req} = 1$ and an event is fired, then either $\txt{req}' = 1$ or $\txt{req}' = 0$; and (ii) whenever $\txt{req} = 0$ and an event is fired, then $b = b'$. Both conditions are trivially satisfied since $\txt{req}' \in \{0, 1\}$ by $\text{inv2}$ and $b = b' = 0$ by definition of $b$.

\item \emph{Premise 4:} $M \vdash \gamma_2$. We must check that 
\[
\text{cvar}_c(t,\; \txt{req} = 1,\; \txt{req}  = 1 \rightarrow \txt{req} = 1), 
\]
defined as in~\eqref{eq-cvar3}, always holds. Since independently of any fired event, by our definition of $b$ it holds that $b = b' = 0$, and furthermore $\txt{req}  = 1 \rightarrow \txt{req} = 1$ is a tautology, this is trivially the case. \qed
\end{itemize}

\subsubsection*{Proof Construction for $ \Box \Diamond (\txt{buf} < N) $ } 

Let $\chi \equiv \txt{buf}=N \wedge \txt{req}=1$. Note that $\neg\chi$ is equivalent to
$\varphi \equiv \txt{buf}<N \vee \txt{req} = 0$.

The term $t(\txt{buf}) = \txt{buf}$ is a c-variant for $\chi$: whenever $\chi$ holds
($\txt{buf}=N$ and $\txt{req}=1$) the only enabled event is \textbf{consumer\_fulfill},
which strictly decreases $t$. Hence $\text{var}_c(t,\chi)$ holds. Moreover, states satisfying $\chi$
are not terminal (the enabled \textbf{consumer\_fulfill} can fire), so $\mathrm{Dlf}(\chi)$ holds.

Applying rule E with $\neg\varphi=\chi$ yields
\[
M \models \Box\Diamond\bigl(\txt{buf}<N \vee \txt{req} = 0 \bigr),
\]
i.e.\ $M\models\Box\Diamond(\neg\chi)$.

Inspecting the events of the machine we obtain the following instances of $\mathrm{Leadsto}$ (Eq.~\eqref{eq-leadsto}):
\[
M \vdash \mathrm{Leadsto}\bigl(\txt{buf}=N\wedge \txt{req} = 0,\; \chi\bigr),
\qquad
M \vdash \mathrm{Leadsto}\bigl(\chi,\; \txt{buf}<N\bigr).
\]
Indeed, from any state with $\txt{buf}=N\wedge \txt{req} = 0$ the only enabled event is \textbf{consumer\_request} whose effect is $\txt{req}'=1$ and $\txt{buf}'=N$, hence the post-state satisfies $\chi$; from any state satisfying $\chi$ the only enabled event is \textbf{consumer\_fulfill} whose effect yields $\txt{buf}'<N$.

From the two $\mathrm{Leadsto}$ facts and the definition of $\varphi$ we obtain, for every state $S$,
\[
S\models (\txt{buf}<N) \Rightarrow S\models \Diamond(\txt{buf}<N),
\]
and
\[
S\models (\txt{buf}=N\wedge\neg\txt{req}) \Rightarrow S\models \Diamond(\txt{buf}<N),
\]
because the first case gives $\txt{buf}<N$ immediately, and the second case reaches $\chi$ in one step and then $\txt{buf}<N$ in a further step (hence $\Diamond\Diamond(\txt{buf}<N)$, which is equivalent to $\Diamond(\txt{buf}<N)$).  

Finally, since we have established that $M\models \Box\Diamond\varphi$ and $M\models \varphi\rightarrow\Diamond(\txt{buf}<N)$, we have that every recurrence of $\varphi$ produces a subsequent occurrence of $\txt{buf}<N$. It follows that $
M \models \Box\Diamond(\txt{buf}<N)$. \qed
\vspace*{0.5cm}

As an additional example, consider next the following \emph{persistence (Eventual Stability) property}: 
\[\Diamond \Box (\texttt{produced\_count} = M),\]
where the producer had a finite quota $M$ of items to create. The objective is to guarantee that the producer eventually completes its quota, after which the system remains in a state where no further items are produced. 

To express this property, we must use a slightly extended version of the single producer/consumer Event-B model introduced above. In particular, we introduce a new variable $\txt{produced\_count}$, initialized to $0$, to track the number of items produced, and we extend the invariants with $\txt{produced\_count} \in  \{0,\dots,M\}$. Moreover, the guard for the event \textbf{produce} must be changed to 
\[\txt{buf} < N \wedge \txt{produced\_count} < M,\] 
ensuring that the quota is not exceeded. Triggering this event now increases not only $\txt{buf}$ but also $\txt{produced\_count}$. 

Naturally, in this modified model the earlier progress property no longer holds unless we also adjust the guard of the event \textbf{consumer\_request} to \[\txt{req} = 0 \wedge \txt{produced\_count} < M.\]

\subsubsection*{Proof Construction for $\Diamond \Box (\mathtt{produced\_count} = M)$}

We use the rule P for persistency to prove that the producer eventually completes its quota of $M$ items and the system persists in a stable state. 

Define the formula
\[
\varphi \equiv \txt{produced\_count} = M
\]
and the lexicographic d-variant:
\[
t = \bigl( M - \txt{produced\_count},\; \txt{buf},\; 1 - \txt{req} \bigr),
\]
with lexicographic order over $\mathbb{N}^3$.  
This variant is \emph{well-founded} as each component is a natural number and decreases only finitely many times.  
\begin{itemize}
    \item Whenever $\neg\varphi$ holds, $\txt{produced\_count} < M$ and at least one component of $t$ strictly decreases when any enabled event fires:
    a) If $\txt{buf} < N$, \textbf{produce} decreases the first component $M - \txt{produced\_count}$.  
  b) If $\txt{buf} = N$ and $\txt{req} = 0$, only \textbf{consumer\_request} is enabled, which decreases the third component $1 - \txt{req}$.  
  c) If $\txt{buf} > 0$ and $\txt{req} = 1$, \textbf{consumer\_fulfill} decreases the second component $\txt{buf}$. Hence, the d-variant strictly decreases whenever $\txt{produced\_count} < M$
  
    \item Whenever $\varphi$ holds, the first component of the d-variant, i.e., $M - \txt{produced\_count}$,  equals $0$
  and this cannot change as the guard of \textbf{produce} is false.  
Likewise, the guard of \textbf{consumer\_request} is also false. The event \textbf{consumer\_fulfill} might still fire (once), but this decreases the second component of the d-variant, i.e., $\txt{buf}$. Thus, either the lexicographic value of $t$ decreases or the run stops as no further event can be fired. 
\end{itemize}

\noindent
Therefore, the d-variant formula $\text{var}_d(t,\varphi)$ is satisfied.  

Regarding deadlock freeness, whenever $\txt{produced\_count} < M$, at least one event is enabled: If $\txt{buf} < N$, \textbf{produce} is enabled.  If $\txt{buf} = N$ and $\txt{req} = 0$, \textbf{consumer\_request} is enabled.  If $\txt{buf} > 0$ and $\txt{req} = 1$, \textbf{consumer\_fulfill} is enabled.  
Therefore, no deadlock occurs while $\neg \varphi$ holds.

By the derivation rule P, we get $M \vdash \lozenge\square (\txt{produced\_count} = M)$, establishing that the system eventually reaches a stable state where the producer has completed its quota.\qed


\section{Relative Completeness}\label{sec:complete}

Taking all our derivation rules together we obtain a sound system $\mathfrak{R}$ of derivation rules for TREBL formula, which comprises a sound and complete set of derivation rules for first-order logic with types and the 63 derivation rules INV$_1$, INV$_2$, INV$_1^1$, INV$_2^1$, E, E$^{11}$, E$^{10}$, E$^{01}$, P, P$^{11}$, P$^{10}$, P$^{01}$, R, R$^1$, PR, PR$^{11}_1$, PR$^{11}_2$, PR$^{11}_3$, PR$^{10}_1$, PR$^{10}_2$, PR$^{10}_3$, PR$^{01}$, PR$_0$, PR$^{10}_0$, PR$^{11}_0$, N, NPR, PRE, PRE$^{11}$, PRP, PRP$^{10}$, PRP$^{01}$, PRP$^{11}$, PR$^1$E, PR$^1$E$^{00}$, PR$^1$E$^{10}$, PR$^1$E$^{01}$, PR$^1$E$^{11}$, PR$^1$P, PR$^1$P$^{00}$, PR$^1$P$^{10}$, PR$^1$P$^{01}$, PR$^1$P$^{11}$, PPR, PPR$^{11}_1$, PPR$^{11}_2$, PPR$^{11}_3$, PPR$^{10}_1$, PPR$^{10}_2$, PPR$^{10}_3$, PPR$^{01}$, P$^1$PR, P$^1$PR$^{11}_1$, P$^1$PR$^{11}_2$, P$^1$PR$^{11}_3$, P$^1$PR$^{10}_1$, P$^1$PR$^{10}_2$, P$^1$PR$^{10}_3$, P$^1$PR$^{01}$, PRPR and PR$^1$PR. 

Taking the lemmata in the previous section together we have already shown the following theorem.

\begin{thm}\label{thm-sound}

The system $\mathfrak{R}$ of derivation rules is sound for the derivation of TREBL formulae, i.e. for any TREBL formula $\varphi$ such that $M \vdash_{\mathfrak{R}} \varphi$ holds we have $M \models \varphi$.

\end{thm}

\noindent
We now want to show the converse, i.e. that for $M \models \varphi$ we can always obtain a derivation with rules in $\mathfrak{R}$. However, whenever we need to apply a derivation rule that involves variant formulae as antecedents, this is only possible, if the corresponding variant terms are defined for the machine $M$. We therefore consider only machines $M$ satisfying the following conditions:

\begin{enumerate}

\item $M$ is consistent, i.e. the initial state $S_0$ satisfies the invariant $\iota$, and whenever a state $S$ satisfies $\iota$, then this is also the case for all successor states of $S$ defined by the events of $M$;

\item Every state formula $\varphi$ in the logic $\mathcal{L}$ that holds in all reachable states is included in the invariant $\iota$;

\item $M$ is sufficiently refined, i.e. all required variant terms can be defined in $M$.

\end{enumerate}

\noindent
A machine $M$ satisfying these three conditions will be called to satisfy the {\em relativity assumptions}.

We already emphasised in Section \ref{sec:trebl} that the consistency condition in (i) is stronger than needed, as it requires the preservation of $\iota$ not only for reachable states, but for all states. This is a work-around for the undecidability of reachability of states. In addition, the assumption in condition (ii) allows us to reduce all invariance properties to the consistency proof obligation, as $\iota \rightarrow \varphi$ will hold for all states. Concerning condition (iii) we exploit the lemmata in the previous section showing that we can always define a conservative refinement of $M$, in which these variants exist. We say that we obtain {\em relative completeness}, if we can show that whenever $M \models \varphi$ holds for a machine $M$ satisfying the conditions above and TREBL formula $\varphi$, then there exists a derivation, i.e. $M \vdash \varphi$ holds.

This will then be the crucial steps in the proof of the following theorem. The rest is just done by induction, for which we use a well-founded order on TREBL formulae. For arbitrary TREBL formulae $\varphi$ we first define a {\em rank} $r(\varphi)$ inductively (with $X \in \{ \square, \square^1 \}$ and $Y \in \{ \lozenge, \lozenge^1 \}$) as follows:
\begin{alignat*}{3}
r(\varphi) &= 0 \;\;\text{for}\; \varphi \in \mathcal{L} \quad&\quad r(X \varphi) &= r(\varphi) + 1 \quad&\quad r(Y \varphi) &= r(\varphi) + 1 \\
r(XY \varphi) &= r(\varphi) + 2 & r(YX \varphi) &= r(\varphi) + 2 &\quad r(X (\varphi \rightarrow Y \psi)) &= r(\psi) + 3
\end{alignat*}

\noindent
We then get the desired order letting $\varphi \le \psi$ for $r(\varphi) \le r(\psi) \neq 0$. In addition, we let $\varphi \le \psi$, $X \varphi \le X \psi$ and $Y \varphi \le Y \psi$ for $\varphi, \psi \in \mathcal{L}$ with $\models \varphi \rightarrow \psi$. Then for all derivation rules except INV$_2$ and INV$_2^1$ the antecedents have smaller rank than the conclusion.

\begin{thm}\label{thm-complete}

The system $\mathfrak{R}$ of derivation rules is relative complete for TREBL, i.e. for any TREBL formula $\varphi$ such that $M \models \varphi$ holds we have $M \vdash_{\mathfrak{R}} \varphi$, provided that $M$ satisfies the relativity assumptions above.

\end{thm}

\proof
Let $\varphi \in \mathcal{LT}$ and assume that $M \models \varphi$ holds. We use induction over $\le$ (where $\le$ is the well-founded order on the rank 
$r(\varphi)$ defined just above) to show that $M \vdash_{\mathfrak{R}} \varphi$ holds. For $r(\varphi) = 0$, i.e. $\varphi \in \mathcal{L}$ we immediately get $M \vdash_{\mathfrak{R}} \varphi$, because $\mathfrak{R}$ contains a complete set of derivation rules for first-order logic with types. In the following assume $r(\varphi) > 0$. Then we distinguish several cases.

\textbf{Invariance.} Consider an invariance formula $\varphi = \square \psi$ with $\psi \in \mathcal{L}$. Then for an arbitrary trace $\tau = S_0, S_1, \dots$ of $M$ we have $S_k \models \psi$ for all $k$. According to our assumption $\psi$ is included in the invariant $\iota$ of $M$, hence $\models \iota \rightarrow \psi$ holds, and by the induction hypothesis we get $\vdash_{\mathfrak{R}} \iota \rightarrow \psi$. Furthermore, as $M$ is consistent, so we have $M \models \square \iota$, hence also $M \vdash_{\mathfrak{R}} \square \iota$. Then we can apply the derivation rule INV$_2$ we get $M \vdash_{\mathfrak{R}} \square \psi$ as claimed.

Analogously, for $\varphi = \square^1 \psi$ with $\psi \in \mathcal{L}$ there exists a trace $\tau = S_0, S_1, \dots$ of $M$ with $S_k \models \psi$ for all $k$. Again we get $\vdash_{\mathfrak{R}} \iota \rightarrow \psi$ and $M \vdash_{\mathfrak{R}} \square^1\iota$. Then we can apply the derivation rule INV$_2^1$ we get $M \vdash_{\mathfrak{R}} \square \psi$ as claimed.


\textbf{Reachability.} Consider a reachability formula $\varphi = \lozenge \psi$ with $\psi \in \mathcal{L}$. As we assume that $M$ is sufficiently refined, Lemma \ref{lem-reach} implies that there exists an e-variant $t = (t_1,b)$ such that $M \models \text{var}_e(t,\neg\psi)$ and $M \models b=0$ hold. by the induction hypothesis we get $M \vdash_{\mathfrak{R}} \text{var}_e(t,\neg\psi)$ and $M \vdash_{\mathfrak{R}} b=0$. Then with the derivation rule R we get $M \vdash_{\mathfrak{R}} \lozenge \psi$ as claimed.

The case of a reachability formula $\varphi = \lozenge^1 \psi$ with $\psi \in \mathcal{L}$ is handled analogously using Lemma \ref{lem-reach} and the derivation rule R$^1$.

\textbf{Existence.} Consider an existence formula $\varphi = \square \lozenge \psi$ with $\psi \in \mathcal{L}$. As $M$ is sufficiently refined, Lemma \ref{lem-conv} implies the existence of a c-variant $t$ such that $M \models \text{var}_c(t, \neg\psi)$ holds. Furthermore, for an arbitrary trace $\tau = S_0, S_1, \dots$ of $M$ we have $S_k \models \lozenge \psi$ for all $k$. In particular, $\tau$ cannot terminate in a state satisfying $\neg \psi$, hence $M \models \Dlf(\neg\psi)$. by the induction hypothesis we get $M \vdash_{\mathfrak{R}} \text{var}_c(t, \neg\psi)$ and $M \vdash_{\mathfrak{R}} \Dlf(\neg\psi)$. Then we can apply the derivation rule E, which yields $M \vdash_{\mathfrak{R}} \varphi$ as claimed.

For an existence formula of the form $\varphi = \square^1 \lozenge^1 \psi$ with $\psi \in \mathcal{L}$ Lemma \ref{lem-conv1} implies the existence of an lc-variant $t$ such that $M \models \text{var}_c^1(t, \neg\psi)$ holds. by the induction hypothesis we get $M \vdash_{\mathfrak{R}} \text{var}_c^1(t, \neg\psi)$, and then we can apply rule E$^{11}$ to get $M \vdash_{\mathfrak{R}} \varphi$.

Analogously, for $\varphi = \square^1 \lozenge \psi$ or $\varphi = \square \lozenge^1 \psi$ with $\psi \in \mathcal{L}$, respectively, we get an lc-variant $t$ from Lemma \ref{lem-conv1} such that $M \models \text{var}_c^1(t, \neg\psi)$ holds. In the former case we must have in addition $M \models \lozenge \psi$, and in the latter case we have $M \models \Dlf(\neg\psi)$. by the induction hypothesis we get $M \vdash_{\mathfrak{R}} \text{var}_c^1(t, \neg\psi)$ and either $M \vdash_{\mathfrak{R}} \lozenge \psi$ or $M \vdash_{\mathfrak{R}} \Dlf(\neg\psi)$, respectively. This allows us to apply rule E$^{10}$ or rule E$^{01}$, respectively, to yield $M \vdash_{\mathfrak{R}} \varphi$.

\textbf{Persistence.} Consider a persistence formula $\varphi = \lozenge\square\psi$ with $\psi \in \mathcal{L}$. For an arbitrary trace $\tau = S_0, S_1, \dots$ of $M$ there exists some $k$ such that $S_\ell \models \psi$ holds for all $\ell \ge k$. In particular, $\tau$ cannot terminate in a state satisfying $\neg\psi$, i.e. $M \models \Dlf(\neg\psi)$ must hold. Furthermore, as $M$ is assumed to be sufficiently refined, Lemma \ref{lem-div} implies the existence of a d-variant $t$ such that $M \models \text{var}_d(t,\psi)$ holds. by the induction hypothesis we get $M \vdash_{\mathfrak{R}} \text{var}_d(t,\psi)$ and $M \vdash_{\mathfrak{R}} \Dlf(\neg\psi)$. We can apply rule P, which yiels $M \vdash_{\mathfrak{R}} \varphi$ as claimed.

For diversified persistence formulae we proceed analogously. For $\varphi = \lozenge^1\square^\psi$ with $\psi \in \mathcal{L}$ we can use Lemma \ref{lem-div1}, hence there exists an ld-variant $t$ such that $M \models \text{var}_d^1(t, \psi)$ holds. For $\varphi = \lozenge\square^1\psi$ with $\psi \in \mathcal{L}$ we can use Lemma \ref{lem-pdiv}, hence there exists a pld-variant $t = (t_1, t_2)$ such that $M \models \text{var}_d^{p1}(t, \psi)$ holds. For $\varphi = \lozenge^1\square \psi$ with $\psi \in \mathcal{L}$ we can also use Lemma \ref{lem-pdiv}, which gives us a pd-variant $t = (t_1, t_2)$ such that $M \models \text{var}_d^p(t, \psi)$ holds. In these three cases we get $M \vdash_{\mathfrak{R}} \text{var}_d^1(t, \psi)$, $M \vdash_{\mathfrak{R}} \text{var}_d^{p1}(t, \psi)$ or $M \vdash_{\mathfrak{R}} \text{var}_d^p(t, \psi)$, respectively, by the induction hypothesis. Then we can apply the rules P$^{11}$, P$^{01}$ and P$^{10}$, respectively, which imply $M \vdash_{\mathfrak{R}} \varphi$.

\textbf{Progress.} Next consider a progress formula $\varphi = \square (\varphi_1 \rightarrow \lozenge \varphi_2)$ with $\varphi_1, \varphi_2 \in \mathcal{L}$. Let $\tau = S_0, S_1, \dots$ be a trace of $M$, and assume that there exists a $k$ such that $S_\ell \models \neg\varphi_2$ holds for all $\ell \ge k$. According to Lemma \ref{lem-progress} other traces can be ignored. We can also assume that at least one such trace exists. Otherwise, $M \models \square\lozenge \varphi_2$ holds, and by the induction hypothesis we get $M \vdash_{\mathfrak{R}} \square\lozenge \varphi_2$. We can apply rule PR$_0$, which gives us $M \vdash_{\mathfrak{R}} \square (\varphi_1 \rightarrow \lozenge \varphi_2)$ as claimed.

Then $\tau$ contains only a finite number of states satisfying $\varphi_2$. We consider all states in these traces that satisfy $\neg \varphi_2$ and for which there exists a later state satisfying $\varphi_2$. Let $\varphi_3$ be a common invariant for these states. More precisely, for a state $S$ satisfying $\neg \varphi_2$ take its type, i.e. the set of all formulae in $\mathcal{L}$ that hold in $S$, let $\varphi_S$ be an isolating formula for this type, and define $\varphi_3$ as the disjunction of all these formulae $\varphi_S$. 

Then by construction $M \models \square^{\neg\varphi_2} \text{leadslocto}(\varphi_3 \wedge \neg\varphi_2, \varphi_3 \vee \varphi_2)$ holds. As $M$ is sufficiently refined, we can apply Lemma \ref{lem-cvar}, hence there exists a conditional c-variant $t = (t_1,b)$ such that $M \models \text{cvar}_c(t, \neg\varphi_2, \text{leadslocto}(\varphi_3 \wedge \neg\varphi_2, \varphi_3 \vee \varphi_2))$ holds. By the induction hypothesis we have $M \vdash_{\mathfrak{R}} \text{cvar}_c(t, \neg\varphi_2, \text{leadslocto}(\varphi_3 \wedge \neg\varphi_2, \varphi_3 \vee \varphi_2))$.

If a trace has an tail of states satisfying $\neg\varphi_2$, these states must also satisfy $\neg\varphi_1$. Then our construction implies that $M \models \square^{\neg\varphi_2} (\varphi_1 \wedge \neg \varphi_2 \rightarrow \varphi_3)$ holds. As $M$ is sufficiently refined, we can apply Lemma \ref{lem-cvar}, hence there exists a conditional c-variant $t = (t_1,b)$ such that $M \models \text{cvar}_c(t, \neg\varphi_2, \varphi_1 \wedge \neg \varphi_2 \rightarrow \varphi_3)$. By the induction hypothesis we get $M \vdash_{\mathfrak{R}} \text{cvar}_c(t, \neg\varphi_2, \varphi_1 \wedge \neg \varphi_2 \rightarrow \varphi_3)$.

Furthermore, as $M$ is sufficiently refined, then for all traces $\tau$ that are divergent in $\neg \varphi_2$ Lemma \ref{lem-div} implies the existence of a variant term $t(\bar{v})$ such that $\text{var}_d(t, \neg \varphi_1 \wedge \neg \varphi_2)$ is valid, in particular, $t(\bar{v}) > v_{\text{min}}$ is included in $\varphi_3$, and $t(\bar{v}) = v_{\text{min}}$ holds in all tailing states satisfying $\neg \varphi_1 \wedge \neg \varphi_2$. This implies $M \models \lozenge^{\neg\varphi_2} \square \neg \varphi_3$. We can apply Lemma \ref{lem-svar} to get a simultaneous cd-variant $t^\prime$ such that $M \models \text{svar}_{cd}(t^\prime, (\neg\varphi_2, \neg\varphi_3))$. Then by the induction hypothesis we get $M \vdash_{\mathfrak{R}} \text{svar}_{cd}(t^\prime, (\neg\varphi_2, \neg\varphi_3))$.
Finally, we can apply rule PR, which yields $M \vdash_{\mathfrak{R}} \square (\varphi_1 \rightarrow \lozenge \varphi_2)$ as claimed.

For a progress formula $\varphi = \square (\varphi_1 \rightarrow \lozenge^1 \varphi_2)$ with $\varphi_1, \varphi_2 \in \mathcal{L}$ we can assume without loss of generality that $M \not\models \square\lozenge^1 \varphi_2$ holds. Otherwise by the induction hypothesis we would get $M \vdash_{\mathfrak{R}} \square\lozenge^1 \varphi_2$, which allows us to apply rule PR$_0$ to obtain $M \vdash_{\mathfrak{R}} \square (\varphi_1 \rightarrow \lozenge^1 \varphi_2)$ as claimed.

We concentrate again on traces $\tau = S_0, S_1, \dots$ of $M$ such that there exists a $k$ such that $S_\ell \models \neg\varphi_2$ holds for all $\ell \ge k$. According to Lemma \ref{lem-progress01} other traces can be ignored. Then we construct the formula $\varphi_3$ analogously, but we only consider states in traces $S_0, S_1, \dots$ such that whenever $S_k \models \varphi_1$ holds, also $S_\ell \models \varphi_2$ holds for some $\ell \ge k$. Every state satisfying $\varphi_2$ appears in at least one such trace. 

By construction we get $M \models \square^{\neg\varphi_2} \text{leadslocto}^1(\varphi_3 \wedge \neg\varphi_2, \varphi_3 \vee \varphi_2)$. Furthermore, we get immediately $M \models \square^{\neg\varphi_2} (\varphi_1 \wedge \neg \varphi_2 \rightarrow \varphi_3)$ and $M \models \lozenge^{\neg\varphi_2} \square \neg \varphi_3$ using the same arguments as above. As $M$ is sufficiently refined, we can apply Lemmata \ref{lem-svar} and \ref{lem-cvar}, i.e. there exists a simulataneous cd-variant $t$ and a conditional c-variant $t^\prime$ such that $M \models \text{svar}_{cd}(t, (\neg\varphi_2, \neg\varphi_3))$ and $M \models \text{cvar}_c(t^\prime_i, \neg\varphi_2, \theta_i)$ for $\theta_1 = \varphi_1 \wedge \neg \varphi_2 \rightarrow \varphi_3$ and $\theta_2 = \text{leadslocto}^1(\varphi_3 \wedge \neg\varphi_2, \varphi_3 \vee \varphi_2)$.

By the induction hypothesis we get that  $M \vdash_{\mathfrak{R}} \text{svar}_{cd}(t, (\neg\varphi_2, \neg\varphi_3))$ and $M \vdash_{\mathfrak{R}} \text{cvar}_c(t^\prime_i, \neg\varphi_2, \theta_i)$ ($i=1,2$), which allows us to apply rule PR$^{01}$ to infer $M \vdash_{\mathfrak{R}} \square (\varphi_1 \rightarrow \lozenge^1 \varphi_2)$ as claimed.

Next consider diversified progress formulae $\varphi = \square^1 (\varphi_1 \rightarrow \lozenge \varphi_2)$. We can assume that there exists at least one trace of $M$, for which there exists a $k$ such that $S_\ell \models \neg\varphi_2$ holds for all $\ell \ge k$. Otherwise $M \models \square\lozenge \psi$ holds, and by the induction hypothesis we would get $M \vdash_{\mathfrak{R}} \square\lozenge \psi$. Then $M \vdash_{\mathfrak{R}} \varphi$ would follow immediately using rule PR$_0$. 

Assuming that $M \models \varphi$ holds there exists a trace $S_0, S_1, \dots$ of $M$ such that $S_i \models \varphi_1 \rightarrow \lozenge\varphi_2$ holds for all $i$. Let us first assume that there is some $k$ with $S_\ell \models \neg \varphi_2$ for all $\ell \ge k$. Then also $S_\ell \models \neg \varphi_1$ holds for all $\ell \ge k$. 

\textbf{\em Case 1.} Assume that there exists a trace $S_0^\prime, S_1^\prime, \dots$ and some $k$ such that $S_k^\prime \models \varphi_1$ and $S_\ell^\prime \models \neg\varphi_2$ hold for all $\ell \ge k$. This cannot be the trace $S_0, S_1, \dots$. Then we consider all finite subtraces $S_i^\prime, \dots, S_j^\prime$ of all traces with $\neg\varphi_2$-tail such that $S_i^\prime \models \varphi_1$, $S_x \models \neg\varphi_2$ for all $i \le x \le j$, $S_{j+1}^\prime \models \varphi_2$ and $S_i^\prime = S_i$. Consider all states $S_x^\prime$ appearing in such subtraces, and let $\varphi_3$ be a common invariant, constructed as above.

Then $M \models \square^1 (\varphi_1 \wedge \neg\varphi_2 \rightarrow \varphi_3)$ holds, because $\varphi_1 \wedge \neg\varphi_2 \rightarrow \varphi_3$ is satisfied by all states $S_i$. Also $M \models \square^{\neg\varphi_2} \text{leadslocto}(\neg\varphi_2 \wedge \varphi_3, \varphi_2 \vee \varphi_3)$ holds by construction, because in the selected set of traces that states satisfying $\neg\varphi_2 \wedge \varphi_3$ are just the states used for the definition of $\varphi_3$, and the defining subtraces have been chosen such that they are always followed by a state satisfying $\varphi_2$, and such a state must exists for all the subtraces with a start state $S_i^\prime = S_i$. Furthermore, $M \models \lozenge^{\neg\varphi_2} \square \neg\varphi_3$ holds by construction of $\varphi_3$.

Then by Lemmata \ref{lem-svar} and \ref{lem-cvar} there exist a simultaneous cd-variant $t$ and a conditional c-variant $t^\prime$ such that $M \models \text{svar}_{cd}(t, (\neg\varphi_2, \neg\varphi_3))$ and $M \models \text{cvar}_c(t^\prime_i, \neg\varphi_2, \theta)$ for $\theta = \text{leadslocto}(\varphi_3 \wedge \neg\varphi_2, \varphi_3 \vee \varphi_2)$. By the induction hypothesis we get $M \vdash_{\mathfrak{R}} \square^1 (\varphi_1 \wedge \neg\varphi_2 \rightarrow \varphi_3)$, $M \vdash_{\mathfrak{R}} \text{svar}_{cd}(t, (\neg\varphi_2, \neg\varphi_3))$ and $M  \vdash_{\mathfrak{R}} \text{cvar}_c(t^\prime_i, \neg\varphi_2, \theta)$. Then we can apply rule PR$^{10}_3$, which yields $M \vdash_{\mathfrak{R}} \varphi$.

\textbf{\em Case 2.} Now assume that no such trace as in Case 1 exists, in particular, the states in a $\neg\varphi_2$-tail also satisfy $\neg\varphi_1$. As in Case 1 consider subtraces $S_i^\prime, \dots, S_j^\prime$ of all traces with $\neg\varphi_2$-tail without requiring $S_i^\prime = S_i$, and construct again $\varphi_3$ as a common invariant of all states $S_x^\prime$ appearing in such subtraces. We consider two subcases.

\textbf{\em Case 2a.} First assume that there is no trace in the set of selected traces that contains a state $S_x^\prime$ satisfying $\varphi_3$ and a successor state $S_{x+1}^\prime$ satisfying $\neg\varphi_3$. Note that such a trace could only exist, if $S_\ell^\prime \models \neg\varphi_2$ holds for all $\ell \ge x$. Then $M \models \square^{\neg\varphi_2} \text{leadslocto}(\neg\varphi_2 \wedge \varphi_3, \varphi_2 \vee \varphi_3)$ still holds by construction. In addition, $M \models \square^{\neg\varphi_2} (\varphi_1 \wedge \neg\varphi_2 \rightarrow \varphi_3)$ holds, because states in the tail of the traces under consideration do not contain states satisfying $\varphi_1$. Furthermore, $M \models \lozenge^1 \square (\neg\varphi_2 \wedge \neg\varphi_3)$ holds, as the defining condition is satisfied by the trace $S_0, S_1, \dots$.

As $M$ is sufficiently refined, we can again exploit Lemma \ref{lem-cvar}. Hence there exist conditional c-variants $t^\prime_i$ such that $M \models \text{cvar}_c(t^\prime_i, \neg\varphi_2, \theta_i)$ for $\theta_1 = \varphi_1 \wedge \neg \varphi_2 \rightarrow \varphi_3$ and $\theta_2 = \text{leadslocto}(\varphi_3 \wedge \neg\varphi_2, \varphi_3 \vee \varphi_2)$. By the induction hypothesis we get $M \vdash_{\mathfrak{R}} \text{cvar}_c(t^\prime_i, \neg\varphi_2, \theta_i)$ for $i = 1,2$ as well as $M \vdash_{\mathfrak{R}} \lozenge^1 \square (\neg\varphi_2 \wedge \neg\varphi_3)$. Hence, we can apply rule PR$^{10}_1$, which implies again $M \vdash_{\mathfrak{R}} \varphi$.

\textbf{\em Case 2b.} Finally, assume that there exists a trace in the set of selected traces that contains a state $S_x^\prime$ satisfying $\varphi_3$ and a successor state $S_{x+1}^\prime$ satisfying $\neg\varphi_3$. Then still $M \models \square^1 \text{leadslocto}(\neg\varphi_2 \wedge \varphi_3, \varphi_2 \vee \varphi_3)$ still holds by construction, as the defining condition holds for the trace $S_0, S_1, \dots$. Also $M \models \square^{\neg\varphi_2} (\varphi_1 \wedge \neg\varphi_2 \rightarrow \varphi_3)$ still holds, because states in the tail of the traces under consideration do not contain states satisfying $\varphi_1$. Furthermore, $M \models \lozenge^{\neg\varphi_2} \square \neg\varphi_3$ holds by construction of $\varphi_3$.

Then by Lemmata \ref{lem-svar} and \ref{lem-cvar} there exist a simultaneous cd-variant $t$ and a conditional c-variant $t^\prime$ such that $M \models \text{svar}_{cd}(t, (\neg\varphi_2, \neg\varphi_3))$ and $M \models \text{cvar}_c(t^\prime_i, \neg\varphi_2, \theta)$ for $\theta = \varphi_1 \wedge \neg\varphi_2 \rightarrow \varphi_3$. By the induction hypothesis we get $M \vdash_{\mathfrak{R}} \text{svar}_{cd}(t, (\neg\varphi_2, \neg\varphi_3))$, $M \vdash_{\mathfrak{R}} \text{cvar}_c(t^\prime_i, \neg\varphi_2, \theta)$ and $M \vdash_{\mathfrak{R}} \square^1 \text{leadslocto}(\neg\varphi_2 \wedge \varphi_3, \varphi_2 \vee \varphi_3)$. Hence, we can apply rule PR$^{10}_2$, which implies again $M \vdash_{\mathfrak{R}} \varphi$.

Finally, if the trace $S_0, S_1, \dots$ terminates in a state satisfying $\varphi_2$ or $S_i \models \varphi_2$ holds for infinitely many $i$, there exists a Boolean-valued variable $b$ such that $M \models (b=1 \leftrightarrow \varphi_1)$, $M \models \text{pre}(\neg \varphi_1, b)$, and $M \models \square^1 \lozenge (b=0 \vee \varphi_2)$ hold. By the induction hypothesis we get $M \vdash_{\mathfrak{R}} (b=1 \leftrightarrow \varphi_1)$, $M \vdash_{\mathfrak{R}} \text{pre}(\neg \varphi_1, b)$, and $M \vdash_{\mathfrak{R}} \square^1 \lozenge (b=0 \vee \varphi_2)$, so we can apply rule PR$^{10}_0$ to get $M \vdash_{\mathfrak{R}} \varphi$.

The case of diversified progress formulae $\varphi = \square^1 (\varphi_1 \rightarrow \lozenge^1 \varphi_2)$ is handled analogously. We only need to replace $\text{leadslocto}$ by $\text{leadslocto}^1$ and apply the rules PR$^{11}_j$ ($1 \le j \le 3$) from Lemma \ref{lem-progress11} instead of the rules from Lemma \ref{lem-progress10}, or rule PR$^{11}_0$.

\textbf{Nested Temporal Formulae.} The extension to TREBL formulae that result from multiple application of the constructors in Definition \ref{def-trebl} is handled in Appendix \ref{appB}.\qed



\section{Discussion and Conclusions}\label{sec:discussion}

In this article we developed a temporal Event-B logic and showed the existence of a sound and relative complete set of derivation rules. First attempts in this direction go back to the work of Hoang and Abrial, who discovered a simple set of sound derivation rules that can be used to prove invariance, existence, progress and persistence conditions that are to hold for all traces of an Event-B machine \cite{hoang:icfem2011}. Their work was grounded on coupling the UNTIL-fragment of LTL \cite{pnueli:focs1977} with the logic of Event-B---the resulting logic was called LTL(EB) in \cite{ferrarotti:foiks2024}---by means of which they could get rid of the intrinsic restriction of LTL being in essence an extension of propositional logic. However, they preserved the semantics of LTL, i.e. the interpretation of temporal formulae over arbitrary traces, so they gained expressiveness and mechanical proof capability at the cost of losing completeness. However, completeness is essential to ensure that a proof by means of derivation rules can actually be conducted.

Completeness for a fragment of LTL(EB) large enough to cover the most important liveness conditions was addressed in our previous work \cite{ferrarotti:foiks2024}, where we could show that under some restrictions the derivation rules discovered by Hoang and Abrial with some small modifications are relative complete for this fragment. That is, if a consistent Event-B machine $M$ is sufficiently refined, and $M$ satisfies a formula in the fragment, then there also exists a derivation for this formula. In other words, if there exists a proof of some liveness condition, the proof can be done by using the derivation rules, provided that certain variant terms are explicitly defined for $M$. The crucial point is to show that this is always possible, which reflects the intuitively clear fact, that any condition that holds for a machine must in some way be reflected explicitly or implicitly in the specification.

Still the work in \cite{ferrarotti:foiks2024} was insufficient, as the relative completeness result was restricted to machines that are tail-homogeneous\footnote{Tail homogeneity requires that either all traces are divergent for state formulae appearing in the liveness conditions to be proven or none.}, and the semantics definition does not reflect the decisive fact that for a given Event-B machine any valid state $S$ determines already the traces starting in $S$. That is, if the building of successor states and successors of successors, etc. is taken into the logic, we can obtain a logic that allows us to express temporal properties, but is interpreted over states rather than traces, which is a significant simplification and also removes the necessity to distinguish between state formulae and temporal formulae. In this article we removed all these restrictions.

A parallel treatment based on Abstract State Machines \cite{ferrarotti:abz2024} defines a temporal extension of the logic of non-deterministic ASMs \cite{ferrarotti:amai2018}, in which update sets defining successor states are used as first class objects\footnote{This defines a complete second-order logic with Henkin semantics.}, so it becomes possible to reason about successor states, successors of successors, etc. In order to obtain formulae that express properties of complete runs it was sufficient to extend the building of update sets to arbitray sequences of steps. In doing so, all relevant temporal operators known from LTL or also CTL \cite{clarke:lop1981} could be defined in the logic, so they are merely shortcuts for rather complex formulae. Also, definable teams of traces (capturing the spirit of team semantics \cite{gutsfeld:lics2022}) can be integrated, though this was only done for restricted cases.

We concluded that it is a very promising idea to integrate the essential ideas of the modal extensions of the logic of ASMs into our treatment of a temporal logic for Event-B. As Event-B is grounded in the theory of sets, it is straightforward to define update sets, and also the inductive definition of update sets expressing the changes of multiple steps is straightforward. In this way we obtained an extension $\mathcal{L}^{ext}$ of the logic of Event-B, in which all temporal operators for all traces, for a single trace or for selected definable sets of traces can be defined. This is the basis for the definition of TREBL as a sound and relative complete fragment of $\mathcal{L}^{ext}$. The definable sets of traces allow us to remove the tail-homogeneity restriction.

We consider the definition of a sound and relative complete temporal logic for Event-B a major breakthrough. The usefulness of known temporal logics such as LTL, CTL or CTL$^*$ for the verification of liveness conditions of real systems is severely restricted by the propositional nature of these logics. The tight coupling of temporal logic with Event-B removes these restrictions, and allows us to obtain not only sound derivation rules, but also relative completeness. Surely, proofs may require the definition of appropriate variant terms as part of an Event-B specification, but proofs of complex conditions are never for free. We only need to exploit two simple facts:

\begin{itemize}

\item States of a machine determine traces starting in that state, so the logic can be defined in a way that formulae are interpreted over states and temporal operators are merely shortcuts\footnote{Leslie Lamport characterised the explicit quantification over the number of steps as evil and recommended to stick to temporal logics as the lesser evil \cite{lamport:pnueli2010}. Indeed, the defining formulae for the temporal operators may appear horrible, but in TREBL they are not used. Once defined the temporal operators as in any other temporal logic, but formulae are nonetheless interpreted over states and express properties of traces starting in such states.}.

\item Variant terms that are needed for the proofs can be made explicit by means of refinement.

\end{itemize}

\noindent
In addition to the theory of TREBL we are in the (almost completed) process of integrating all our derivation rules into the RODIN proof tool using the EB4EB meta-theory, which will be reported in a forthcoming Part II of our work. In doing so we provide a desirable mechanical proof suppport for TREBL. We further demonstrated the power of TREBL on various examples, in particular examples adressing security aspects. Some of these conditions such as non-interference were considered to be very challenging, but in the TREBL context they become almost trivial.

During the development of this research we often discovered that we most likely could further extend TREBL. For instance, it should be possible to also include UNTIL formulae in the logic or try to integrate arbitrary definable teams of traces. Also the quantification over traces is limited, and it seems likely that it can be handled more generally. As our treatment until now has already grown significantly, we keep further extensions to TREBL including sound derivation rules and a proof of relative completeness for future research.

As a final remark it is clear that while our work here is based on Event-B, it should be possible to conduct analogous research for any other state-based rigorous method. In particular, the two decisive facts mentioned above are not bound to Event-B.

\bibliographystyle{alphaurl}
\bibliography{trebl}

\appendix

\section{Variant Existence Proofs}\label{appA}

\proof[Proof (of Lemma \ref{lem-conv}).]
In the following, if $F$ is a formula or set of actions and $v, w$ are variables, we use $F(v/w)$ to denote the formula or set of actions obtained by substituting in $F$ each occurrence of $v$ by $w$. Furthermore, we use list and list operators, with $[]$ denoting the empty list and $l_1 \cdot l_2$ denoting concatenation of lists $l_1$ and $l_2$. Of course, with proper encoding we could use sets instead of lists, but lists give us a cleaner presentation for this proof. 

Let $M'$ be the Event-B machine built form $M$ as follows: 

\begin{itemize}

\item The initial event $\mathit{init}'$ of $M'$ is the result of first applying the substitution $\bar{v}/\bar{w}$ to the initial event $\mathit{init}$ of $M$, and then adding the following actions: $s, u := 0$, $\bar{v} :\!| \, \neg\varphi(\bar{v}')$ and $l := []$.

\item We add to the set of events $E'$ of $M'$ the following event, which is triggered only once and only if the corresponding trace of $M$ is of length $1$. \\ 
    $e_{\mathit{firstA}} = {\bf any} \, \bar{x} \, {\bf where} \, u = 0 \wedge \neg \big( \bigvee_{e_i \in E} G_{i}(\bar{x}, \bar{v}/\bar{w}) \big) \, {\bf then} \, \{ s:= 1, u:= 2, \bar{v}:=\bar{w}\}$.

\item We add to the set of events $E'$ of $M'$ the following event, which is triggered only once and only if the corresponding trace of $M$ is of length $> 1$. \\ 
    $e_{\mathit{firstB}} = {\bf any} \, \bar{x} \, {\bf where} \, u = 0 \wedge \big( \bigvee_{e_i \in E} G_{i}(\bar{x}, \bar{v}/\bar{w}) \big) \, {\bf then} \, \{u:= 1\}$.
    
\item For each $e_i = {\bf any} \, \bar{x} \, {\bf where} \, G_i(\bar{x}, \bar{v}) \, {\bf then} \, A_i(\bar{x}, \bar{v}, \bar{v}') \, {\bf end}$ in the set of events $E$ of $M$, we add the following two corresponding events to $E'$ (the set of events of $M'$).\\ 
    $e_{ia} = {\bf any} \, \bar{x} \, {\bf where} \, G_{ia}(\bar{x}, \bar{v}, \bar{w}, l, s, u) \, {\bf then} \, A_{ia}(\bar{x}, \bar{v}, \bar{w}, l, s, u, \bar{v}', \bar{w}', l', s', u')$\\
    $e_{ib} = {\bf any} \, \bar{x} \, {\bf where} \, G_{ib}(\bar{x}, \bar{v}, \bar{w}, l, s, u) \, {\bf then} \, A_{ib}(\bar{x}, \bar{v}, \bar{w}, l, s, u, \bar{v}', \bar{w}', l', s', u')$\\
    $e_{ic} = {\bf any} \, \bar{x} \, {\bf where} \, G_{ic}(\bar{x}, \bar{v}, \bar{w}, l, s, u) \, {\bf then} \, A_{ic}(\bar{x}, \bar{v}, \bar{w}, l, s, u, \bar{v}', \bar{w}', l', s', u')$\\
    where
    
\begin{itemize}

\item $G_{ia}(\bar{x}, \bar{v}, \bar{w}, l, s, u) \equiv G_i(\bar{x}, \bar{v}/\bar{w}) \wedge \varphi(\bar{v}/\bar{w}) \wedge s = 0 \wedge u = 1$\\
        $A_{ia}(\bar{x}, \bar{v}, \bar{w}, l, s, u, \bar{v}', \bar{w}', l', s', u') =$ \\ 
        \hspace*{2cm} $A_i(\bar{x}, \bar{v}/\bar{w}, \bar{v}'/\bar{w}') \cup \{l := l \cdot [\bar{w}]\}$ 

\item $G_{ib}(\bar{x}, \bar{v}, \bar{w}, l, s, u) \equiv G_i(\bar{x}, \bar{v}/\bar{w}) \wedge \neg \varphi(\bar{v}/\bar{w}) \wedge s = 0 \wedge u = 1 \wedge l \neq []$\\
        $A_{ib}(\bar{x}, \bar{v}, \bar{w}, l, s, u, \bar{v}', \bar{w}', l', s', u' ) = $\\ \hspace*{2cm} $A_i(\bar{x}, \bar{v}/\bar{w}, \bar{v}'/\bar{w}') \cup \{s := 1, \bar{v} := \mathit{head}(l), l := \mathit{tail}(l) \cdot [\bar{w}] \}$

\item $G_{ic}(\bar{x}, \bar{v}, \bar{w}, l, s, u) \equiv G_i(\bar{x}, \bar{v}/\bar{w}) \wedge \neg \varphi(\bar{v}/\bar{w}) \wedge s = 0 \wedge u = 1 \wedge l = []$\\
        $A_{ic}(\bar{x}, \bar{v}, \bar{w}, l, s, u,  \bar{v}', \bar{w}', l', s', u') = $\\ \hspace*{2cm} $A_i(\bar{x}, \bar{v}/\bar{w}, \bar{v}'/\bar{w}') \cup \{s := 1, \bar{v} := \bar{w}\}$
        
\end{itemize}

\item We add the following two additional events $e_r$ and $e_s$ to the set $E'$ of $M'$: 

\begin{itemize}

\item $G_{r}(\bar{x}, \bar{v}, \bar{w}, l, s, u) \equiv l \neq [] \wedge s = 1 \wedge u = 1$\\
        $A_{r}(\bar{x}, \bar{v}, \bar{w}, l, s, \bar{v}', \bar{w}', l', s') = \{\bar{v} := head(l), l := tail(l) \}$ 

\item $G_{s}(\bar{x}, \bar{v}, \bar{w}, l, s, u) \equiv l = [] \wedge s = 1 \wedge u = 1$\\
        $A_{s}(\bar{x}, \bar{v}, \bar{w}, l, s, u, \bar{v}', \bar{w}', l', s', u') =  \{s := 0\}$ 

\end{itemize}

\end{itemize}

\noindent
Let $t(\bar{v}, \bar{w}, l, s, u)$ be the term $\mathit{len}(l)$, which gives us the length of the list $l$. We show next that properties (i) and (ii) of the lemma hold. Let us start with (i), which can be proven by showing that if $M$ is convergent in $\varphi$, $\sigma$ is a trace of $M'$ and $0 \leq k \leq \ell(\sigma)$, then $\sigma^{(k)} \models \mathit{var}_c(M, \mathit{len}(l), \varphi)$. Let us assume without loss of generality that $\sigma = S_0, S_1, \ldots$ is infinite, i.e., $\ell(\sigma) = \omega$, and proceed by induction on $k$. Note that our proof below also holds when $\sigma$ is finite.

Any trace $\sigma$ of $M'$ will have at least $2$ states. This follows from the fact that $\mathit{init}'$ sets the value of $u$ in $S_0$ to $0$, and thus either event $e_\mathit{firstA}$ or $e_\mathit{firstB}$ will be enabled in $S_0$. Therefore, we start the induction with $k=1$. By definition of $\mathit{init}'$, we have that $S_0 \not\models \varphi(\bar{v})$ and thus that $\sigma^{(0)} \models \mathit{var}_c(M, \mathit{len}(l), \varphi)$. 
If state $S_1$ is the result of triggering $e_\mathit{firstA}$, then by construction of $M'$ from $M$ and the fact that $M$ is convergent in $\varphi$, we get that $S_0 \not\models \varphi(\bar{v}/\bar{w})$. We also get by the actions of $e_\mathit{firstA}$ that the value of $\bar{v}$ in $S_1$ is the same as the value of $\bar{w}$ in $S_0$. Then, $S_1 \not\models \varphi(\bar{v})$. 
If state $S_1$ is the result of triggering $e_\mathit{firstB}$ instead, then the value of $\bar{v}$ is not updated and thus it is the same in both $S_0$ and $S_1$. This again gives us that $S_1 \not\models \varphi(\bar{v})$. We can then conclude that  $\sigma^{(1)} \models \mathit{var}_c(M, \mathit{len}(l), \varphi)$.

If $k+1 > 1$, we know by inductive hypothesis that  $\sigma^{(k)} \models \mathit{var}_c(M, \mathit{len}(l), \varphi)$. Thus, either (a) $S_{k} \not\models \varphi(\bar{v})$ or (b) $S_k \models \forall \bar{x} (\neg \bigvee_{e_i \in E} G_{i}(\bar{x}, \bar{v}))$ or (c) $S_{k} \models \mathit{len}(l) \neq 0 \wedge \mathit{len}(l') < \mathit{len}(l)$, where $l'$ is the value of $l$ in state $S_{k+1}$. In order to show that $\sigma^{(k+1)} \models \mathit{var}_c(M, \mathit{len}(l), \varphi)$, we consider the four different cases depending on the values of $s$ in $S_k$ and $S_{k+1}$. 

\begin{itemize}

\item If $s = 0$ in both $S_{k}$ and $S_{k+1}$, then an  event of the form $e_{ia}$ must have been triggered in $S_{k}$, as otherwise $s$ would be $1$ in $S_{k+1}$ or $S_k$ would be the initial state and thus $k+1 = 1$. We know that $\mathit{len}(l') > \mathit{len}(l)$ whenever an event of the form $e_{ia}$ is triggered. Therefore, option (c) above is false and it must be the case by the inductive hypothesis that $S_{k} \not\models \varphi(\bar{v})$ or $S_k \models \forall \bar{x} (\neg \bigvee_{e_i \in E} G_{i}(\bar{x}, \bar{v}))$. Since the action of an event $e_{ia}$ does not update the value of $\bar{v}$, it follows that $S_{k+1} \not\models \varphi(\bar{v})$ or $S_k \models \forall \bar{x} (\neg \bigvee_{e_i \in E} G_{i}(\bar{x}, \bar{v}))$, and thus $\sigma^{(k+1)} \models \mathit{var}_c(M, \mathit{len}(l), \varphi)$.
    
\item If $s = 1$ in $S_{k}$ and $s = 0$ in $S_{k+1}$, then event $e_s$ was triggered in $S_{k}$, as otherwise $s$ would be $1$ in $S_{k+1}$. We know that $\mathit{len}(l') = \mathit{len}(l)$ since $e_s$ does not update the value of $l$. Therefore, option (c) above is false and it must be the case by the inductive hypothesis that $S_{k} \not\models \varphi(\bar{v})$ or $S_k \models \forall \bar{x} (\neg \bigvee_{e_i \in E} G_{i}(\bar{x}, \bar{v}))$. Since $e_{s}$ neither updates the value of $\bar{v}$, it follows that $S_{k+1} \not\models \varphi(\bar{v})$ or $S_k \models \forall \bar{x} (\neg \bigvee_{e_i \in E} G_{i}(\bar{x}, \bar{v}))$, and thus $\sigma^{(k+1)} \models \mathit{var}_c(M, \mathit{len}(l), \varphi)$.

\item If $s = 0$ in $S_{k}$ and $s = 1$ in $S_{k+1}$, then  event $e_\mathit{last}$ or an event of the form $e_{ib}$ or $e_{ic}$ was triggered in $S_{k}$ (notice that $e_\mathit{firstA}$ can only be triggered in state $S_0$). Otherwise $s$ would be $0$ in $S_{k+1}$. 

We know that if event $e_\mathit{last}$ was triggered in $S_k$, then $S_k \models \forall \bar{x} (\neg  (\bigvee_{e_i \in E} G_{i}(\bar{x}, \bar{v}/\bar{w})))$ and that the value of $\bar{v}$ in $S_{k+1}$ equals the value of $\bar{w}$ in $S_k$. Thus, $S_{k+1} \models \forall x (\neg  (\bigvee_{e_i \in E} G_{i}(\bar{x}, \bar{v})))$. Therefore, $\sigma^{(k+1)} \models \mathit{var}_c(M, \mathit{len}(l), \varphi)$. 
  
If instead an event of the form $e_{ib}$ was triggered in $S_k$, then $l \neq []$ in $S_{k+1}$. Then, the only event that can be triggered in $S_{k+1}$ is $e_r$ and thus $\mathit{len}(l') < \mathit{len}(l)$. Consequently,  $\sigma^{(k+1)} \models \mathit{var}_c(M, \mathit{len}(l), \varphi)$.

Finally, if an event of the form $e_{ic}$ was triggered in $S_k$, then we know that $S_k \not\models \varphi(\bar{v}/\bar{w})$ and that the value of $\bar{v}$ in $S_{k+1}$ is equal to the value $\bar{w}$ in $S_k$. Therefore, $S_{k+1} \not\models \varphi(\bar{v})$, which implies that $\sigma^{(k+1)} \models \mathit{var}_c(M, \mathit{len}(l), \varphi)$. 

\item If $s = 1$ in both $S_{k}$ and $S_{k+1}$, then event $e_r$ was triggered in $S_{k}$, as otherwise $s$ would be $0$ in $S_{k+1}$. If $\mathit{tail}(l) \neq []$ in $S_{k}$, then only event $e_r$ can be triggered in $S_{k+1}$. Triggering event $e_r$ always results in $\mathit{len}(l') < \mathit{len}(l)$. Consequently,  $\sigma^{(k+1)} \models \mathit{var}_c(M, \mathit{len}(l), \varphi)$.

On the other hand, if $\mathit{tail}(l) = []$ in $S_{k}$, then the value of $\bar{v}$ in $S_{k+1}$ equals the value of $\mathit{head}(l)$ in $S_k$. This value corresponds to the value $w$ in the latest preceding state $S_i$ in the computation $\sigma$ where an event of the form $e_{ib}$ was triggered, meaning that $S_i \not\models \varphi(\bar{v}/\bar{w})$. Thus,  $S_{k+1} \not\models \varphi(\bar{v})$ and consequently $\sigma^{(k+1)} \models \mathit{var}_c(M, \mathit{len}(l), \varphi)$. 

\end{itemize}

\noindent
Next we show that property (ii) also holds, i.e. $\sigma$ is a trace of $M$ iff $\sigma = \sigma'|_{s = 1,\bar{v}}$ for some trace $\sigma'$ of $M'$. We use the fact that $M$ is convergent in $\varphi$ to divide every trace $\sigma$ of $M$ into contiguous segments $s_0, s_1, \ldots$, where for each segment $s_j = S_{j0} \ldots S_{jk_j}$ the following holds:
\begin{itemize}
    \item The first state in the segment $s_{j+1}$ (if it exists) is the immediate successor in $\sigma$ of the last state $S_{jk_j}$ in segment $s_j$.
    \item $S_{jk_j} \not\models \varphi(\bar{v})$.
    \item $S_{jm} \models \varphi(\bar{v})$ for all $0 \leq m < k_j$.
\end{itemize}

\noindent
It follows that for every segment $s_j$ of a trace $\sigma$ of $M$ there is a corresponding segment $s_j'$ of a computation $\sigma'$ of $M'$ such that $s_j = s_j'|_{s=1,\bar{v}}$, and vice versa. Next we prove the first direction of this fact by induction on $j$. 

For $j = 0$, let the initial segment $s_0 = S_{00} \ldots S_{0k_0}$. By construction of $\mathit{init}'$, we get that in the initial state $S_{00}'$ of $M'$ the value of $\bar{w}$ coincides with that of $\bar{v}$ in $S_{00}$. If $k_0 = 0$, then depending on whether  $\ell(\sigma) = 1$ or $\ell(\sigma) > 1$, either event $e_\mathit{firstA}$ or $e_\mathit{firstB}$ can be triggered in $S_{00}'$. In the former case, $s = 1$ in the next state $S_{01}'$ in the trace of $M'$ and the value of $\bar{v}$ coincides with its value in $S_0$. A last step in the trace of $M'$ triggers $e_s$, which leads to a further state $S_{02}'$ where $s = 0$. Thus, $s_0 = s_0'|_{s = 1, \bar{v}}$ for $s_0' = S_{00}' S_{01}' S_{02}'$. In the latter case, the value of $\bar{w}$ and $s$ in state $S_{01}'$ coincide with the respective values in $S_{00}'$ and  an event of the form $e_{ic}$ will be triggered (notice that $l = []$ and an event $e_i$ was triggered by $M$ in state $S_{00}$ as $\ell(\sigma) > 1$). Thus, in the next state $S'_{02}$ in the trace of $M'$ we get that $s = 1$ and the value of $\bar{v}$ coincides with its value in $S_{00}$. An additional step of $M'$ is triggered in state $S_{02}'$ by the event $e_s$ leading to state $S_{03}'$ where $s = 0$. Thus $s_0 = s_0'|_{s = 1, \bar{v}}$ for $s_0' = S_{00}'S_{01}'S_{02}'S_{03}'$.  

On the other hand, if in the initial segment $s_0$ of $\sigma$ we have that $k_0 > 0$, then we know by construction of $M'$ that in its initial state $S_{00}'$ event $e_\mathit{firstB}$ will be triggered, and that there is a trace of $M'$ in which it holds that the value of $\bar{w}$ in $S_{0m}'$ for each $1 \leq m \leq k_0+1$ coincides with the value of $\bar{v}$ in $S_{0m-1}$. Moreover, we know that an event of the form $e_{ia}$ is then triggered in each of these states except for the last one $S_{0k_0+1}$. Consequently, in state $S_{0k_0+1}'$ the list $l$ is of length $k_0$ and the element in the $m$-th position of $l$ corresponds to the value taken by $\bar{v}$ in state $S_{0m-1}$. Furthermore, the state $S_{0k_0+1}'$ in this traceof $M'$ will necessarily trigger an event of the form $e_{ib}$, resulting in a state $S_{0k_0+2}'$ where $s = 1$ and the value of $\bar{v}$ corresponds to its value in $S_{00}$. Clearly, the values that $\bar{v}$ will take in the next $k_0$ states following $S_{0k_0+2}'$ in this trace of $M'$ will correspond to its values in $S_{01}, \ldots S_{0k_0}$, respectively. In the last of these states, $M'$ will trigger event $e_s$ leading to a further state where $s$ takes again he value $0$. Thus $s_0 = s_0'|_{s = 1, \bar{v}}$, where $s_0'$ is the trace of $M'$ of length $2k_0+3$ described above.

For the inductive step we assume that $j + 1 > 0$. The inductive hypothesis gives us that for the segments  $s_j$ and $s_j'$ of $M$ and $M'$, respectively, it holds that $s_j = s_j'|_{s = 1, \bar{v}}$. It is easy to see that the values of $\bar{w}$ in the last state of $S_{jk_j'}'$ of $s_j'$ coincide with the value of $\bar{v}$ in the last state $S_{jk_j}$ of $s_j$. Since furthermore $s = 0$ in $S_{jk_j'}'$, we can use a similar argument as before to show that $s_{j+1} = s_{j+1}'|_{s = 1, \bar{v}}$. We omit the tedious but simple technical details.

Likewise, we can prove (by induction on $j$) that for every segment $s_j'$ of a trace $\sigma'$ of $M'$ there is a corresponding segment $s_j$ of a trace $\sigma$ of $M$ such that $s_j = s_j'|_{s=1,\bar{v}}$. Again, we omit the tedious, but simple technical details.\qed


\proof[Proof (of Lemma \ref{lem-conv1}).]
We slightly modify the construction of $M'$ from the proof of Lemma \ref{lem-conv}. As long as $s=0$ holds, we also count the number of steps using the state variable $c$ for the step counter. Then let $k$ be the constant in the definition of $\Conv^1(\varphi)$. When the value of $c$ exceeds $k$, $M'$ sets $s$ to 1 and resets $c$ to 0. This guarantees that traces of $M$ with infinitely many states satisfying $\varphi$ are properly simulated by $M'$.

Then the proof of (ii) remains the same. For the proof of (i) we need to consider only the one trace defining the local convergence property. For this trace we obtain a sequence of events. As there is no infinite sequence of states all satisfying $\varphi$ the value of the variant term strictly decreases until a state satisfying $\neg\varphi$ is reached. As the variant term is defined by the length of this sequence we get the validity of $\mathit{var}_c^1(M, t, \varphi)$.\qed


\proof[Proof (of Lemma \ref{lem-div}).]
As $M \models \Div(\varphi)$ holds, in any trace $\sigma$ there can only be finitely many states satisfying $\neg\varphi$. Consequently, for each state variable $v_i$ there are only finitely many values that $v_i$ can take in any state satisfying $\neg\varphi$ of any trace. Let $M_i \in \text{HF}(A)$ be the set of all these values.

We define the machine $M'$ with additional state variables $\bar{c} = (c_0, \ldots, c_m)$ and $\bar{b} = (b_0, \ldots, b_m)$. The initialisation event of $M'$ simply extends the event \textit{init} of $M$ by the assigments $c_i := \emptyset$ and $b_i := M_i$. Then the projection of the initial state $S_0$ of $M'$ to the state variables $\bar{v}$ is the initial state of $M$.

For each event $e$ of $M$ with guard $G_e(\bar{x}, \bar{v})$ and action $A_e(\bar{x}, \bar{v}, \bar{v}')$ we define two events $e_+$ and $e_-$ of $M'$. 
The guard of $e_+$ is $G_{e_+}(\bar{x}, \bar{v}, \bar{c}, \bar{b}) = G_e(\bar{x}, \bar{v}) \wedge \varphi(\bar{v})$, and the action $A_{e_+}(\bar{x}, \bar{v}, \bar{c}, \bar{b}, \bar{v}', \bar{c}', \bar{b}')$ is the same as $A_e(\bar{x}, \bar{v}, \bar{v}')$.
The guard of $e_-$ is $G_{e_-}(\bar{x}, \bar{v}, \bar{c}, \bar{b}) = G_e(\bar{x}, \bar{v}) \wedge \neg\varphi(\bar{v})$, and the action $A_{e_-}(\bar{x}, \bar{v}, \bar{c}, \bar{b}, \bar{v}', \bar{c}', \bar{b}')$ results from adding the assignments $c_i := c_i \cup \{ v_i^\prime \}$ for all $i=1,\dots,m$ to the action $A_e(\bar{x}, \bar{v}, \bar{v}')$.

Regardless if $M'$ is in a state satisfying $\varphi$ or in a state satisfying $\neg\varphi$, the projection of the action of either $e_+$ or $e_-$ to the state variables $\bar{v}$ yields the action of $e$. It follows that property (ii) of the lemma holds.

For property (i) we define the variant term $t(\bar{v}, \bar{c}, \bar{b}) = (b_0 - c_0, \dots, b_m - c_m)$. The values of the state variables $\bar{b}$ are never updated, and the values of $\bar{c}$ are only updated by events $e_-$, i.e. only in states satisfying $\neg\varphi$. This implies that
\begin{equation}\label{eq-div-3}
\varphi(\bar{v}) \wedge G_e(\bar{x}, \bar{v}, \bar{c}, \bar{b}) \wedge P_{A_e}(\bar{x}, \bar{v}, \bar{c}, \bar{b}, \bar{v}', \bar{c}', \bar{b}') \rightarrow t(\bar{v}', \bar{c}', \bar{b}') = t(\bar{v}, \bar{c}, \bar{b})
\end{equation}
holds for all events $e$ of $M'$ and all $\bar{x}$, $\bar{v}$, $\bar{c}$, $\bar{b}$, $\bar{v}'$, $\bar{c}'$, $\bar{b}'$.

As every event $e_-$ inserts additional values into the sets $c_i$, the sequence of values for the tuple $(c_0, \dots, c_m)$ is strictly increasing in states satisfying $\neg\varphi$. We cannot have equality, because otherwise we would have a state $S_i$ satisfying $\neg\varphi$ with a possible successor state $S_{i+1} = S_i$ resulting from firing some event $e_-$. Then this event could be fired again and again, thus leading to a trace with an infinite subsequence of states satisfying $\neg\varphi$ contradicting $M \models \Div(\varphi)$. Then the sequence of tuples $(b_0 - c_0, \dots, b_m - c_m)$ is strictly decreasing in states satisfying $\neg\varphi$, i.e. 
\begin{equation}\label{eq-div-1}
\neg\varphi(\bar{v}) \wedge G_e(\bar{x}, \bar{v}, \bar{c}, \bar{b}) \wedge P_{A_e}(\bar{x}, \bar{v}, \bar{c}, \bar{b}, \bar{v}', \bar{c}', \bar{b}') \rightarrow t(\bar{v}', \bar{c}', \bar{b}') < t(\bar{v}, \bar{c}, \bar{b})
\end{equation}
holds for all events $e$ of $M'$ and all $\bar{x}$, $\bar{v}$, $\bar{c}$, $\bar{b}$, $\bar{v}'$, $\bar{c}'$, $\bar{b}'$. 

Furthermore, the choice of the constants $M_i$ stored in the state variable $b_i$ ensures that always $c_i \subsetneq b_i$ holds, which implies 
\begin{equation}\label{eq-div-2}
\neg\varphi(\bar{v}) \wedge G_e(\bar{x}, \bar{v}, \bar{c}, \bar{b}) \wedge P_{A_e}(\bar{x}, \bar{v}, \bar{c}, \bar{b}, \bar{v}', \bar{c}', \bar{b}') \rightarrow t(\bar{v}, \bar{c}, \bar{b}) \neq \emptyset
\end{equation}
for all events $e$ of $M'$ and all $\bar{x}$, $\bar{v}$, $\bar{c}$, $\bar{b}$, $\bar{v}'$, $\bar{c}'$, $\bar{b}'$. 

Finally, (\ref{eq-div-1}), (\ref{eq-div-2}) and (\ref{eq-div-3}) together are equivalent to claim (i), which concludes the proof.\qed


\proof[Proof (of Lemma \ref{lem-div1}).]
The refined machine $M'$ is constructed in the same way as in the proof of Lemma \ref{lem-div} with the only difference that the set values $b_i$ are taken from the one trace with a tail of states satisfying $\varphi$. The ld-variant term $t$ is also defined in the same way. Then we get that conditions (\ref{eq-div-1}), (\ref{eq-div-2}) and (\ref{eq-div-3}) hold for the states in one trace, hence $\mathit{var}_d^1(M', t, \varphi)$ is valid for $M'$.\qed


\proof[Proof (of Lemma \ref{lem-reach}).]
For e-variants the construction of $M'$ for is analogous to the construction in the proof of Lemma \ref{lem-conv} with the difference that an additional Boolean-valued state variable $b$ is used, and there is only once a switch from $s=0$ to $s=1$, and in this transition also $b$ is set to $1$. When the list $\ell$ has been emptied, $s$ is set to $2$ and $M'$ continues in the same way as $M$ ignoring the additional state variables. Then the sought e-variant term is $(t_1, b)$, where $t_1$ is the length of $\ell$. The detailed arguments are then exactly the same as in the proof of Lemma \ref{lem-conv}.

For e1-variants the construction of $M'$ for is analogous to the construction in the proof of Lemma \ref{lem-conv1} with the same modifications as in the proof of Lemma \ref{lem-reach} exploiting the constant $k$ that is guaranteed by $M \models \lozenge^1 \varphi$. $M'$ uses an additional Boolean-valued state variable $b$, and there is only once a switch from $s=0$ to $s=1$, and in this transition also $b$ is set to $1$. When the list $\ell$ has been emptied, $s$ is set to $2$ and $M'$ continues in the same way as $M$ ignoring the additional state variables. Then the sought e1-variant term is $(t_1, b)$, where $t_1$ is the length of $\ell$. The detailed arguments are then exactly the same as in the proof of Lemma \ref{lem-conv1}.\qed


\proof[Proof (of Lemma \ref{lem-simultaneous}).]
We construct $M'$ in the same way as in the proof on Lemma \ref{lem-div} with the only difference that the values $b_0, \dots, b_m$ are the maximum set values in traces satisfying the selection condition $\langle \psi \rangle$. For these traces $S_0, S_1, \dots$ there exists a $k$ such that $S_\ell \models \psi$ holds for all $\ell \ge k$, i.e. there are only finitely many states $S_i$ satisfying $\neg\psi$ and hence the maximum values $b_0, \dots, b_m$ exist. 

Then with the simultaneous cd-variant term $t(\bar{v}, \bar{c}, \bar{b}) = (b_0 - c_0, \dots, b_m - c_m)$ we can use  the same arguments as in the proof of Lemma \ref{lem-div} to show the required conditions (i) and (ii).\qed


\proof[Proof (of Lemma \ref{lem-conditional}).]
The construction of $M'$ is basically the same as in the proof of Lemma \ref{lem-conv} with the following modifications:

\begin{itemize}

\item We add a Boolean-valued state variable $b$. In the case that $\theta$ is $\varphi \wedge \neg\psi \rightarrow \chi$ we initialise $b$ as 0, if $\theta$ holds in the initial state of $M$, otherwise as 1. In the other two cases for $\theta$ we initialise $b$ as 1 iff $M$ satisfies $\Conv(\neg\psi)$.

\item As long as $M'$ is in a state satisfying $b=0 \wedge \theta$, $M'$ executes the events of $M$ directly. A switch from $b=0$ to $b=1$ is added to every event of $M'$, if the invariant $\theta$ is violated.

\item With a switch to $b=1$ $M'$ operates in exactly the same way as constructed in the proof of Lemma \ref{lem-conv}.

\end{itemize}

\noindent
Then the claims (i) and (ii) follow using the same arguments as in the proof of Lemma \ref{lem-conv}.\qed



\section{Nested Temporal Operators}\label{appB}

In Section \ref{sec:fragment} we dealt with sound derivation rules for TREBL formulae of the form $X \varphi$, $Y \varphi$, $XY \varphi$, $YX \varphi$ and $X (\psi \rightarrow Y \varphi)$ with $X \in \{ \square, \square^1 \}$, $Y \in \{ \lozenge, \lozenge^1 \}$, and $\varphi, \psi \in \mathcal{L}$, but the definition of TREBL also allows formulae with $\varphi \in \mathcal{LT}$, which we will address now in this appendix.

For arbitrary $\varphi \in \mathcal{LT}$---in fact, we could even consider $\varphi \in \mathcal{L}^{ext}$---the equivalences $\square\square\varphi \leftrightarrow \square\varphi$, $\square^1\square\varphi \leftrightarrow \square\varphi$, $\square\square^1\varphi \leftrightarrow \square\varphi$ and $\square^1\square^1\varphi \leftrightarrow \square^1\varphi$ are easily proven, which implies (for $X_i \in \{ \square, \square^1 \}$ for $1 \le i \le m$)
\[ X_1 \dots X_m \varphi \leftrightarrow \begin{cases} \square \varphi &\text{if}\; X_i = \square \;\text{holds for at least one}\; i \\
\square^1 \varphi &\text{if}\; X_i = \square^1 \;\text{holds for all}\; i \end{cases} \; . \]

By duality we further get (for $Y_i \in \{ \lozenge, \lozenge^1 \}$ for $1 \le i \le m$)
\[ Y_1 \dots Y_m \varphi \leftrightarrow \begin{cases} \lozenge \varphi &\text{if}\; Y_i = \lozenge \;\text{holds for all}\; i \\
\lozenge^1 \varphi &\text{if}\; Y_i = \lozenge^1 \;\text{holds for at least one}\; i \end{cases} \; . \]

For alternating sequences of temporal operators the following equivalences are likewise easily proven:
\begin{alignat*}{4}
\square\lozenge\square\varphi &\leftrightarrow \lozenge\square\varphi \quad&\quad
\square\lozenge\square^1\varphi &\leftrightarrow \lozenge\square^1\varphi \quad&\quad
\lozenge\square\lozenge\varphi &\leftrightarrow \square\lozenge\varphi \quad&\quad
\lozenge\square\lozenge^1\varphi &\leftrightarrow \square\lozenge^1\varphi \\
\lozenge^1\square^1\lozenge^1\varphi &\leftrightarrow \square^1\lozenge^1\varphi &
\lozenge^1\square^1\lozenge\varphi &\leftrightarrow \square^1\lozenge\varphi &
\square^1\lozenge^1\square^1\varphi &\leftrightarrow \lozenge^1\square^1\varphi &\quad
\square^1\lozenge^1\square\varphi &\leftrightarrow \lozenge^1\square\varphi \\
\square\lozenge^1\square^1\varphi &\leftrightarrow \lozenge\square^1\varphi &
\square\lozenge^1\square\varphi &\leftrightarrow \lozenge\square\varphi &
\lozenge\square^1\lozenge^1\varphi &\leftrightarrow \square\lozenge^1\varphi &
\lozenge\square^1\lozenge\varphi &\leftrightarrow \square\lozenge\varphi \\
\lozenge^1\square\lozenge\varphi &\leftrightarrow \square^1\lozenge\varphi &
\lozenge^1\square\lozenge^1\varphi &\leftrightarrow \square^1\lozenge^1\varphi &
\square^1\lozenge\square\varphi &\leftrightarrow \lozenge^1\square\varphi &\quad
\square^1\lozenge\square^1\varphi &\leftrightarrow \lozenge^1\square^1\varphi
\end{alignat*}

\noindent
We can consider all these eqivalences above as a confluent set of term rewriting rules (with the longer sequence of temporal operators on the left). Then any TREBL formula of the form $X_1 \dots X_m \varphi$ with $X_i \in \{ \square, \square^1, \lozenge, \lozenge^1 \}$ has a unique normal form $N(X_1 \dots X_m) \varphi$, where $N(X_1 \dots X_m)$ is a sequence of at most two temporal operators. As we are dealing with equivalences, we get a sound derivation rule
\begin{center}
\begin{minipage}{14.5cm}
\begin{prooftree}
\AxiomC{$M \vdash N(X_1 \dots X_m) \,\varphi$}
\LeftLabel{$\mathrm{N}$: \;}
\UnaryInfC{$M \vdash X_1 \dots X_m \,\varphi$}
\end{prooftree}
\end{minipage}
\end{center}

\noindent
It remains to investigate TREBL formulae, in which at least once a constructor for a progress formula is used. First consider formulae of the form $X_1 \dots X_n (\varphi \rightarrow Y_1 \dots Y_m \psi)$ with $X_n \in \{ \square, \square^1 \}$ and $Y_1 \in \{ \lozenge, \lozenge^1 \}$. The equivalences above imply directly that for $X_i, Y_j \in \{ \square, \square^1 , \lozenge, \lozenge^1 \}$, $\varphi \in \mathcal{L}$ and $\psi \in \mathcal{LT}$ the following derivation rule is sound:
\begin{center}
\begin{minipage}{14.5cm}
\begin{prooftree}
\AxiomC{$M \vdash N(X_1 \dots X_n) (\varphi \rightarrow N(Y_1 \dots Y_m) \psi)$}
\LeftLabel{$\mathrm{NPR}$: \;}
\UnaryInfC{$M \vdash X_1 \dots X_n (\varphi \rightarrow Y_1 \dots Y_m \psi)$}
\end{prooftree}
\end{minipage}
\end{center}

\subsection{Prefixed Variants}

With the results above we look at TREBL formulae of the form $\square (\varphi \rightarrow XY \psi)$ and $\square (\varphi \rightarrow YX \psi)$ with $X \in \{ \square, \square^1 \}$, $Y \in \{ \lozenge, \lozenge^1 \}$ and $\varphi, \psi \in \mathcal{L}$. In all cases, if $\tau = S_0, S_1, \dots$ is a trace of $M$ let $k$ be minimal with $S_k \models \varphi$---if no such $k$ exists, let $k = \omega$. 

Clearly, for $k = \omega$ we have $S_i \models \varphi \rightarrow XY \psi$ for all $i$, and for $k < \omega$ we have $S_i \models \varphi \rightarrow XY \psi$ for $i \ge k$ iff $S_i \models XY \psi$. That is, every trace has a prefix, in which the required condition trivially holds, while for the remainder of the trace it suffices to show that $XY \psi$ holds. 

Therefore, with a Boolean-valued state variable $b$ we can exploit the prefix formula $\text{pre}(\neg\varphi, b)$ defined in (\ref{eq-prefix}), and define {\em prefixed variant formulae} by modification of c-variants and lc-variants as defined in (\ref{eq-conv}) and (\ref{eq-conv1}):
\begin{gather}
\text{pvar}_c(\varphi, (t,b), \neg\psi) \equiv \text{pre}(\neg\varphi, b) \wedge \text{var}_c(t, \psi \wedge b=1) \;\text{and} \notag\\
\text{pvar}_c^1(\varphi, (t,b), \neg\psi) \equiv \text{pre}(\neg\varphi, b) \wedge \text{var}_c^1(t, \psi \wedge b=1)
\label{eq-pvar-conv}
\end{gather}

\noindent
For TREBL formulae of the form $\square (\varphi \rightarrow XY \psi)$ we obtain sound derivation rules by simple modifications of the rules E, E$^{11}$, E$^{10}$ and E$^{01}$:
\begin{center}
\begin{minipage}{14.5cm}
\begin{prooftree}
\AxiomC{$M \vdash \text{pvar}_c(\varphi, (t,b), \neg \psi)$}
\AxiomC{\hspace*{-5mm}$M \vdash \, ( \varphi \leftrightarrow b=1 )$}
\AxiomC{\hspace*{-5mm}$M \vdash \, \square \, \Dlf(\neg \psi \wedge b=1)$}
\LeftLabel{{\rm PRE}: \quad}
\TrinaryInfC{$M \vdash \square (\varphi \rightarrow \square\lozenge \, \psi)$}
\end{prooftree}
\end{minipage}\\
\begin{minipage}{14.5cm}
\begin{prooftree}
\AxiomC{$M \vdash \text{pvar}_c^1(\varphi, (t,b), \neg \psi)$}
\AxiomC{\hspace*{-5mm}$M \vdash \, ( \varphi \leftrightarrow b=1 )$}
\LeftLabel{{\rm PRE$^{11}$}: \quad}
\BinaryInfC{$M \vdash \square (\varphi \rightarrow \square^1\lozenge^1 \, \psi)$}
\end{prooftree}
\end{minipage}\\
\begin{minipage}{14.5cm}
\begin{prooftree}
\AxiomC{$M \vdash \text{pvar}_c^1(\varphi, (t,b), \neg \psi)$}
\AxiomC{\hspace*{-5mm}$M \vdash \, ( \varphi \leftrightarrow b=1 )$}
\AxiomC{\hspace*{-5mm}$M \vdash \, \square \, \Dlf(\neg \psi \wedge b=1)$}
\LeftLabel{{\rm PRE$^{01}$}: \quad}
\TrinaryInfC{$M \vdash \square (\varphi \rightarrow \square\lozenge^1 \, \psi)$}
\end{prooftree}
\end{minipage}\\
\begin{minipage}{14.5cm}
\begin{prooftree}
\AxiomC{$M \vdash \text{pvar}_c^1(\varphi, (t,b), \neg \psi)$}
\AxiomC{\hspace*{-5mm}$M \vdash \, ( \varphi \leftrightarrow b=1 )$}
\AxiomC{\hspace*{-5mm}$M \vdash \, \square( b=1 \rightarrow \lozenge \psi )$}
\LeftLabel{{\rm PRE$^{10}$}: \quad}
\TrinaryInfC{$M \vdash \square (\varphi \rightarrow \square^1\lozenge \, \psi)$}
\end{prooftree}
\end{minipage}
\end{center}

\noindent
The soundness proofs are the same as in Lemmata \ref{lem-conv-rule} and \ref{lem-conv1-rule}, and the existence of the variant terms follows as in Lemmata \ref{lem-conv} and \ref{lem-conv1} with the small add-on of a prefix of states all satisfying $\neg\varphi$.

Analogously, for TREBL formulae of the form $\square (\varphi \rightarrow YX \psi)$ we obtain {\em prefixed variant formulae} by modification of d-, ld-, pld- and pd-variants as defined in (\ref{eq-div}), (\ref{eq-div1}), (\ref{eq-pdiv1}) and (\ref{eq-pdiv}):
\begin{gather}
\text{pvar}_d(\varphi, (t,b), \psi) \equiv \text{pre}(\neg\varphi, b) \wedge \text{var}_d(t, \psi \wedge b=1) \; , \notag\\
\text{pvar}_d^1(\varphi, (t,b), \psi) \equiv \text{pre}(\neg\varphi, b) \wedge \text{var}_d^1(t, \psi \wedge b=1) \; , \notag\\
\text{pvar}_d^{p1}(\varphi, (t,b), \psi) \equiv \text{pre}(\neg\varphi, b) \wedge \text{var}_d^{p1}(t, \psi \wedge b=1) \;\text{and} \notag\\
\text{pvar}_d^p(\varphi, (t,b), \psi) \equiv \text{pre}(\neg\varphi, b) \wedge \text{var}_d^p(t, \psi \wedge b=1)
\label{eq-pvar-div}
\end{gather}

\noindent
We then obtain the following sound derivation rules by simple modifications of the rules P, P$^{11}$, P$^{01}$ and P$^{10}$:
\begin{center}
\begin{minipage}{14.5cm}
\begin{prooftree}
\AxiomC{$M \vdash \, \text{pvar}_d(\varphi, (t,b), \psi)$}
\AxiomC{\hspace*{-5mm}$M \vdash \, ( \varphi \leftrightarrow b=1 )$}
\AxiomC{\hspace*{-5mm}$M \vdash \, \square \, \Dlf(b=1 \wedge \neg\psi)$}
\LeftLabel{$\mathrm{PRP}$: \quad}
\TrinaryInfC{$M \vdash \square (\varphi \rightarrow \lozenge \square \, \psi)$}
\end{prooftree}
\end{minipage}\\
\begin{minipage}{14.5cm}
\begin{prooftree}
\AxiomC{$M \vdash \, \text{pvar}_d^1(\varphi, (t,b), \psi)$}
\AxiomC{\hspace*{-5mm}$M \vdash \, ( \varphi \leftrightarrow b=1 )$}
\LeftLabel{$\mathrm{PRP}^{11}$: \quad}
\BinaryInfC{$M \vdash \square (\varphi \rightarrow \lozenge^1 \square^1 \, \psi)$}
\end{prooftree}
\end{minipage}\\
\begin{minipage}{14.5cm}
\begin{prooftree}
\AxiomC{$M \vdash \, \text{pvar}_d^{p1}(\varphi, (t,b), \psi)$}
\AxiomC{\hspace*{-5mm}$M \vdash \, ( \varphi \leftrightarrow b=1 )$}
\LeftLabel{$\mathrm{PRP}^{01}$: \quad}
\BinaryInfC{$M \vdash \square (\varphi \rightarrow \lozenge \square^1 \, \psi)$}
\end{prooftree}
\end{minipage}\\
\begin{minipage}{14.5cm}
\begin{prooftree}
\AxiomC{$M \vdash \, \text{pvar}_d^p(\varphi, (t,b), \psi)$}
\AxiomC{\hspace*{-5mm}$M \vdash \, ( \varphi \leftrightarrow b=1 )$}
\LeftLabel{$\mathrm{PRP}^{10}$: \quad}
\BinaryInfC{$M \vdash \square (\varphi \rightarrow \lozenge^1 \square \, \psi)$}
\end{prooftree}
\end{minipage}    
\end{center}

\noindent
The soundness proofs are the same as in Lemmata \ref{lem-div-rule}, \ref{lem-div1-rule}, \ref{lem-divp1-rule} and \ref{lem-divp-rule}, and the existence of the variant terms follows as in Lemmata \ref{lem-div}, \ref{lem-div1} and \ref{lem-pdiv} with the small add-on of a prefix of states all satisfying $\neg\varphi$.

\subsection{Prefixes in a Single Trace}

For formulae of the form $\square^1 (\varphi \rightarrow XY \psi)$ we proceed analogously. The only difference is that need to consider only a prefix with states satisfying $\neg\varphi$ of one trace, which is captured by a sentence $\text{pre}^1(\varphi, b)$ defined as
\begin{gather*}
\forall \bar{v} \Big( \varphi(\bar{v}) \rightarrow \bigvee_{e_i \in E} \exists \bar{x} \bar{v}^\prime \big( G_i(\bar{x}, \bar{v}) \wedge P_{A_i}(\bar{x}, \bar{v}, \bar{v}^\prime) \wedge b^\prime = b \big) \\
\wedge \Big( \neg\varphi(\bar{v}) \rightarrow \bigwedge_{e_i \in E} \forall \bar{x} \bar{v}^\prime \big( G_i(\bar{x}, \bar{v}) \wedge P_{A_i}(\bar{x}, \bar{v}, \bar{v}^\prime) \rightarrow b^\prime = 1 \big) \Big)
\end{gather*}

\noindent
Then we get {\em locally prefixed variant formulae} again by modification of c-variants and lc-variants as defined in (\ref{eq-conv}) and (\ref{eq-conv1}):
\begin{gather}
\text{lpvar}_c(\varphi, (t,b), \neg\psi) \equiv \text{pre}^1(\neg\varphi, b) \wedge \text{var}_c(t, \psi \wedge b=1) \;\text{and} \notag\\
\text{lpvar}_c^1(\varphi, (t,b), \neg\psi) \equiv \text{pre}^1(\neg\varphi, b) \wedge \text{var}_c^1(t, \psi \wedge b=1)
\label{eq-pvar-conv2}
\end{gather}

\noindent
Thise gives rise to the following two sound derivation rules:
\begin{center}
\begin{minipage}{14.5cm}
\begin{prooftree}
\AxiomC{$M \vdash \text{lpvar}_c(\varphi, (t,b), \neg \psi)$}
\AxiomC{\hspace*{-5mm}$M \vdash \, ( \varphi \leftrightarrow b=1 )$}
\AxiomC{\hspace*{-5mm}$M \vdash \, \square \, \Dlf(\neg \psi \wedge b=1)$}
\LeftLabel{{\rm PR$^1$E$^{00}$}: \quad}
\TrinaryInfC{$M \vdash \square^1 (\varphi \rightarrow \square\lozenge \, \psi)$}
\end{prooftree}
\end{minipage}\\
\begin{minipage}{14.5cm}
\begin{prooftree}
\AxiomC{$M \vdash \text{lpvar}_c^1(\varphi, (t,b), \neg \psi)$}
\AxiomC{\hspace*{-5mm}$M \vdash \, ( \varphi \leftrightarrow b=1 )$}
\AxiomC{\hspace*{-5mm}$M \vdash \, \square \, \Dlf(\neg \psi \wedge b=1)$}
\LeftLabel{{\rm PR$^1$E$^{01}$}: \quad}
\TrinaryInfC{$M \vdash \square^1 (\varphi \rightarrow \square\lozenge^1 \, \psi)$}
\end{prooftree}
\end{minipage}
\end{center}

\noindent
Again, the soundness proofs are the same as in Lemmata \ref{lem-conv-rule} and \ref{lem-conv1-rule}, and the existence of the variant terms follows as in Lemmata \ref{lem-conv} and \ref{lem-conv1} with the small add-on of a prefix of states all satisfying $\neg\varphi$.

The remaining two cases are simpler. If there exists a trace $S_0, S_1, \dots$ with $S_i \models \neg\varphi$ for all $i$, then the progress formula $\square^1 (\varphi \rightarrow XY \psi)$ is trivially satisfied for all $X \in \{ \square, \square^1 \}$ and $Y \in \{ \lozenge, \lozenge^1 \}$. If all traces contain a state $S_i$ with $S_i \models \varphi$, then the condition $\varphi$ can be ignored. Therefore, the following derivation rules are sound:
\begin{center}
\begin{minipage}{14.5cm}
\begin{prooftree}
\AxiomC{$M \vdash \square^1 \neg\varphi$}
\LeftLabel{{\rm PR$^1$E}: \quad}
\UnaryInfC{$M \vdash \square^1 (\varphi \rightarrow X Y \, \psi)$}
\end{prooftree}
\end{minipage}\\
\begin{minipage}{14.5cm}
\begin{prooftree}
\AxiomC{$M \vdash \lozenge \varphi$}
\AxiomC{\hspace*{-5mm}$M \vdash \, \square^1 \lozenge \, \psi $}
\LeftLabel{{\rm PR$^1$E$^{10}$}: \quad}
\BinaryInfC{$M \vdash \square^1 (\varphi \rightarrow \square^1 \lozenge \, \psi)$}
\end{prooftree}
\end{minipage}\\
\begin{minipage}{14.5cm}
\begin{prooftree}
\AxiomC{$M \vdash \lozenge \varphi$}
\AxiomC{\hspace*{-5mm}$M \vdash \, \square^1 \lozenge^1 \, \psi $}
\LeftLabel{{\rm PR$^1$E$^{11}$}: \quad}
\BinaryInfC{$M \vdash \square^1 (\varphi \rightarrow \square^1 \lozenge^1 \, \psi)$}
\end{prooftree}
\end{minipage}
\end{center}

\noindent
For formulae of the form $\square^1 (\varphi \rightarrow Y X \psi)$ we obtain sound derivation rules in the same way.   If there exists a trace $S_0, S_1, \dots$ with $S_i \models \neg\varphi$ for all $i$, then the progress formula $\square^1 (\varphi \rightarrow YX \psi)$ is trivially satisfied for all $X \in \{ \square, \square^1 \}$ and $Y \in \{ \lozenge, \lozenge^1 \}$. If all traces contain a state $S_i$ with $S_i \models \varphi$, then for $Y = \lozenge^1$ the condition $\varphi$ can be ignored. Therefore, the following derivation rules are sound:
\begin{center}
\begin{minipage}{14.5cm}
\begin{prooftree}
\AxiomC{$M \vdash \square^1 \neg\varphi$}
\LeftLabel{{\rm PR$^1$P}: \quad}
\UnaryInfC{$M \vdash \square^1 (\varphi \rightarrow Y X \, \psi)$}
\end{prooftree}
\end{minipage}\\
\begin{minipage}{14.5cm}
\begin{prooftree}
\AxiomC{$M \vdash \lozenge \varphi$}
\AxiomC{\hspace*{-5mm}$M \vdash \, \lozenge^1 \square \, \psi $}
\LeftLabel{{\rm PR$^1$P$^{10}$}: \quad}
\BinaryInfC{$M \vdash \square^1 (\varphi \rightarrow \lozenge^1 \square \, \psi)$}
\end{prooftree}
\end{minipage}\\
\begin{minipage}{14.5cm}
\begin{prooftree}
\AxiomC{$M \vdash \lozenge \varphi$}
\AxiomC{\hspace*{-5mm}$M \vdash \, \lozenge^1 \square^1 \, \psi $}
\LeftLabel{{\rm PR$^1$P$^{11}$}: \quad}
\BinaryInfC{$M \vdash \square^1 (\varphi \rightarrow \lozenge^1 \square^1 \, \psi)$}
\end{prooftree}
\end{minipage} 
\end{center}

\noindent
This leaves the cases with $Y = \lozenge$. We get {\em locally prefixed variant formulae} again by modification of d-variants and pld-variants as defined in (\ref{eq-div}) and (\ref{eq-pdiv1}):
\begin{gather}
\text{lpvar}_d(\varphi, (t,b), \neg\psi) \equiv \text{pre}^1(\neg\varphi, b) \wedge \text{var}_d(t, \psi \wedge b=1) \;\text{and} \notag\\
\text{lpvar}_d^{p1}(\varphi, (t,b), \neg\psi) \equiv \text{pre}^1(\neg\varphi, b) \wedge \text{var}_d^{p1}(t, \psi \wedge b=1)
\label{eq-pvar-conv3}
\end{gather}

\noindent
Thise gives rise to the following two sound derivation rules:
\begin{center}
\begin{minipage}{14.5cm}
\begin{prooftree}
\AxiomC{$M \vdash \text{lpvar}_d(\varphi, (t,b), \neg \psi)$}
\AxiomC{\hspace*{-5mm}$M \vdash \, ( \varphi \leftrightarrow b=1 )$}
\AxiomC{\hspace*{-5mm}$M \vdash \, \square \, \Dlf(\neg \psi \wedge b=1)$}
\LeftLabel{{\rm PR$^1$P$^{00}$}: \quad}
\TrinaryInfC{$M \vdash \square^1 (\varphi \rightarrow \lozenge\square \, \psi)$}
\end{prooftree}
\end{minipage}\\
\begin{minipage}{14.5cm}
\begin{prooftree}
\AxiomC{$M \vdash \text{lpvar}_d^{p1}(\varphi, (t,b), \neg \psi)$}
\AxiomC{\hspace*{-5mm}$M \vdash \, ( \varphi \leftrightarrow b=1 )$}
\LeftLabel{{\rm PR$^1$P$^{01}$}: \quad}
\BinaryInfC{$M \vdash \square^1 (\varphi \rightarrow \lozenge\square^1 \, \psi)$}
\end{prooftree}
\end{minipage}
\end{center}

\noindent
Again, the soundness proofs are the same as in Lemmata \ref{lem-div-rule} and \ref{lem-div1-rule}, and the existence of the variant terms follows as in Lemmata \ref{lem-div} and \ref{lem-pdiv} with the small add-on of a prefix of states all satisfying $\neg\varphi$.

\subsection{Prefixes for Simultaneous and Conditional Variants}

The rules above are for the derivation of TREL formulae, where we use invariance, existence, persistence and reachability constructors inside a progress rule constructor. We now investigate rules for formulae, where a progress rule constructor is used inside invariance, existence, persistence and reachability constructors. Due to rule NPR we have to consider formulae of the form $\lozenge X (\varphi \rightarrow Y \psi)$ or $\lozenge^1 X (\varphi \rightarrow Y \psi)$ with $X \in \{ \square, \square^1 \}$ and $Y \in \{ \lozenge, \lozenge^1 \}$. 

In the former case for all traces $S_0, S_1, \dots$ there exists some $k$ such that $S_k \models X (\varphi \rightarrow Y \psi)$. In the latter case there needs to exist just one such trace. For $X (\varphi \rightarrow Y \psi)$ we have discovered the sound derivation rules PR, PR$^{11}_j$, PR$^{01}$ and PR$^{10}_j$ ($1 \le j \le 3$). In addition we need to consider a prefix, i.e. $S_0, \dots, S_{k-1}$, and this prefix may contain only finitely many states $S_i$ with $S_i \models \varphi$. 

That is, the prefix---in the obvious sense---behaves like a trace satisfying $\Div(\neg\varphi)$. We therefore consider a term $t$ taking values in some well-founded set $V$ and a Boolean-valued state variable $b$ and define the sentence $\text{pre}(\varphi, (t,b))$ as
\begin{gather*}
  \forall \bar{v} \bigwedge_{e_i \in E} \forall \bar{x} \bar{v}^\prime \Big( G_i(\bar{x}, \bar{v}) \wedge P_{A_i}(\bar{x}, \bar{v}, \bar{v}^\prime) \rightarrow \big( b=1 \rightarrow b^\prime = 1 \big) \Big) \wedge \\
  \big( \varphi(\bar{v}) \wedge b=0 \rightarrow t(\bar{v}) \neq 0 \wedge t(\bar{v}^\prime) < t(\bar{v}) \big)
  \wedge \big( \neg\varphi(\bar{v}) \wedge b=0 \rightarrow t(\bar{v}^\prime) \le t(\bar{v}) \big)
  \end{gather*}

\noindent
With this prefix formula we can then define {\em prefixed simultaneous variant formulae}
$\text{psvar}_{cd}(\varphi$, $(t^\prime,b),t,(\psi,\chi))$
and (with $\theta$ as in (\ref{eq-cvar2}), (\ref{eq-cvar3}) and (\ref{eq-cvar1}))
{\em prefixed conditional variant formulae}
$\text{pcvar}_c(\varphi$, $(t^\prime,b),t,\psi,\theta)$
as follows:
\begin{align}
\text{psvar}_{cd}(\varphi, (t^\prime,b), t, (\psi, \chi))
  &\equiv \text{pre}(\varphi, (t^\prime,b)) \wedge \text{svar}_{cd}(t, (\psi \wedge b=1, \chi))
  \label{eq-psvar} \\
\text{pcvar}_c(\varphi, (t^\prime,b), t, \psi, \theta)
  &\equiv \text{pre}(\varphi, (t^\prime,b)) \wedge \text{cvar}_c(t, \psi \wedge b=1, \theta)
  \label{eq-pcvar}
\end{align}

\noindent
Now we only need to replace the simultaneous and conditional variant formulae in the antecedents of the rules PR, PR$^{11}_j$, PR$^{01}$ and PR$^{10}_j$ ($1 \le j \le 3$) by their prefixed versions (\ref{eq-psvar}) and (\ref{eq-pcvar}), respectively, which gives us derivation rules PPR, PPR$^{11}_j$, PPR$^{01}$ and PPR$^{10}_j$ ($1 \le j \le 3$) for the derivation of TREBL formulae $\lozenge\square(\varphi \rightarrow \lozenge \psi)$, $\lozenge\square^1(\varphi \rightarrow \lozenge^1 \psi)$, $\lozenge\square(\varphi \rightarrow \lozenge^1 \psi)$ and $\lozenge\square^1(\varphi \rightarrow \lozenge \psi)$, respectively.

Using the same arguments as in the proofs of Lemmata \ref{lem-progress}, \ref{lem-progress10}, \ref{lem-progress01}, \ref{lem-progress11} in combination with the proofs of Lemmata \ref{lem-svar} and \ref{lem-cvar} we can show the soundness of these derivation rules. We only need to take into account that the involved prefix formula $\text{pre}(\varphi, (t^\prime,b))$ defines a prefix of traces with states satisfying $b=0$, which contain only finitely many states satisfying $\varphi$ and do not terminate in a state satisfying $\varphi$. Using the same arguments as in the proofs of Lemmata \ref{lem-simultaneous} and \ref{lem-conditional} we can show that prefixed simultaneous and prefixed conditional variants exist in sufficiently refined machines. The only necessary modification is to add a machine for the prefix. We omit the tedious technical details.

For TREBL formulae of the form $\lozenge^1 X (\varphi \rightarrow Y \psi)$ we proceed analogously. Again we need a prefix with only finitely many states satisfying $\varphi$, but such a prefix is only required for one trace and not necessarily for all traces. For this we consider a term $t$ taking values in some well-founded set $V$ and a Boolean-valued state variable $b$ and define the prefix formula $\text{pre}^1(\varphi, (t,b))$ as
\begin{gather*}
  \forall \bar{v} \bigg( \Big( \varphi(\bar{v}) \wedge b=0 \rightarrow t(\bar{v}) \neq 0 \wedge \bigvee_{e_i \in E} \exists \bar{x} \bar{v}^\prime \big( G_i(\bar{x}, \bar{v}) \wedge P_{A_i}(\bar{x}, \bar{v}, \bar{v}^\prime) \wedge t(\bar{v}^\prime) < t(\bar{v}) \big) \Big) \\
  \wedge \bigwedge_{e_i \in E} \forall \bar{x} \bar{v}^\prime \Big( \neg\varphi(\bar{v}) \wedge b=0 \rightarrow \big( G_i(\bar{x}, \bar{v}) \wedge P_{A_i}(\bar{x}, \bar{v}, \bar{v}^\prime) \rightarrow t(\bar{v}^\prime) \le t(\bar{v}) \big) \Big) \\
  \wedge \big( b=1 \rightarrow b^\prime = 1 \big) \bigg)
  \end{gather*}
  
\noindent
With this prefix formula we can define again {\em prefixed simultaneous variant formulae}
$\text{psvar}_{cd}^1(\varphi$, $(t^\prime,b),t,(\psi,\chi))$
and (with $\theta$ as in (\ref{eq-cvar2}), (\ref{eq-cvar3}) and (\ref{eq-cvar1}))
{\em prefixed conditional variant formulae}
$\text{pcvar}_c^1(\varphi$, $(t^\prime,b),t,\psi,\theta)$
as follows:
\begin{align}
\text{psvar}_{cd}^1(\varphi, (t^\prime,b), t, (\psi, \chi))
  &\equiv \text{pre}^1(\varphi, (t^\prime,b)) \wedge \text{svar}_{cd}(t, (\psi \wedge b=1, \chi))
  \label{eq-psvar1} \\
\text{pcvar}_c^1(\varphi, (t^\prime,b), t, \psi, \theta)
  &\equiv \text{pre}^1(\varphi, (t^\prime,b)) \wedge \text{cvar}_c(t, \psi \wedge b=1, \theta)
  \label{eq-pcvar1}
\end{align}

\noindent
Replacing the simultaneous and conditional variant formulae in the antecedents of the rules PR, PR$^{11}_j$, PR$^{01}$ and PR$^{10}_j$ ($1 \le j \le 3$) by their prefixed versions (\ref{eq-psvar1}) and (\ref{eq-pcvar1}), respectively, which gives us derivation rules P$^1$PR, P$^1$PR$^{11}_j$, P$^1$PR$^{01}$ and P$^1$PR$^{10}_j$ ($1 \le j \le 3$) for the derivation of TREBL formulae $\lozenge^1\square(\varphi \rightarrow \lozenge \psi)$, $\lozenge^1\square^1(\varphi \rightarrow \lozenge^1 \psi)$, $\lozenge^1\square(\varphi \rightarrow \lozenge^1 \psi)$ and $\lozenge^1\square^1(\varphi \rightarrow \lozenge \psi)$, respectively. 

The soundness of these rules follows in the same way as in the proofs of Lemmata \ref{lem-progress}, \ref{lem-progress10}, \ref{lem-progress01}, \ref{lem-progress11} in combination with the proofs of Lemmata \ref{lem-svar} and \ref{lem-cvar}.

\begin{rem}

Note that the rules for nested temporal TREBL formulae that we discovered so far involve prefixed variant formulae, which are conjunctions of variant formulae and prefix formulae. Thus, the antecedents of the rules can be split, and rules could be optimised so that prefix formulae just need to be added as additional antecedents. We do not consider such optimisation of the derivation rules in this article.

\end{rem}

\subsection{Nested Progress Operators}

The only remaining cases concern nested formulae, in which at least two progress operators are involved. It suffices to consider formulae of the form $\square (\varphi \rightarrow YX (\chi \rightarrow Y^\prime \psi))$ or $\square^1 (\varphi \rightarrow YX (\chi \rightarrow Y^\prime \psi))$. 

The former case is analogous to case handled above, where a variant of a persistence formula is nested inside a progress formula with the difference that $\chi \rightarrow Y^\prime \psi$ is not a formula in $\mathcal{L}$. Nonetheless, the argument that traces require a prefix of states satisfying $\neg\varphi$ such that the remainder of the trace satisfies $YX (\chi \rightarrow Y^\prime \psi)$ emains the same. If we use a Boolean-valued state variable $b$ to capture such a prefix, it follows that
\begin{center}
\begin{minipage}{14.5cm}
\begin{prooftree}
\AxiomC{$M \vdash \text{pre}( \neg\varphi, b)$}
\AxiomC{\hspace*{-5mm}$M \vdash \, YX ((\chi \wedge b=1) \rightarrow Y^\prime \psi)$}
\AxiomC{\hspace*{-5mm}$M \vdash \, (\varphi \leftrightarrow b=1)$}
\LeftLabel{{\rm PRPR}: \quad}
\TrinaryInfC{$M \vdash \square (\varphi \rightarrow YX (\chi \rightarrow Y^\prime \psi))$}
\end{prooftree}
\end{minipage}
\end{center}
is a sound derivation rule for TREBL formulae of the form $\square (\varphi \rightarrow YX (\chi \rightarrow Y^\prime \psi))$ with $X \in \{ \square, \square^1 \}$ and $Y, Y^\prime \in \{ \lozenge, \lozenge^1 \}$. While $\varphi, \chi \in \mathcal{L}$, we may have $\psi \in \mathcal{LT}$, so multiple nesting of progress operators is supported.

The case of nested TREBL formulae of the form $\square^1 (\varphi \rightarrow YX (\chi \rightarrow Y^\prime \psi))$ can be handled in the same way using a prefix for a single trace rather than all traces. That is, we obtain the following sound derivation rule:
\begin{center}
\begin{minipage}{14.5cm}
\begin{prooftree}
\AxiomC{$M \vdash \text{pre}^1( \neg\varphi, b)$}
\AxiomC{\hspace*{-5mm}$M \vdash \, YX ((\chi \wedge b=1) \rightarrow Y^\prime \psi)$}
\AxiomC{\hspace*{-5mm}$M \vdash \, (\varphi \leftrightarrow b=1)$}
\LeftLabel{{\rm PR$^1$PR}: \quad}
\TrinaryInfC{$M \vdash \square^1 (\varphi \rightarrow YX (\chi \rightarrow Y^\prime \psi))$}
\end{prooftree}
\end{minipage}
\end{center}

\noindent
Concerning the proof of Theorem \ref{thm-complete} we left out the cases of TREBL formulae that result from multiple application of the constructors in Definition \ref{def-trebl}. These are handled here.

\proof[Proof (extension of the proof of Theorem \ref{thm-complete}).]
If only the constructors for invariance, reachability, existence and persistence formulae are applied, we get a formula $\varphi = X_1 \dots X_m \psi$ with $X_i \in \{ \square, \square^1, \lozenge, \lozenge^1 \}$ and $\psi \in \mathcal{L}$. We know that this formula is equivalent to $\varphi^\prime = N(X_1 \dots X_m) \psi$. Therefore, if $M \models \varphi$ holds, then also $M \models \varphi^\prime$ holds. By the induction hypothesis we get $M \vdash_{\mathfrak{R}} \varphi^\prime$, and then the application of rule N gives us $M \vdash_{\mathfrak{R}} \varphi$.

Analogously, a TREBL formula $\varphi = X_1 \dots X_n (\chi Y_1 \dots Y_m \psi)$ with $X_i, Y_j \in \{ \square, \square^1$, $\lozenge, \lozenge^1 \}$ and $\psi, \chi \in \mathcal{L}$---such a formula arises from the application of the constructors for invariance, reachability, existence and persistence plus a single progress operator---is equivalent to a formula $\varphi ^\prime= N(X_1 \dots X_n) (\chi N(Y_1 \dots Y_m) \psi)$. Hence, if $M \models \varphi$ holds, then also $M \models \varphi^\prime$ holds. By the induction hypothesis we get $M \vdash_{\mathfrak{R}} \varphi^\prime$, and then the application of rule NPR gives us $M \vdash_{\mathfrak{R}} \varphi$.

For TREBL formulae of the form $\varphi = \square (\varphi_1 \rightarrow XY \varphi_2)$ with $X \in \{ \square, \square^1 \}$, $Y \in \{ \lozenge, \lozenge^1 \}$ and $\varphi_1, \varphi_2 \in \mathcal{L}$ there exist appropriate prefixed variant terms such that $M$ satisfies the antecedents of the rules PRE, PRE$^{11}$, PRE$^{01}$ or PRE$^{10}$, respectively,  By the induction hypothesis, these antecedents are derivable by rules in $\mathfrak{R}$, and then the application of the corresponding rule yields $M \vdash_{\mathfrak{R}} \varphi$.

The same arguments apply for TREBL formulae of the form $\varphi = \square (\varphi_1 \rightarrow YX \varphi_2)$ with the difference that we use the rules PRP, PRP$^{11}$, PRP$^{01}$ or PRP$^{10}$, respectively. Analogously, for TREBL formulae of the form $\varphi = \square^1 (\varphi_1 \rightarrow XY \varphi_2)$ we use the rules PR$^1$E, PR$^1$E$^{00}$, PR$^1$E$^{11}$, PR$^1$E$^{01}$ or PR$^1$E$^{10}$, respectively, and for TREBL formulae of the form $\varphi = \square^1 (\varphi_1 \rightarrow YX \varphi_2)$ we use the rules PR$^1$P, PR$^1$P$^{00}$, PR$^1$P$^{11}$, PR$^1$P$^{01}$ or PR$^1$P$^{10}$, respectively.

For TREBL formulae of the form $\varphi = \lozenge^1 X (\chi \rightarrow Y \psi)$ with $X \in \{ \square, \square^1 \}$, $Y \in \{ \lozenge, \lozenge^1 \}$ and $\psi, \chi \in \mathcal{L}$ there exist appropriate prefixed simultaneous and conditional variants such that $M$ satisfies the antecedents of the rules P$^1$PR, P$^1$PR$^{01}$, P$^1$PR$^{10}_j$ or P$^1$PR$^{11}_j$ ($1 \le j \le 3$), respectively. By the induction hypothesis, these antecedents are derivable by rules in $\mathfrak{R}$, and then the application of the corresponding rule yields $M \vdash_{\mathfrak{R}} \varphi$.

Finally, for TREBL formulae with multiply nested progress operators we can concentrate on $\varphi = \square (\varphi_1 \rightarrow YX (\varphi_2 \rightarrow Y^\prime \varphi_3))$ or $\varphi = \square^1 (\varphi_1 \rightarrow YX (\varphi_2 \rightarrow Y^\prime \varphi_3))$ with $X \in \{ \square, \square^1 \}$, $Y, Y^\prime \in \{ \lozenge, \lozenge^1 \}$ and $\varphi_i \in \mathcal{L}$. Then the appropriate prefixes exists, and $M$ satisfies the antecedents of the rule PRPR (or PR$^1$PR, respectively). By the induction hypothesis, these antecedents are derivable by rules in $\mathfrak{R}$, and then the application of the rule PRPR (or PR$^1$PR, respectively) yields $M \vdash_{\mathfrak{R}} \varphi$.\qed


\end{document}